\documentclass[11pt, a4paper,onecolumn,accepted=2023-07-26]{quantumarticle}
\pdfoutput=1
\usepackage{revsymb, amsmath, amsfonts, amssymb, enumerate, fullpage, amsthm, graphicx, relsize, bbm, mathrsfs,dsfont}
\usepackage{subcaption,comment,multirow}

\usepackage[linktocpage=true, colorlinks=true, linkcolor=blue, urlcolor=blue, citecolor=blue]{hyperref}

\usepackage{cite}

\usepackage[normalem]{ulem}

\usepackage{stackrel}

\newtheorem{conj}{Conjecture}
\newtheorem{hiercond}{Hierarchy constraints -- Level}
\newtheorem*{hiercondarb}{Hierarchy constraints -- Level k}

\makeatletter
\newcommand{\neutralize}[1]{\expandafter\let\csname c@#1\endcsname\count@}
\makeatother

\newtheorem{prob}{Optimisation Problem}

\newenvironment{prob'}[1]
  {\renewcommand{\theprob}{\ref*{#1}$'$}%
       \neutralize{prob}\phantomsection
   \begin{prob}}
  {\end{prob}}

\usepackage[export]{adjustbox}

\usepackage{paralist}

\usepackage{dsfont}
\usepackage{rotating}
\usepackage{floatpag}
\rotfloatpagestyle{empty}
\renewcommand{\theenumi}{\arabic{enumi}}
\usepackage{amsmath,amsthm,amssymb}
\usepackage{xspace,enumerate,color,epsfig}
\usepackage{graphicx}
\graphicspath{{.}{./figures/}}
\usepackage{marvosym}

\usepackage{tikzfig}
\usepackage{stmaryrd}
\usepackage{docmute}
\usepackage{keycommand}

\usepackage{pifont}

\usepackage{enumitem}

\newkeycommand{\pointmap}[style=point][1]{\,\tikz{\node[style=\commandkey{style}] (x) at (0,-0.3) {$#1$};\node [style=none] (3) at (0, 1.0) {};\node [style=none] (3) at (0, -1.0) {};\draw (x)--(0,0.7);}\,}

\newkeycommand{\typedkpoint}[style=kpoint,edgestyle=][2]{%
\,\begin{tikzpicture}
  \begin{pgfonlayer}{nodelayer}
    \node [style=none] (0) at (0, 1) {};
    \node [style=\commandkey{style}] (1) at (0, -0.75) {$#2$};
    \node [style=right label] (2) at (0.25, 0.5) {$#1$};
  \end{pgfonlayer}
  \begin{pgfonlayer}{edgelayer}
    \draw [\commandkey{edgestyle}] (1) to (0.center);
  \end{pgfonlayer}
\end{tikzpicture}\,}

\newkeycommand{\typedkpointdag}[style=kpoint adjoint,edgestyle=][2]{%
\,\begin{tikzpicture}
  \begin{pgfonlayer}{nodelayer}
    \node [style=none] (0) at (0, -1) {};
    \node [style=\commandkey{style}] (1) at (0, 0.75) {$#2$};
    \node [style=right label] (2) at (0.25, -0.5) {$#1$};
  \end{pgfonlayer}
  \begin{pgfonlayer}{edgelayer}
    \draw [\commandkey{edgestyle}] (0.center) to (1);
  \end{pgfonlayer}
\end{tikzpicture}\,}

\newcommand{\bistateadj}[1]{\,\begin{tikzpicture}[yshift=-3mm]
  \begin{pgfonlayer}{nodelayer}
    \node [style=kpointadj, minimum width=1 cm, inner sep=2pt] (0) at (0, 0.5) {$\psi$};
    \node [style=none] (1) at (-0.75, 0.25) {};
    \node [style=none] (2) at (0.75, 0.25) {};
    \node [style=none] (3) at (-0.75, -0.5) {};
    \node [style=none] (4) at (0.75, -0.5) {};
  \end{pgfonlayer}
  \begin{pgfonlayer}{edgelayer}
    \draw (1.center) to (3.center);
    \draw (2.center) to (4.center);
  \end{pgfonlayer}
\end{tikzpicture}\,}

\newkeycommand{\pointketbra}[style1=copoint,style2=point][2]{\,%
\begin{tikzpicture}
  \begin{pgfonlayer}{nodelayer}
    \node [style=none] (0) at (0, -1.5) {};
    \node [style=\commandkey{style1}] (1) at (0, -0.75) {$#1$};
    \node [style=\commandkey{style2}] (2) at (0, 0.75) {$#2$};
    \node [style=none] (3) at (0, 1.5) {};
  \end{pgfonlayer}
  \begin{pgfonlayer}{edgelayer}
    \draw (0.center) to (1);
    \draw (2) to (3.center);
  \end{pgfonlayer}
\end{tikzpicture}\,}

\newkeycommand{\twocopointketbra}[style1=copoint,style2=copoint,style3=point][3]{\,%
\begin{tikzpicture}
  \begin{pgfonlayer}{nodelayer}
    \node [style=none] (0) at (-0.75, -1.5) {};
    \node [style=none] (1) at (0.75, -1.5) {};
    \node [style=\commandkey{style1}] (2) at (-0.75, -0.75) {$#1$};
    \node [style=\commandkey{style2}] (3) at (0.75, -0.75) {$#2$};
    \node [style=\commandkey{style3}] (4) at (0, 0.75) {$#3$};
    \node [style=none] (5) at (0, 1.5) {};
  \end{pgfonlayer}
  \begin{pgfonlayer}{edgelayer}
    \draw (0.center) to (2);
    \draw (1.center) to (3);
    \draw (4) to (5.center);
  \end{pgfonlayer}
\end{tikzpicture}\,}

\newkeycommand{\twopointketbra}[style1=copoint,style2=point,style3=point][3]{\,%
\begin{tikzpicture}
  \begin{pgfonlayer}{nodelayer}
    \node [style=none] (0) at (0, -1.5) {};
    \node [style=none] (1) at (-0.75, 1.5) {};
    \node [style=\commandkey{style1}] (2) at (0, -0.75) {$#1$};
    \node [style=\commandkey{style2}] (3) at (-0.75, 0.75) {$#2$};
    \node [style=\commandkey{style3}] (4) at (0.75, 0.75) {$#3$};
    \node [style=none] (5) at (0.75, 1.5) {};
  \end{pgfonlayer}
  \begin{pgfonlayer}{edgelayer}
    \draw (0.center) to (2);
    \draw (1.center) to (3);
    \draw (4) to (5.center);
  \end{pgfonlayer}
\end{tikzpicture}\,}

\newkeycommand{\pointbraket}[style1=point,style2=copoint][2]{\,%
\begin{tikzpicture}
  \begin{pgfonlayer}{nodelayer}
    \node [style=\commandkey{style1}] (0) at (0, -0.5) {$#1$};
    \node [style=\commandkey{style2}] (1) at (0, 0.5) {$#2$};
  \end{pgfonlayer}
  \begin{pgfonlayer}{edgelayer}
    \draw (0) to (1);
  \end{pgfonlayer}
\end{tikzpicture}\,}

\newkeycommand{\kpointbraket}[style1=kpoint,style2=kpoint adjoint][2]{\,%
\begin{tikzpicture}
  \begin{pgfonlayer}{nodelayer}
    \node [style=\commandkey{style1}] (0) at (0, -0.75) {$#1$};
    \node [style=\commandkey{style2}] (1) at (0, 0.75) {$#2$};
  \end{pgfonlayer}
  \begin{pgfonlayer}{edgelayer}
    \draw (0) to (1);
  \end{pgfonlayer}
\end{tikzpicture}\,}

\newkeycommand{\kpointmap}[style1=kpoint,style2=map][2]{\,%
\begin{tikzpicture}
  \begin{pgfonlayer}{nodelayer}
    \node [style=\commandkey{style1}] (0) at (0, -1.1) {$#1$};
    \node [style=\commandkey{style2}] (1) at (0, 0.2) {$#2$};
    \node [style=none] (2) at (0, 1.5) {};
  \end{pgfonlayer}
  \begin{pgfonlayer}{edgelayer}
    \draw (0) to (1);
    \draw (1) to (2.center);
  \end{pgfonlayer}
\end{tikzpicture}%
\,}

\newkeycommand{\sandwichtwo}[style1=point,style2=map,style3=map,style4=copoint][4]{\,%
\begin{tikzpicture}
  \begin{pgfonlayer}{nodelayer}
    \node [style=\commandkey{style2}] (0) at (0, -0.75) {$#2$};
    \node [style=\commandkey{style3}] (1) at (0, 0.75) {$#3$};
    \node [style=\commandkey{style1}] (2) at (0, -2) {$#1$};
    \node [style=\commandkey{style4}] (3) at (0, 2) {$#4$};
  \end{pgfonlayer}
  \begin{pgfonlayer}{edgelayer}
    \draw (2) to (0);
    \draw (0) to (1);
    \draw (1) to (3);
  \end{pgfonlayer}
\end{tikzpicture}\,}

\newkeycommand{\longbraket}[style1=point,style2=copoint][2]{\,%
\begin{tikzpicture}
  \begin{pgfonlayer}{nodelayer}
    \node [style=\commandkey{style1}] (0) at (0, -2) {$#1$};
    \node [style=\commandkey{style2}] (1) at (0, 2) {$#2$};
  \end{pgfonlayer}
  \begin{pgfonlayer}{edgelayer}
    \draw (0) to (1);
  \end{pgfonlayer}
\end{tikzpicture}\,}

\newkeycommand{\onb}[style=point]{\ensuremath{\left\{\tikz{\node[style=\commandkey{style}] (x) at (0,-0.3) {$j$};\draw (x)--(0,0.7);}\right\}}\xspace}

\newkeycommand{\copointmap}[style=copoint][1]{\,\tikz{\node[style=\commandkey{style}] (x) at (0,0.3) {$#1$};\draw (x)--(0,-0.7);}\,}

\newcommand{\ket}[1]{\ensuremath{\left|  #1 \right\rangle}}

\tikzstyle{cdiag}=[matrix of math nodes, row sep=3em, column sep=3em, text height=1.5ex, text depth=0.25ex,inner sep=0.5em]
\tikzstyle{arrow above}=[transform canvas={yshift=0.5ex}]
\tikzstyle{arrow below}=[transform canvas={yshift=-0.5ex}]

\usetikzlibrary{shapes.multipart}
\tikzset{XOR/.style={draw,fill=white,circle,append after command={
        [shorten >=\pgflinewidth, shorten <=\pgflinewidth,]
        (\tikzlastnode.north) edge (\tikzlastnode.south)
        (\tikzlastnode.east) edge (\tikzlastnode.west)
        }
    }
}

\tikzset{->-/.style={decoration={
  markings,
  mark=at position .5 with {\arrow{>}}},postaction={decorate}}}
\tikzset{-<-/.style={decoration={
  markings,
  mark=at position .5 with {\arrow{<}}},postaction={decorate}}}

\tikzstyle{bwSpider}=[
       rectangle split,
       rectangle split parts=2,
       rectangle split part fill={black,white},
 minimum size=3.6 mm, inner sep=-2mm, draw=black,scale=0.5,rounded corners=0.8 mm
       ]
 \tikzstyle{wbSpider}=[
       rectangle split,
       rectangle split parts=2,
       rectangle split part fill={white,black},
 minimum size=3.6 mm, inner sep=-2mm, draw=black,scale=0.5,rounded corners=0.8 mm
       ]

\tikzstyle{epiCopoint}=[regular polygon,regular polygon sides=3,draw,scale=0.75,inner sep=-0.5pt,minimum width=5mm,fill=white,regular polygon rotate=0,line width=1pt]
\tikzstyle{epiPoint}=[regular polygon,regular polygon sides=3,draw,scale=0.75,inner sep=-0.5pt,minimum width=5mm,fill=white,regular polygon rotate=180,line width=1pt]
\tikzstyle{epiPointWide}=[regular polygon,regular polygon sides=3,draw,scale=0.75,inner sep=-0.5pt,minimum width=8mm,fill=white,regular polygon rotate=180,line width=1pt]
\tikzstyle{epiBox}=[fill=white,draw, line width = 1pt,inner sep=0.6mm,font=\footnotesize,minimum height=3mm,minimum width=3mm]
\tikzstyle{epiBoxWide}=[fill=white,draw, line width = 1pt,inner sep=0.6mm,font=\footnotesize,minimum height=3mm,minimum width=5mm]
\tikzstyle{epiBoxVeryWide}=[fill=white,draw, line width = 1pt,inner sep=0.6mm,font=\footnotesize,minimum height=3mm,minimum width=7mm]
\tikzstyle{oWire}=[line width = .75pt, color=green!40!black!70!]
\tikzstyle{qWire}=[line width = 1.5pt, color=red!40!black]
\tikzstyle{cWire}=[color=gray!80,line width = .75pt]
\tikzstyle{CqWire}=[color=gray,line width = .75pt,->-]
\tikzstyle{CcWire}=[cWire]
\tikzstyle{RqWire}=[line width = 1pt, color=black,-<-]
\tikzstyle{RcWire}=[cWire]
\tikzstyle{env}=[copoint,regular polygon rotate=0,minimum width=0.2cm, fill=black]

\tikzstyle{probs}=[shape=semicircle,fill=white,draw=black,shape border rotate=180,minimum width=1.2cm]

\tikzstyle{every picture}=[baseline=-0.25em,scale=0.5]
\tikzstyle{dotpic}=[] 
\tikzstyle{diredges}=[every to/.style={diredge}]
\tikzstyle{math matrix}=[matrix of math nodes,left delimiter=(,right delimiter=),inner sep=2pt,column sep=1em,row sep=0.5em,nodes={inner sep=0pt},text height=1.5ex, text depth=0.25ex]

\tikzstyle{inline text}=[text height=1.5ex, text depth=0.25ex,yshift=0.5mm]
\tikzstyle{label}=[font=\footnotesize,text height=1.5ex, text depth=0.25ex,yshift=0.5mm]
\tikzstyle{left label}=[label,anchor=east,xshift=1.5mm]
\tikzstyle{right label}=[label,anchor=west,xshift=-1mm]
\tikzstyle{up label}=[label,anchor=south,yshift=-1mm]

\tikzstyle{braceedge}=[decorate,decoration={brace,amplitude=2mm,raise=-1mm}]
\tikzstyle{small braceedge}=[decorate,decoration={brace,amplitude=1mm,raise=-1mm}]

\tikzstyle{doubled}=[line width=1.6pt] 
\tikzstyle{boldedge}=[doubled,shorten <=-0.17mm,shorten >=-0.17mm]
\tikzstyle{boldedgegray}=[doubled,gray,shorten <=-0.17mm,shorten >=-0.17mm]
\tikzstyle{singleedgegray}=[gray]

\tikzstyle{semidoubled}=[line width=1.4pt] 
\tikzstyle{semiboldedgegray}=[semidoubled,gray,shorten <=-0.17mm,shorten >=-0.17mm]

\tikzstyle{boxedge}=[semiboldedgegray]

\tikzstyle{boldedgedashed}=[very thick,dashed,shorten <=-0.17mm,shorten >=-0.17mm]
\tikzstyle{vboldedgedashed}=[doubled,dashed,shorten <=-0.17mm,shorten >=-0.17mm]
\tikzstyle{left hook arrow}=[left hook-latex]
\tikzstyle{right hook arrow}=[right hook-latex]
\tikzstyle{sembracket}=[line width=0.5pt,shorten <=-0.07mm,shorten >=-0.07mm]

\tikzstyle{causal edge}=[->,thick,gray]
\tikzstyle{causal nondir}=[thick,gray]
\tikzstyle{timeline}=[thick,gray, dashed]

\tikzstyle{cedge}=[<->,thick,gray!70!white]

\tikzstyle{empty diagram}=[draw=gray!40!white,dashed,shape=rectangle,minimum width=1cm,minimum height=1cm]
\tikzstyle{empty diagram small}=[draw=gray!50!white,dashed,shape=rectangle,minimum width=0.6cm,minimum height=0.5cm]

\tikzstyle{dot}=[inner sep=0mm,minimum width=2mm,minimum height=2mm,draw,shape=circle]
\tikzstyle{bigdot}=[inner sep=0mm,minimum width=5mm,minimum height=5mm,draw,shape=circle]
\tikzstyle{leak}=[white dot, shape=regular polygon, minimum size=3.3 mm, regular polygon sides=3, outer sep=-0.2mm, regular polygon rotate=270]
\tikzstyle{proj}=[regular polygon,regular polygon sides=4,draw,scale=0.75,inner sep=-0.5pt,minimum width=6mm,fill=white]
\tikzstyle{projOut}=[regular polygon,regular polygon sides=3,draw,scale=0.75,inner sep=-0.5pt,minimum width=7.5mm,fill=white,regular polygon rotate=180]
\tikzstyle{projIn}=[regular polygon,regular polygon sides=3,draw,scale=0.75,inner sep=-0.5pt,minimum width=7.5mm,fill=white]
\tikzstyle{Vleak}=[white dot, shape=regular polygon, minimum size=3.3 mm, regular polygon sides=3, outer sep=-0.2mm, regular polygon rotate=90]
\tikzstyle{dleak}=[white dot, line width=1.6pt, shape=regular polygon, minimum size=3.3 mm, regular polygon sides=3, outer sep=-0.2mm, regular polygon rotate=270]

\tikzstyle{Wsquare}=[white dot, shape=regular polygon, rounded corners=0.8 mm, minimum size=3.3 mm, regular polygon sides=3, outer sep=-0.2mm]
\tikzstyle{Wsquareadj}=[white dot, shape=regular polygon, rounded corners=0.8 mm, minimum size=3.3 mm, regular polygon sides=3, outer sep=-0.2mm, regular polygon rotate=180]
\tikzstyle{ddot}=[inner sep=0mm, doubled, minimum width=2.5mm,minimum height=2.5mm,draw,shape=circle]

\tikzstyle{clear dot}=[dot,fill=none,text depth=-0.2mm,draw=gray, line width = .75pt]
\tikzstyle{tall clear dot}=[dot,fill=none,text depth=-0.2mm,draw=gray, line width = .75pt,shape=ellipse, minimum height=5mm]
\tikzstyle{wide clear dot}=[dot,fill=none,text depth=-0.2mm,draw=gray, line width = .75pt, shape=ellipse, minimum width = 5mm]
\tikzstyle{very wide clear dot}=[dot,fill=none,text depth=-0.2mm,draw=gray, line width = .75pt, shape=ellipse, minimum width = 7mm ]

\tikzstyle{black dot}=[dot,fill=black]
\tikzstyle{white dot}=[dot,fill=white,,text depth=-0.2mm]
\tikzstyle{white Wsquare}=[Wsquare,fill=gray,,text depth=-0.2mm]
\tikzstyle{white Wsquareadj}=[Wsquareadj,fill=white,,text depth=-0.2mm]
\tikzstyle{green dot}=[white dot] 
\tikzstyle{gray dot}=[dot,fill=gray!40!white,,text depth=-0.2mm]
\tikzstyle{red dot}=[gray dot] 

\tikzstyle{black ddot}=[ddot,fill=black]
\tikzstyle{white ddot}=[ddot,fill=white]
\tikzstyle{gray ddot}=[ddot,fill=gray!40!white]

\tikzstyle{gray edge}=[gray!60!white]

\tikzstyle{small dot}=[inner sep=0.2mm,minimum width=0pt,minimum height=0pt,draw,shape=circle]

\tikzstyle{small black dot}=[small dot,fill=black]
\tikzstyle{small white dot}=[small dot,fill=white]
\tikzstyle{small gray dot}=[small dot,fill=gray,draw=gray]

\tikzstyle{causal dot}=[inner sep=0.4mm,minimum width=0pt,minimum height=0pt,draw=white,shape=circle,fill=gray!40!white]

\tikzstyle{phase dimensions}=[minimum size=5mm,font=\footnotesize,rectangle,rounded corners=2.5mm,inner sep=0.2mm,outer sep=-2mm]
\tikzstyle{dphase dimensions}=[minimum size=5mm,font=\footnotesize,rectangle,rounded corners=2.5mm,inner sep=0.2mm,outer sep=-2mm]

\tikzstyle{white phase dot}=[dot,fill=white,phase dimensions]
\tikzstyle{white phase ddot}=[ddot,fill=white,dphase dimensions]

\tikzstyle{white rect ddot}=[draw=black,fill=white,doubled,minimum size=5mm,font=\footnotesize,rectangle,rounded corners=2.5mm,inner sep=0.2mm]
\tikzstyle{gray rect ddot}=[draw=black,fill=gray!40!white,doubled,minimum size=6mm,font=\footnotesize,rectangle,rounded corners=3mm]

\tikzstyle{gray phase dot}=[dot,fill=gray!40!white,phase dimensions]
\tikzstyle{gray phase ddot}=[ddot,fill=gray!40!white,dphase dimensions]
\tikzstyle{grey phase dot}=[gray phase dot]
\tikzstyle{grey phase ddot}=[gray phase ddot]

\tikzstyle{small phase dimensions}=[minimum size=4mm,font=\tiny,rectangle,rounded corners=2mm,inner sep=0.2mm,outer sep=-2mm]
\tikzstyle{small dphase dimensions}=[minimum size=4mm,font=\tiny,rectangle,rounded corners=2mm,inner sep=0.2mm,outer sep=-2mm]

\tikzstyle{small gray phase dot}=[dot,fill=gray!40!white,small phase dimensions]
\tikzstyle{small gray phase ddot}=[ddot,fill=gray!40!white,small dphase dimensions]

\tikzstyle{small map}=[draw,shape=rectangle,minimum height=4mm,minimum width=4mm,fill=white]

\tikzstyle{cnot}=[fill=white,shape=circle,inner sep=-1.4pt]

\tikzstyle{asym hadamard}=[fill=white,draw,shape=NEbox,inner sep=0.6mm,font=\footnotesize,minimum height=4mm]
\tikzstyle{asym hadamard conj}=[fill=white,draw,shape=NWbox,inner sep=0.6mm,font=\footnotesize,minimum height=4mm]
\tikzstyle{asym hadamard dag}=[fill=white,draw,shape=SEbox,inner sep=0.6mm,font=\footnotesize,minimum height=4mm]

\tikzstyle{hadamard}=[fill=white,draw,inner sep=0.6mm,font=\footnotesize,minimum height=4mm,minimum width=4mm]
\tikzstyle{small hadamard}=[fill=white,draw,inner sep=0.6mm,minimum height=1.5mm,minimum width=1.5mm]
\tikzstyle{small hadamard rotate}=[small hadamard,rotate=45]
\tikzstyle{dhadamard}=[hadamard,doubled]
\tikzstyle{small dhadamard}=[small hadamard,doubled]
\tikzstyle{small dhadamard rotate}=[small hadamard rotate,doubled]
\tikzstyle{antipode}=[white dot,inner sep=0.3mm,font=\footnotesize]

\tikzstyle{scalar}=[diamond,draw,inner sep=0.5pt,font=\small]
\tikzstyle{dscalar}=[diamond,doubled, draw,inner sep=0.5pt,font=\small]

\tikzstyle{small box}=[rectangle,inline text,fill=white,draw,minimum height=5mm,yshift=-0.5mm,minimum width=5mm,font=\small]
\tikzstyle{small gray box}=[small box,fill=gray!30]
\tikzstyle{medium box}=[rectangle,inline text,fill=white,draw,minimum height=5mm,yshift=-0.5mm,minimum width=10mm,font=\small]
\tikzstyle{square box}=[small box] 
\tikzstyle{medium gray box}=[small box,fill=gray!30]
\tikzstyle{semilarge box}=[rectangle,inline text,fill=white,draw,minimum height=5mm,yshift=-0.5mm,minimum width=12.5mm,font=\small]
\tikzstyle{large box}=[rectangle,inline text,fill=white,draw,minimum height=5mm,yshift=-0.5mm,minimum width=15mm,font=\small]
\tikzstyle{large gray box}=[small box,fill=gray!30]

\tikzstyle{Bayes box}=[rectangle,fill=black,draw, minimum height=3mm, minimum width=3mm]

\tikzstyle{gray square point}=[small box,fill=gray!50]

\tikzstyle{dphase box white}=[dhadamard]
\tikzstyle{dphase box gray}=[dhadamard,fill=gray!50!white]
\tikzstyle{phase box white}=[hadamard]
\tikzstyle{phase box gray}=[hadamard,fill=gray!50!white]

\tikzstyle{point}=[regular polygon,regular polygon sides=3,draw,scale=0.75,inner sep=-0.5pt,minimum width=9mm,fill=white,regular polygon rotate=180]
\tikzstyle{infpoint}=[regular polygon,regular polygon sides=3,draw,scale=0.75,inner sep=-0.5pt,minimum width=9mm,fill=white,regular polygon rotate=90]
\tikzstyle{point nosep}=[regular polygon,regular polygon sides=3,draw,scale=0.75,inner sep=-2pt,minimum width=9mm,fill=white,regular polygon rotate=180]
\tikzstyle{infcopoint}=[regular polygon,regular polygon sides=3,draw,scale=0.75,inner sep=-0.5pt,minimum width=9mm,fill=white,regular polygon rotate=270]
\tikzstyle{copoint}=[regular polygon,regular polygon sides=3,draw,scale=0.75,inner sep=-0.5pt,minimum width=9mm,fill=white]
\tikzstyle{dpoint}=[point,doubled]
\tikzstyle{dcopoint}=[copoint,doubled]

\tikzstyle{pointgrow}=[shape=cornerpoint,kpoint common,scale=0.75,inner sep=3pt]
\tikzstyle{pointgrow dag}=[shape=cornercopoint,kpoint common,scale=0.75,inner sep=3pt]

\tikzstyle{wide copoint}=[fill=white,draw,shape=isosceles triangle,shape border rotate=90,isosceles triangle stretches=true,inner sep=0pt,minimum width=1.5cm,minimum height=6.12mm]
\tikzstyle{wide point}=[fill=white,draw,shape=isosceles triangle,shape border rotate=-90,isosceles triangle stretches=true,inner sep=0pt,minimum width=1.5cm,minimum height=6.12mm,yshift=-0.0mm]
\tikzstyle{wide point plus}=[fill=white,draw,shape=isosceles triangle,shape border rotate=-90,isosceles triangle stretches=true,inner sep=0pt,minimum width=1.74cm,minimum height=7mm,yshift=-0.0mm]

\tikzstyle{wide dpoint}=[fill=white,doubled,draw,shape=isosceles triangle,shape border rotate=-90,isosceles triangle stretches=true,inner sep=0pt,minimum width=1.5cm,minimum height=6.12mm,yshift=-0.0mm]

\tikzstyle{tinypoint}=[regular polygon,regular polygon sides=3,draw,scale=0.55,inner sep=-0.15pt,minimum width=6mm,fill=white,regular polygon rotate=180]

\tikzstyle{white point}=[point]
\tikzstyle{white dpoint}=[dpoint]
\tikzstyle{green point}=[white point] 
\tikzstyle{white copoint}=[copoint]
\tikzstyle{gray point}=[point,fill=gray!40!white]
\tikzstyle{gray dpoint}=[gray point,doubled]
\tikzstyle{red point}=[gray point] 
\tikzstyle{gray copoint}=[copoint,fill=gray!40!white]
\tikzstyle{gray dcopoint}=[gray copoint,doubled]

\tikzstyle{white point guide}=[regular polygon,regular polygon sides=3,font=\scriptsize,draw,scale=0.65,inner sep=-0.5pt,minimum width=9mm,fill=white,regular polygon rotate=180]

\tikzstyle{black point}=[point,fill=black,font=\color{white}]
\tikzstyle{black copoint}=[copoint,fill=black,font=\color{white}]

\tikzstyle{tiny gray point}=[tinypoint,fill=gray!40!white]

\tikzstyle{diredge}=[->]
\tikzstyle{ddiredge}=[<->]
\tikzstyle{rdiredge}=[<-]
\tikzstyle{thickdiredge}=[->, very thick]
\tikzstyle{pointer edge}=[->,very thick,gray]
\tikzstyle{pointer edge part}=[very thick,gray]
\tikzstyle{dashed edge}=[dashed]
\tikzstyle{thick dashed edge}=[very thick,dashed]
\tikzstyle{thick gray dashed edge}=[thick dashed edge,gray!40]
\tikzstyle{thick map edge}=[very thick,|->]

\makeatletter
\newcommand{\boxshape}[3]{%
\pgfdeclareshape{#1}{
\inheritsavedanchors[from=rectangle] 
\inheritanchorborder[from=rectangle]
\inheritanchor[from=rectangle]{center}
\inheritanchor[from=rectangle]{north}
\inheritanchor[from=rectangle]{south}
\inheritanchor[from=rectangle]{west}
\inheritanchor[from=rectangle]{east}
\backgroundpath{
\southwest \pgf@xa=\pgf@x \pgf@ya=\pgf@y
\northeast \pgf@xb=\pgf@x \pgf@yb=\pgf@y

\@tempdima=#2
\@tempdimb=#3

\pgfpathmoveto{\pgfpoint{\pgf@xa - 5pt + \@tempdima}{\pgf@ya}}
\pgfpathlineto{\pgfpoint{\pgf@xa - 5pt - \@tempdima}{\pgf@yb}}
\pgfpathlineto{\pgfpoint{\pgf@xb + 5pt + \@tempdimb}{\pgf@yb}}
\pgfpathlineto{\pgfpoint{\pgf@xb + 5pt - \@tempdimb}{\pgf@ya}}
\pgfpathlineto{\pgfpoint{\pgf@xa - 5pt + \@tempdima}{\pgf@ya}}
\pgfpathclose
}
}}

\boxshape{NEbox}{0pt}{3pt}
\boxshape{SEbox}{0pt}{-3pt}
\boxshape{NWbox}{5pt}{0pt}
\boxshape{SWbox}{-5pt}{0pt}
\boxshape{EBox}{-3pt}{3pt}
\boxshape{WBox}{3pt}{-3pt}
\makeatother

\tikzstyle{cloud}=[shape=cloud,draw,minimum width=1.5cm,minimum height=1.5cm]

\tikzstyle{map}=[draw,shape=NEbox,inner sep=1pt,minimum height=4mm,fill=white]
\tikzstyle{dashedmap}=[draw,dashed,shape=NEbox,inner sep=2pt,minimum height=6mm,fill=white]
\tikzstyle{mapdag}=[draw,shape=SEbox,inner sep=1pt,minimum height=4mm,fill=white]
\tikzstyle{mapadj}=[draw,shape=SEbox,inner sep=2pt,minimum height=6mm,fill=white]
\tikzstyle{maptrans}=[draw,shape=SWbox,inner sep=2pt,minimum height=6mm,fill=white]
\tikzstyle{mapconj}=[draw,shape=NWbox,inner sep=2pt,minimum height=6mm,fill=white]

\tikzstyle{medium map}=[draw,shape=NEbox,inner sep=2pt,minimum height=6mm,fill=white,minimum width=7mm]
\tikzstyle{medium map dag}=[draw,shape=SEbox,inner sep=2pt,minimum height=6mm,fill=white,minimum width=7mm]
\tikzstyle{medium map adj}=[draw,shape=SEbox,inner sep=2pt,minimum height=6mm,fill=white,minimum width=7mm]
\tikzstyle{medium map trans}=[draw,shape=SWbox,inner sep=2pt,minimum height=6mm,fill=white,minimum width=7mm]
\tikzstyle{medium map conj}=[draw,shape=NWbox,inner sep=2pt,minimum height=6mm,fill=white,minimum width=7mm]
\tikzstyle{semilarge map}=[draw,shape=NEbox,inner sep=2pt,minimum height=6mm,fill=white,minimum width=9.5mm]
\tikzstyle{semilarge map trans}=[draw,shape=SWbox,inner sep=2pt,minimum height=6mm,fill=white,minimum width=9.5mm]
\tikzstyle{semilarge map adj}=[draw,shape=SEbox,inner sep=2pt,minimum height=6mm,fill=white,minimum width=9.5mm]
\tikzstyle{semilarge map dag}=[draw,shape=SEbox,inner sep=2pt,minimum height=6mm,fill=white,minimum width=9.5mm]
\tikzstyle{semilarge map conj}=[draw,shape=NWbox,inner sep=2pt,minimum height=6mm,fill=white,minimum width=9.5mm]
\tikzstyle{large map}=[draw,shape=NEbox,inner sep=2pt,minimum height=6mm,fill=white,minimum width=12mm]
\tikzstyle{large map conj}=[draw,shape=NWbox,inner sep=2pt,minimum height=6mm,fill=white,minimum width=12mm]
\tikzstyle{very large map}=[draw,shape=NEbox,inner sep=2pt,minimum height=6mm,fill=white,minimum width=17mm]

\tikzstyle{medium dmap}=[draw,doubled,shape=NEbox,inner sep=2pt,minimum height=6mm,fill=white,minimum width=7mm]
\tikzstyle{medium dmap dag}=[draw,doubled,shape=SEbox,inner sep=2pt,minimum height=6mm,fill=white,minimum width=7mm]
\tikzstyle{medium dmap adj}=[draw,doubled,shape=SEbox,inner sep=2pt,minimum height=6mm,fill=white,minimum width=7mm]
\tikzstyle{medium dmap trans}=[draw,doubled,shape=SWbox,inner sep=2pt,minimum height=6mm,fill=white,minimum width=7mm]
\tikzstyle{medium dmap conj}=[draw,doubled,shape=NWbox,inner sep=2pt,minimum height=6mm,fill=white,minimum width=7mm]
\tikzstyle{semilarge dmap}=[draw,doubled,shape=NEbox,inner sep=2pt,minimum height=6mm,fill=white,minimum width=9.5mm]
\tikzstyle{semilarge dmap trans}=[draw,doubled,shape=SWbox,inner sep=2pt,minimum height=6mm,fill=white,minimum width=9.5mm]
\tikzstyle{semilarge dmap adj}=[draw,doubled,shape=SEbox,inner sep=2pt,minimum height=6mm,fill=white,minimum width=9.5mm]
\tikzstyle{semilarge dmap dag}=[draw,doubled,shape=SEbox,inner sep=2pt,minimum height=6mm,fill=white,minimum width=9.5mm]
\tikzstyle{semilarge dmap conj}=[draw,doubled,shape=NWbox,inner sep=2pt,minimum height=6mm,fill=white,minimum width=9.5mm]
\tikzstyle{large dmap}=[draw,doubled,shape=NEbox,inner sep=2pt,minimum height=6mm,fill=white,minimum width=12mm]
\tikzstyle{large dmap conj}=[draw,doubled,shape=NWbox,inner sep=2pt,minimum height=6mm,fill=white,minimum width=12mm]
\tikzstyle{large dmap trans}=[draw,doubled,shape=SWbox,inner sep=2pt,minimum height=6mm,fill=white,minimum width=12mm]
\tikzstyle{large dmap adj}=[draw,doubled,shape=SEbox,inner sep=2pt,minimum height=6mm,fill=white,minimum width=12mm]
\tikzstyle{large dmap dag}=[draw,doubled,shape=SEbox,inner sep=2pt,minimum height=6mm,fill=white,minimum width=12mm]
\tikzstyle{very large dmap}=[draw,doubled,shape=NEbox,inner sep=2pt,minimum height=6mm,fill=white,minimum width=19.5mm]

\tikzstyle{muxbox}=[draw,shape=rectangle,minimum height=3mm,minimum width=3mm,fill=white]
\tikzstyle{dmuxbox}=[muxbox,doubled]

\tikzstyle{box}=[draw,shape=rectangle,inner sep=2pt,minimum height=6mm,minimum width=6mm,fill=white]
\tikzstyle{dbox}=[draw,doubled,shape=rectangle,inner sep=2pt,minimum height=6mm,minimum width=6mm,fill=white]
\tikzstyle{dmap}=[draw,doubled,shape=NEbox,inner sep=2pt,minimum height=6mm,fill=white]
\tikzstyle{dmapdag}=[draw,doubled,shape=SEbox,inner sep=2pt,minimum height=6mm,fill=white]
\tikzstyle{dmapadj}=[draw,doubled,shape=SEbox,inner sep=2pt,minimum height=6mm,fill=white]
\tikzstyle{dmaptrans}=[draw,doubled,shape=SWbox,inner sep=2pt,minimum height=6mm,fill=white]
\tikzstyle{dmapconj}=[draw,doubled,shape=NWbox,inner sep=2pt,minimum height=6mm,fill=white]

\tikzstyle{ddmap}=[draw,doubled,dashed,shape=NEbox,inner sep=2pt,minimum height=6mm,fill=white]
\tikzstyle{ddmapdag}=[draw,doubled,dashed,shape=SEbox,inner sep=2pt,minimum height=6mm,fill=white]
\tikzstyle{ddmapadj}=[draw,doubled,dashed,shape=SEbox,inner sep=2pt,minimum height=6mm,fill=white]
\tikzstyle{ddmaptrans}=[draw,doubled,dashed,shape=SWbox,inner sep=2pt,minimum height=6mm,fill=white]
\tikzstyle{ddmapconj}=[draw,doubled,dashed,shape=NWbox,inner sep=2pt,minimum height=6mm,fill=white]

\boxshape{sNEbox}{0pt}{3pt}
\boxshape{sSEbox}{0pt}{-3pt}
\boxshape{sNWbox}{3pt}{0pt}
\boxshape{sSWbox}{-3pt}{0pt}
\tikzstyle{smap}=[draw,shape=sNEbox,fill=white]
\tikzstyle{smapdag}=[draw,shape=sSEbox,fill=white]
\tikzstyle{smapadj}=[draw,shape=sSEbox,fill=white]
\tikzstyle{smaptrans}=[draw,shape=sSWbox,fill=white]
\tikzstyle{smapconj}=[draw,shape=sNWbox,fill=white]

\tikzstyle{dsmap}=[draw,dashed,shape=sNEbox,fill=white]
\tikzstyle{dsmapdag}=[draw,dashed,shape=sSEbox,fill=white]
\tikzstyle{dsmaptrans}=[draw,dashed,shape=sSWbox,fill=white]
\tikzstyle{dsmapconj}=[draw,dashed,shape=sNWbox,fill=white]

\boxshape{mNEbox}{0pt}{10pt}
\boxshape{mSEbox}{0pt}{-10pt}
\boxshape{mNWbox}{10pt}{0pt}
\boxshape{mSWbox}{-10pt}{0pt}
\tikzstyle{mmap}=[draw,shape=mNEbox]
\tikzstyle{mmapdag}=[draw,shape=mSEbox]
\tikzstyle{mmaptrans}=[draw,shape=mSWbox]
\tikzstyle{mmapconj}=[draw,shape=mNWbox]

\tikzstyle{mmapgray}=[draw,fill=gray!40!white,shape=mNEbox]
\tikzstyle{smapgray}=[draw,fill=gray!40!white,shape=sNEbox]

\makeatletter

\pgfdeclareshape{cornerpoint}{
\inheritsavedanchors[from=rectangle] 
\inheritanchorborder[from=rectangle]
\inheritanchor[from=rectangle]{center}
\inheritanchor[from=rectangle]{north}
\inheritanchor[from=rectangle]{south}
\inheritanchor[from=rectangle]{west}
\inheritanchor[from=rectangle]{east}
\backgroundpath{
\southwest \pgf@xa=\pgf@x \pgf@ya=\pgf@y
\northeast \pgf@xb=\pgf@x \pgf@yb=\pgf@y

\pgfmathsetmacro{\pgf@shorten@left}{\pgfkeysvalueof{/tikz/shorten left}}
\pgfmathsetmacro{\pgf@shorten@right}{\pgfkeysvalueof{/tikz/shorten right}}

\pgfpathmoveto{\pgfpoint{0.5 * (\pgf@xa + \pgf@xb)}{\pgf@ya - 5pt}}
\pgfpathlineto{\pgfpoint{\pgf@xa - 8pt + \pgf@shorten@left}{\pgf@yb - 1.5 * \pgf@shorten@left}}
\pgfpathlineto{\pgfpoint{\pgf@xa - 8pt + \pgf@shorten@left}{\pgf@yb}}
\pgfpathlineto{\pgfpoint{\pgf@xb + 8pt - \pgf@shorten@right}{\pgf@yb}}
\pgfpathlineto{\pgfpoint{\pgf@xb + 8pt - \pgf@shorten@right}{\pgf@yb - 1.5 * \pgf@shorten@right}}
\pgfpathclose
}
}

\pgfdeclareshape{cornercopoint}{
\inheritsavedanchors[from=rectangle] 
\inheritanchorborder[from=rectangle]
\inheritanchor[from=rectangle]{center}
\inheritanchor[from=rectangle]{north}
\inheritanchor[from=rectangle]{south}
\inheritanchor[from=rectangle]{west}
\inheritanchor[from=rectangle]{east}
\backgroundpath{
\southwest \pgf@xa=\pgf@x \pgf@ya=\pgf@y
\northeast \pgf@xb=\pgf@x \pgf@yb=\pgf@y

\pgfmathsetmacro{\pgf@shorten@left}{\pgfkeysvalueof{/tikz/shorten left}}
\pgfmathsetmacro{\pgf@shorten@right}{\pgfkeysvalueof{/tikz/shorten right}}

\pgfpathmoveto{\pgfpoint{0.5 * (\pgf@xa + \pgf@xb)}{\pgf@yb + 5pt}}
\pgfpathlineto{\pgfpoint{\pgf@xa - 8pt + \pgf@shorten@left}{\pgf@ya + 1.5 * \pgf@shorten@left}}
\pgfpathlineto{\pgfpoint{\pgf@xa - 8pt + \pgf@shorten@left}{\pgf@ya}}
\pgfpathlineto{\pgfpoint{\pgf@xb + 8pt - \pgf@shorten@right}{\pgf@ya}}
\pgfpathlineto{\pgfpoint{\pgf@xb + 8pt - \pgf@shorten@right}{\pgf@ya + 1.5 * \pgf@shorten@right}}
\pgfpathclose
}
}

\makeatother

\pgfkeyssetvalue{/tikz/shorten left}{0pt}
\pgfkeyssetvalue{/tikz/shorten right}{0pt}

\tikzstyle{kpoint common}=[draw,fill=white,inner sep=1pt,minimum height=4mm]
\tikzstyle{kpoint sc}=[shape=cornerpoint,kpoint common]
\tikzstyle{kpoint adjoint sc}=[shape=cornercopoint,kpoint common]
\tikzstyle{kpoint}=[shape=cornerpoint,shorten left=5pt,kpoint common]
\tikzstyle{kpoint adjoint}=[shape=cornercopoint,shorten left=5pt,kpoint common]
\tikzstyle{kpoint conjugate}=[shape=cornerpoint,shorten right=5pt,kpoint common]
\tikzstyle{kpoint transpose}=[shape=cornercopoint,shorten right=5pt,kpoint common]
\tikzstyle{kpoint symm}=[shape=cornerpoint,shorten left=5pt,shorten right=5pt,kpoint common]

\tikzstyle{wide kpoint sc}=[shape=cornerpoint,kpoint common, minimum width=1 cm]
\tikzstyle{wide kpointdag sc}=[shape=cornercopoint,kpoint common, minimum width=1 cm]

\tikzstyle{black kpoint}=[shape=cornerpoint,shorten left=5pt,kpoint common,fill=black,font=\color{white}]

\tikzstyle{black kpoint sm}=[shape=cornerpoint,shorten left=5pt,kpoint common,fill=black,font=\color{white},scale=0.75]

\tikzstyle{black kpoint adjoint}=[shape=cornercopoint,shorten left=5pt,kpoint common,fill=black,font=\color{white}]
\tikzstyle{black kpointadj}=[shape=cornercopoint,shorten left=5pt,kpoint common,fill=black,font=\color{white}]

\tikzstyle{black kpointadj sm}=[shape=cornercopoint,shorten left=5pt,kpoint common,fill=black,font=\color{white},scale=0.75]

\tikzstyle{black dkpoint}=[shape=cornerpoint,shorten left=5pt,kpoint common,fill=black, doubled,font=\color{white}]
\tikzstyle{black dkpoint adjoint}=[shape=cornercopoint,shorten left=5pt,kpoint common,fill=black, doubled,font=\color{white}]
\tikzstyle{black dkpointadj}=[shape=cornercopoint,shorten left=5pt,kpoint common,fill=black, doubled,font=\color{white}]

\tikzstyle{black dkpoint sm}=[shape=cornerpoint,shorten left=5pt,kpoint common,fill=black, doubled,font=\color{white},scale=0.75]
\tikzstyle{black dkpointadj sm}=[shape=cornercopoint,shorten left=5pt,kpoint common,fill=black, doubled,font=\color{white},scale=0.75]

\tikzstyle{kpointdag}=[kpoint adjoint]
\tikzstyle{kpointadj}=[kpoint adjoint]
\tikzstyle{kpointconj}=[kpoint conjugate]
\tikzstyle{kpointtrans}=[kpoint transpose]

\tikzstyle{big kpoint}=[kpoint, minimum width=1.2 cm, minimum height=8mm, inner sep=4pt, text depth=3mm]

\tikzstyle{wide kpoint}=[kpoint, minimum width=1 cm, inner sep=2pt]
\tikzstyle{wide kpointdag}=[kpointdag, minimum width=1 cm, inner sep=2pt]
\tikzstyle{wide kpointconj}=[kpointconj, minimum width=1 cm, inner sep=2pt]
\tikzstyle{wide kpointtrans}=[kpointtrans, minimum width=1 cm, inner sep=2pt]

\tikzstyle{wider kpoint}=[kpoint, minimum width=1.25 cm, inner sep=2pt]
\tikzstyle{wider kpointdag}=[kpointdag, minimum width=1.25 cm, inner sep=2pt]
\tikzstyle{wider kpointconj}=[kpointconj, minimum width=1.25 cm, inner sep=2pt]
\tikzstyle{wider kpointtrans}=[kpointtrans, minimum width=1.25 cm, inner sep=2pt]

\tikzstyle{gray kpoint}=[kpoint,fill=gray!50!white]
\tikzstyle{gray kpointdag}=[kpointdag,fill=gray!50!white]
\tikzstyle{gray kpointadj}=[kpointadj,fill=gray!50!white]
\tikzstyle{gray kpointconj}=[kpointconj,fill=gray!50!white]
\tikzstyle{gray kpointtrans}=[kpointtrans,fill=gray!50!white]

\tikzstyle{gray dkpoint}=[kpoint,fill=gray!50!white,doubled]
\tikzstyle{gray dkpointdag}=[kpointdag,fill=gray!50!white,doubled]
\tikzstyle{gray dkpointadj}=[kpointadj,fill=gray!50!white,doubled]
\tikzstyle{gray dkpointconj}=[kpointconj,fill=gray!50!white,doubled]
\tikzstyle{gray dkpointtrans}=[kpointtrans,fill=gray!50!white,doubled]

\tikzstyle{white label}=[draw,fill=white,rectangle,inner sep=0.7 mm]
\tikzstyle{gray label}=[draw,fill=gray!50!white,rectangle,inner sep=0.7 mm]
\tikzstyle{black label}=[draw,fill=black,rectangle,inner sep=0.7 mm]

\tikzstyle{dkpoint}=[kpoint,doubled]
\tikzstyle{wide dkpoint}=[wide kpoint,doubled]
\tikzstyle{dkpointdag}=[kpoint adjoint,doubled]
\tikzstyle{wide dkpointdag}=[wide kpointdag,doubled]
\tikzstyle{dkcopoint}=[kpoint adjoint,doubled]
\tikzstyle{dkpointadj}=[kpoint adjoint,doubled]
\tikzstyle{dkpointconj}=[kpoint conjugate,doubled]
\tikzstyle{dkpointtrans}=[kpoint transpose,doubled]

\tikzstyle{kscalar}=[kpoint common, shape=EBox, inner xsep=-1pt, inner ysep=3pt,font=\small]
\tikzstyle{kscalarconj}=[kpoint common, shape=WBox, inner xsep=-1pt, inner ysep=3pt,font=\small]

\tikzstyle{spekpoint}=[kpoint sc,minimum height=5mm,inner sep=3pt]
\tikzstyle{spekcopoint}=[kpoint adjoint sc,minimum height=5mm,inner sep=3pt]

\tikzstyle{dspekpoint}=[spekpoint,doubled]
\tikzstyle{dspekcopoint}=[spekcopoint,doubled]

 \tikzstyle{upground}=[circuit ee IEC,thick,ground,rotate=90,scale=2.5]
 \tikzstyle{downground}=[circuit ee IEC,thick,ground,rotate=-90,scale=2.5]
 \tikzstyle{infupground}=[circuit ee IEC,thick,ground,rotate=0,scale=2.5]
 \tikzstyle{infdownground}=[circuit ee IEC,thick,ground,rotate=180,scale=2.5]
 \tikzstyle{bigground}=[regular polygon,regular polygon sides=3,draw=gray,scale=0.50,inner sep=-0.5pt,minimum width=10mm,fill=gray]

\tikzstyle{arrs}=[-latex,font=\small,auto]
\tikzstyle{arrow plain}=[arrs]
\tikzstyle{arrow dashed}=[dashed,arrs]
\tikzstyle{arrow bold}=[very thick,arrs]
\tikzstyle{arrow hide}=[draw=white!0,-]
\tikzstyle{arrow reverse}=[latex-]
\tikzstyle{cdnode}=[]

\let\olddagger\dagger
\renewcommand{\dagger}{\ensuremath{\olddagger}\xspace}

\usepackage{makeidx}
\makeindex

\theoremstyle{plain}
\newtheorem*{main theorem}{Main Theorem}
\newtheorem{theorem}{Theorem}[section]

\newtheorem{definition}[theorem]{Definition}

\newtheorem{example*}[theorem]{Example*}
\newtheorem{examples*}[theorem]{Examples*}

\newtheorem{remark*}[theorem]{Remark*}

\newtheorem*{search problem}{Search Problem}

\usepackage{color}
\def\bR{\begin{color}{red}}
\def\bB{\begin{color}{blue}}
\def\bM{\begin{color}{magenta}}
\def\bC{\begin{color}{cyan}}
\def\bW{\begin{color}{white}}
\def\bBl{\begin{color}{black}}
\def\bG{\begin{color}{green}}
\def\bY{\begin{color}{yellow}}
\def\e{\end{color}\xspace}
\newcommand{\bit}{\begin{itemize}}
\newcommand{\eit}{\end{itemize}\par\noindent}
\newcommand{\ben}{\begin{enumerate}}
\newcommand{\een}{\end{enumerate}\par\noindent}
\newcommand{\beq}{\begin{equation}}
\newcommand{\eeq}{\end{equation}\par\noindent}
\newcommand{\beqa}{\begin{eqnarray*}}
\newcommand{\eeqa}{\end{eqnarray*}\par\noindent}
\newcommand{\beqn}{\begin{eqnarray}}
\newcommand{\eeqn}{\end{eqnarray}\par\noindent}

\def\jR{\begin{color}{black}}
\def\jB{\begin{color}{black}}
\def\jM{\begin{color}{magenta}}
\def\jC{\begin{color}{cyan}}
\def\jW{\begin{color}{white}}
\def\jBl{\begin{color}{black}}
\def\jG{\begin{color}{green}}
\def\jY{\begin{color}{yellow}}

\newcommand{\<}{\langle}

\newcommand{\PRM}{\mathrm{PRM}}
\newcommand{\Id}{\mathbbm{1}}

\newcommand{\nSet}{{\boldsymbol{\eta}}}
\newcommand{\bSet}{{\boldsymbol{\beta}}}
\newcommand{\iSet}{{\boldsymbol{\iota}}}
\newcommand{\Bell}{{\color{white}$\mathcal{I}$}}

\definecolor{quantumviolet}{RGB}{79, 4, 134}
\definecolor{quantumgray}{RGB}{134,79,4}

\def\blk{\end{color}\xspace}

\def\bel{\begin{color}{magenta}}
\def\john{\begin{color}{magenta}}
\def\mh{\begin{color}{red}}

\usepackage{float}

\usepackage{tikz}
\usepackage{color}
\usetikzlibrary{calc,decorations.pathreplacing}
\usetikzlibrary{arrows,shapes}
\usetikzlibrary{patterns,intersections}

\definecolor{panblue}{RGB}{0,24,150}
\definecolor{carmine}{RGB}{150, 0, 24}
\definecolor{darkgreen}{rgb}{0,.5,0}

\begin{document}

\title{Correlations constrained by composite  measurements}

\author{John H.~Selby}
\affiliation{International Centre for Theory of Quantum Technologies, University of Gda\'nsk, 80-308 Gda\'nsk, Poland}
\author{Ana Bel\'en Sainz}
\affiliation{International Centre for Theory of Quantum Technologies, University of Gda\'nsk, 80-308 Gda\'nsk, Poland}
\author{Victor Magron}
\affiliation{LAAS-CNRS and Institute of Mathematics, University of Toulouse, LAAS, 7 Avenue du Colonel Roche, 31077 Toulouse C\'edex 4, France}
\author{{\L}ukasz Czekaj}
\affiliation{International Centre for Theory of Quantum Technologies, University of Gda\'nsk, 80-308 Gda\'nsk, Poland}
\author{Micha{\l} Horodecki}
\affiliation{International Centre for Theory of Quantum Technologies, University of Gda\'nsk, 80-308 Gda\'nsk, Poland}

\begin{abstract}
How to understand the set of correlations admissible in nature is one outstanding open problem in the core of the foundations of quantum theory. Here we take a complementary viewpoint to the device-independent approach, and explore the correlations that physical theories may feature when restricted by some particular constraints on their measurements. We show that demanding that a theory exhibits {a composite} measurement imposes a hierarchy of constraints on the structure of its sets of states and effects, which translate to a hierarchy of constraints on the allowed correlations themselves. We moreover focus on the particular case where one demands the existence of  a correlated measurement  that reads out the parity of local fiducial measurements. By formulating a non-linear Optimisation Problem, and semidefinite relaxations of it, we explore the consequences of the existence of such a parity reading measurement for violations of Bell inequalities. In particular, we show that in certain situations this assumption has surprisingly strong consequences, namely, that Tsirelson's bound can be recovered.
\end{abstract}

\maketitle
\tableofcontents

\section{Introduction}

Bell nonclassicality is a well-known phenomenon featured by quantum theory, and attests that correlations observed in nature are not always compatible with a classical common cause shared among the distant wings of an experiment \cite{Bell64}. That is, non-classical common causes are necessary to explain our observational data \cite{Wolfe2020quantifyingbell}. Bell's theorem not only teaches us a valuable lesson about the foundational aspects of nature, but also underpins a variety of current technological applications. For example, non-classical correlations enable cryptographic applications, such as key distribution \cite{BHK,Acin2006QKD,Scarani2006QKD,Acin2007QKD,vazirani14,Kaniewski2016chsh} and  randomness generation \cite{colbeckamp,Pironio2010,Pivoluska2015,Dhara2013DIRNG}, and provide an information-theoretic advantage in other families of so-called non-local games \cite{Broadbent2006,Palazuelos2016,Johnston2016}.

Understanding quantum correlations -- in particular their limitations -- is therefore an important open problem within quantum information theory. Research on these lines has recently been carried out within the device-independent formalism, that is, where the only information used to reason about nature are the classical variables that denote measurement choices and their outcomes, together with the observed outcome statistics. Within this paradigm, quantum correlations are studied ``from the outside'', by exploring the constraints that physical or information-theoretical principles impose on the observed correlations \cite{popescu1994quantum,brassard2006limit,pawlowski2009information,navascues2010glance,MNC,LO,linden2007quantum}. In the device-independent framework, hence, such proposed constraints are therefore formulated at the level of the correlations themselves.

In this work we take a complementary viewpoint to the problem of characterising quantum correlations, by examining the possible correlations that may arise when constraints are imposed on the underlying physical theory. From this perspective, hence, one asks how various elements of the physical theory constrain or enable particular correlations. The particular objects we are interested in here are the \textit{measurements} that the theory may feature.  Even though in principle one could also impose constraints on the states as well, our curiosity on measurements arises from the results of Ref.~\cite{czekaj2018bell} -- if demanding such parity constraints on measurement outcomes at the level of the statistics yields such substantial constraints on correlations (see below), what can we expect if we open the box and demand conditions on the measurement themselves?
On the one hand, it is well known that the theory known colloquially as `Boxworld' \cite{barrett2007information}, which was formulated in order to realise arbitrary no-signalling correlations, only features local measurements and wirings thereof. That is, Boxworld does not display entangled measurements. This is in contrast to quantum theory, where entangled measurements are ubiquitous -- you may for instance think of the so-called Bell measurements. A natural question then arises: is there any relationship between the types of measurements a theory features and the correlations it may produce. 
Even seemingly simple at first sight, this question is far from trivial: by enlarging the set of allowed measurements one necessarily needs to shrink down the set of allowed states, since the state space featured by a given physical theory is constrained by the dual space of the effects\footnote{The so-called effects in a physical theory may be thought of as its dichotomic measurements.} space. How the set of allowed correlations (measurements on states) changes in consequence is therefore not straightforward.

Progress on this question was made in Ref.~\cite{czekaj2018bell}, where it was shown that demanding the existence of a particular entangled effect would constrain the correlations admissible in a bipartite Bell scenario to those realisable by an entangled pair of qubit quantum systems. That is, by demanding that the theory features a particular entangled measurement, it was shown that the allowed correlations in the so-called Clauser-Horne-Shimony-Holt (CHSH) scenario \cite{CHSH} was indeed the set of quantum ones. 

In this paper we explore what types of constraints the existence of bipartite effects impose on the possible correlations that a theory may feature. The framework we use to describe the underlying physical theory is that of General Probabilistic Theories (GPTs) \cite{hardy2001quantum,barrett2007information,Ludwig,davies1970operational,randall1970approach,Piron64,Mackey,chiribella2010probabilistic,hardy2011reformulating,schmid2020structure} . First, we take a compositional perspective and show how the existence of one (arbitrary) bipartite effect imposes not one but an infinite hierarchy of constraints which must be satisfied by the states and effects of the GPT. This hierarchy of constraints on bipartite states immediately translates {to} a hierarchy of constraints on the correlations realisable within the theory. Inspired by Ref.~\cite{czekaj2018bell}, we then consider a particular setup where we demand that there exists a measurement in the GPT which can measure the parity of fiducial measurements {(or a subset thereof), which we call a \textit{(partial)} parity reading measurement.} {We say that the observables which appear in such partial parity reading measurement are \emph{parity-readable observables}.} 
We then define an Optimisation Problem that computes the maximum violation of a Bell inequality by the corresponding GPT, provided that such a parity reading measurement exists. Such an Optimisation Problem provides a way to characterise the set of correlations allowed by such a GPT  when the parties choose among these parity-readable fiducial measurements in the Bell tests. The solution to this optimisation problem is, however, computationally complex, given that the problem itself is polynomial on the optimised variables. We hence present a series of relaxations that upper-bound the solution to the Optimisation problem. We finish by applying our techniques to a variety of Bell inequalities, {and discussing the necessity of the Local Tomography assumption.} 

\bigskip

{Inspired by our results, we moreover formulate a conjecture:}
\begin{conj}
{\it Under {the assumption of local tomography,}
the local observables that are {\bf parity-readable} satisfy Tsirelson's bound, i.e., they cannot violate Bell inequalities better than quantum mechanics does (with arbitrary measurements).} 
\end{conj}
 The inspiration came from the fact that various of our numerical explorations do indeed satisfy this property. {Moreover, we have a counterexample demonstrating the necessity of the assumption of local tomography within the conjecture. That is, if we have a GPT which does not satisfy local tomography, then it is possible to create a PR box using observables which are parity-readable.}

%

Finally, as we comment in the Discussion section, our work opens the door to further conjecturing that Quantum Theory, within the landscape of possible locally-tomographic physical theories, is the theory that features the best balance between allowed states and effects, in the sense that it yields the largest violation of any Bell inequality by parity-readable fiducial measurements.

\section{Descriptive summary of the results}

{Suppose that two parties, the ubiquitous Alice and Bob, each have three binary observables, $X$, $Y$, and $Z$, that they can measure on some shared system. Moreover, suppose that there exists some joint measurement that they could perform on their joint system if they were to get together, whose outcome would determine the parity of $XX$ and $ZZ$. That is, the joint measurement does not necessarily reveal the values that they would have obtained had they measured $X$ and $Z$ individually, but just whether or not their $XX$ and $ZZ$ measurements would have been correlated or not.\footnote{Such is the case in quantum theory, where $X$ and $Z$ are incompatible measurements, but their parity can nonetheless be measured by performing a suitable coarse graining of a standard Bell measurement.} We say that such observables {are \emph{parity-readable}, and illustrate this idea in Fig.~\ref{fig:PRM}.}

\begin{figure}
\begin{subfigure}[b]{0.3\textwidth}
  \centering
  \includegraphics[width=.9\linewidth]{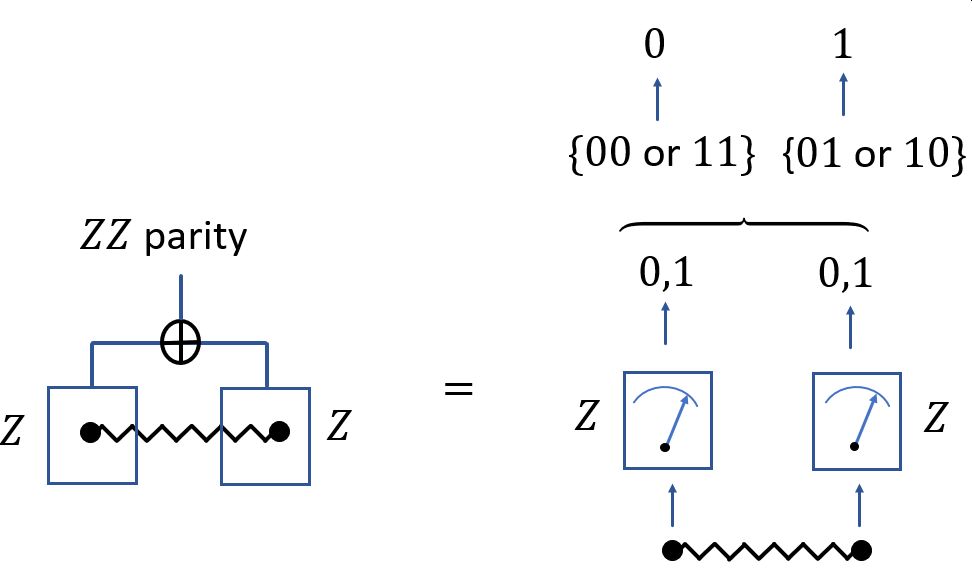}
  \caption{Measuring parity of local observables.}
  \label{fig:local-parity}
\end{subfigure}\hspace{5mm}
\centering
\begin{subfigure}[b]{0.3\textwidth}
  \centering
  \includegraphics[width=.9\linewidth]{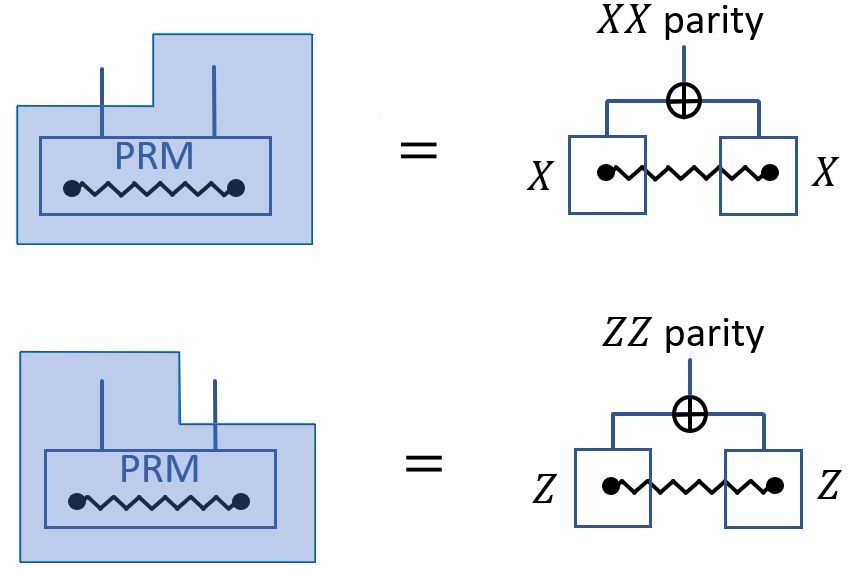}
  \caption{Reading parity of two pairs of observables: $XX$ and $ZZ$.}
  \label{fig:sub1}
\end{subfigure}\hspace{5mm}
\begin{subfigure}[b]{.3\textwidth}
  \centering
  \includegraphics[width=.9\linewidth]{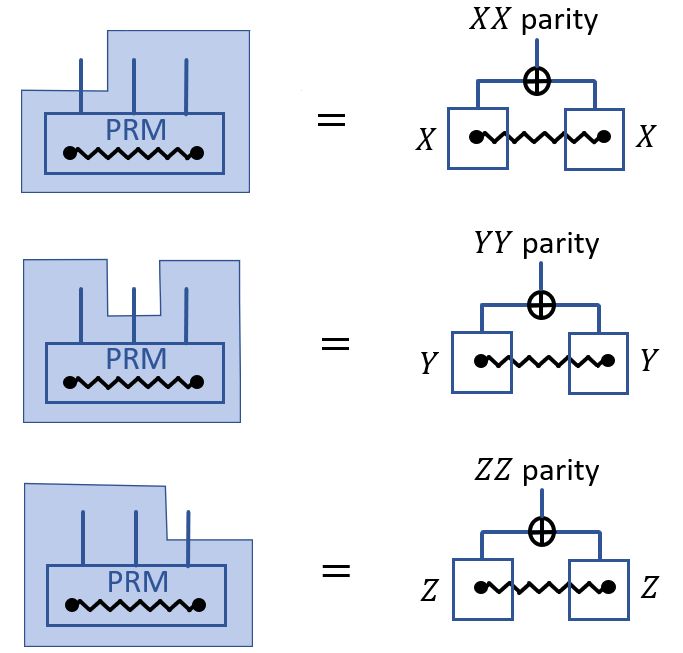}
  \caption{Reading parity of three pairs of observables $XX$, $ZZ$ and $YY$.}
  \label{fig:sub2}
\end{subfigure}
\caption{The idea of {a measurement that reads the parity of the parity-readable observables (denoted by PRM):} a composite measurement that measures parities of several pairs of local observables in one go.}
\label{fig:PRM}
\end{figure}

In this manuscript, we consider the impact that parity readability has on the correlations that can be generated in a Bell scenario {when measuring} parity-readable observables. For example, is it possible to create PR-box correlations via parity-readable observables? 

It is clear that, even within the standard quantum mechanical formalism, this imposes a restriction on the correlations which can be observed -- not all observables are parity-readable, and some correlations can only be achieved by those that are not. However, what about if we go beyond quantum mechanics?

We conjecture that, within this landscape of parity-readable observables, quantum theory is always optimal. That is, any correlation that can be generated by parity-readable observables, independently of which underlying  (tomographically-complete)  physical theory they belong to, can also be generated by parity reading observables within quantum theory.

If this conjecture is true, then this would be in stark contrast to the landscape of arbitrary observables, in which there are correlations which cannot be realised within our quantum world. It would therefore show, for the first time, a way in which quantum theory is an optimal physical theory for an information theoretic task.

Utilising techniques and insight coming from the field of generalised probabilistic theories we specify the constraints that the existence of a parity-reading measurement imposes on the possible correlations those observables may generate in a Bell test. These constraints are formulated as a hierarchy of convex optimisation problems which can be tackled using standard numerical methods. We apply this technique to and numerically explore various Bell scenarios and Bell inequalities, whose results lead us to formulating the conjecture discussed above. 
More precisely, we consider scenarios in which each party has at most three observables $X$, $Y$ and $Z$, and in which either two or three of these are parity-readable. The Bell inequalities explored include the CHSH \cite{CHSH}, AMP \cite{AMP-inequality-2012}, and AQ \cite{Navascues2015-AQ} inequalities. Further development of both the numerical methods, as well as analytical convex optimisation techniques, are necessary to explore this conjecture further.

{ It is worth highlighting that the main technique that we develop and apply here, relies on demanding the parity-readable observables to yield valid probabilities when applied both to a composite state as well as to products of steered states that can be generated from it (see Figs.~\ref{fig:cons}a, \ref{fig:cons}b.). These constraints are indeed phrased as positivity conditions on variables we optimise over. However, to capture the full power of the constraints that these parity-readable observables impose, one needs to take into account infinitely many conditions (examples of which are presented in Figs.~\ref{fig:cons}c and
\ref{fig:cons}d), described in the article, and which deserve further exploration. }
\begin{figure}
    \centering
    \includegraphics[width=0.7\linewidth]{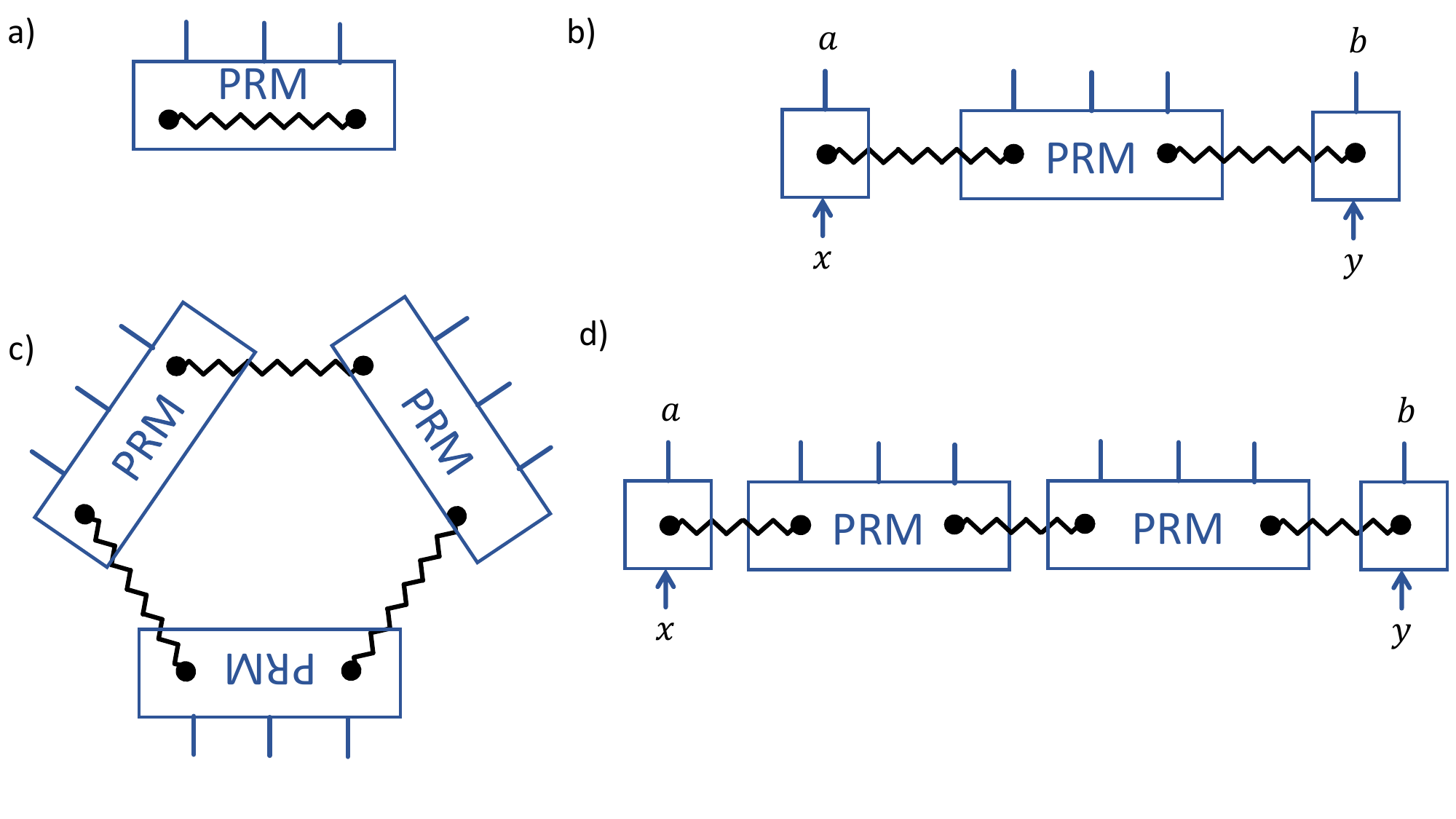}
    \caption{The existence of a measurement that reads the parity of the parity-readable observables, implies an infinite hierarchy of positivity constraints. For instance, the probabilities of outcomes should be positive on (a) a single copy of a state, (b) all pairs of  products of steered states that can be obtained from the original state,  (c) three copies of the state in a triangle network, and (d) a single copy of the state times any two pairs of steered states. }
    \label{fig:cons}
\end{figure}

\section{Warm-up: non-existence of Popescu-Rohrlich correlations for parity measurable $X$ and $Z$ observables.}\label{se:3}

While the general problem
of finding a  bound for Bell inequalities  for parity-readable observables is a complex one (as we will see further in this paper), one can relatively easily show that
the CHSH inequality cannot achieve its maximal algebraic bound. Namely, we shall show that parity-readable observables cannot exhibit so called Popescu-Rohrlich correlations \cite{popescu1994quantum}. 

In this section we shall present such reasoning, as a simple warm-up exercise in anticipation of the rest of the paper. 
In this warm-up we will take a device-independent approach,  in the sense of relying only on the conditional probability distributions for the argument (these black boxes are the only information we leverage for describing the underlying states of the system). In the remainder of the paper we will utilise the language of Generalised Probabilistic Theories (GPTs)
and, in particular, we will express the question using a formal diagrammatic language. Full justification of some formulas will be found later in the paper. For self-consistency of the main part of the paper, we will repeat there some definitions used here. 

\noindent
\subsection{States} 

\noindent
The so called {\it PR-box} \cite{popescu1994quantum} is defined as the set of conditional probabilities: 
\begin{equation}
    p(ab|xy)=
    \left \{\begin{array}{lll}
         \frac12 &\text{if} & a\oplus b= xy  \\
          0 & \text{else} & 
    \end{array}\right. \,,
\end{equation}
where $x,y\in\{0,1\}$ are inputs, and $a,b\in\{0,1\}$ are outputs. The observables $X,Z$ from the previous section are now encoded by the classical variables $x=0,1$ for Alice, and $y=0,1$ for Bob. 
The PR-box is a maximally nonlocal non-signaling box, since it violates the CHSH inequality up to its maximal algebraic value. Moreover, PR-box correlations have the feature that the pairs of observables $XX$, $XZ$, and $ZX$ are perfectly correlated, while the pair $ZZ$ is perfectly anticorrelated. 

The principles of Local Tomography  (see Sec.~\ref{se:GPTmain}) and No Signaling allow one to characterise any bipartite state $s$ by means of the following 
table: 
\begin{equation}
    \label{eq:ps-matrixform-warmup}
    \mathbf{p_s}=\left[ \begin{array}{ccc}
p_s(00|00)&p_s(00|01)   & p_s^{\mathrm{(1)}}(0|0)   \\
p_s(00|10)& p_s(00|11)  &  p_s^\mathrm{(1)}(0|1)  \\
p_s^\mathrm{(2)}(0|0)& p_s^\mathrm{(2)}(0|1)  & 1  
    \end{array}
    \right]\,,
\end{equation}
where $(i)$ specifies the party (i.e., (1) for Alice and (2) for Bob). This is indeed similar to the representation of non-signalling correlations in the CHSH scenario known as Collins-Gisin \cite{CG2004}, which shows how nine parameters are enough to fully specify the 16 components of the full conditional probability distribution.

Since a PR box has perfect correlations or anticorrelations for each pair of observables, it is natural that the steered states obtained from it 
are all the states that have well defined value for both observables. Thus, we have that, for each party, there are  four steered state, the state $s_1$ with $X=0,Z=0$, the state $s_2$ with   $X=0,Z=1$, 
the state $s_3$ with $X=1,Z=0$ and the state $s_4$ with $X=1,Z=1$. These four states are depicted in Fig. \ref{fig:square-bit-warmup}.
\begin{figure}[h]
    \centering
    \includegraphics[width=0.4\linewidth]{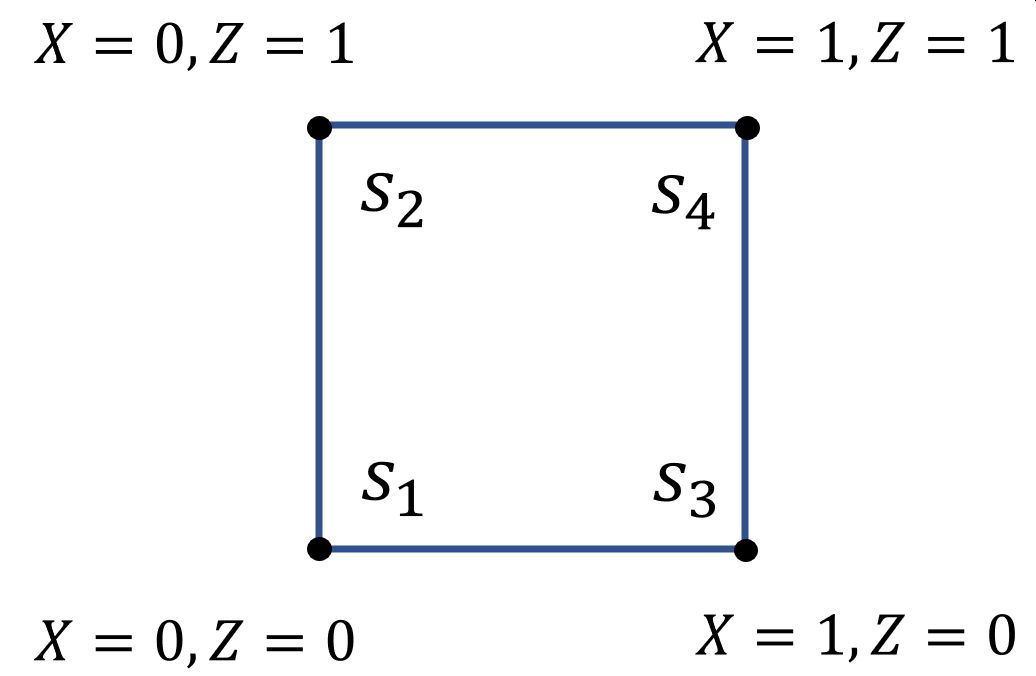}
    \caption{Steered states coming from PR box.}
    \label{fig:square-bit-warmup}
\end{figure}
We shall present here  the four pairs of steered states which we will use in the proof. 
These are products of all four steered states on Bob's site and a single fixed state on Alice's side, namely the state with $X_A=Z_A=1$, as is depicted Fig.~\ref{fig:pairs-of-steered-warmup}.
\begin{figure}[h]
    \centering
    \includegraphics[width=0.4\linewidth]{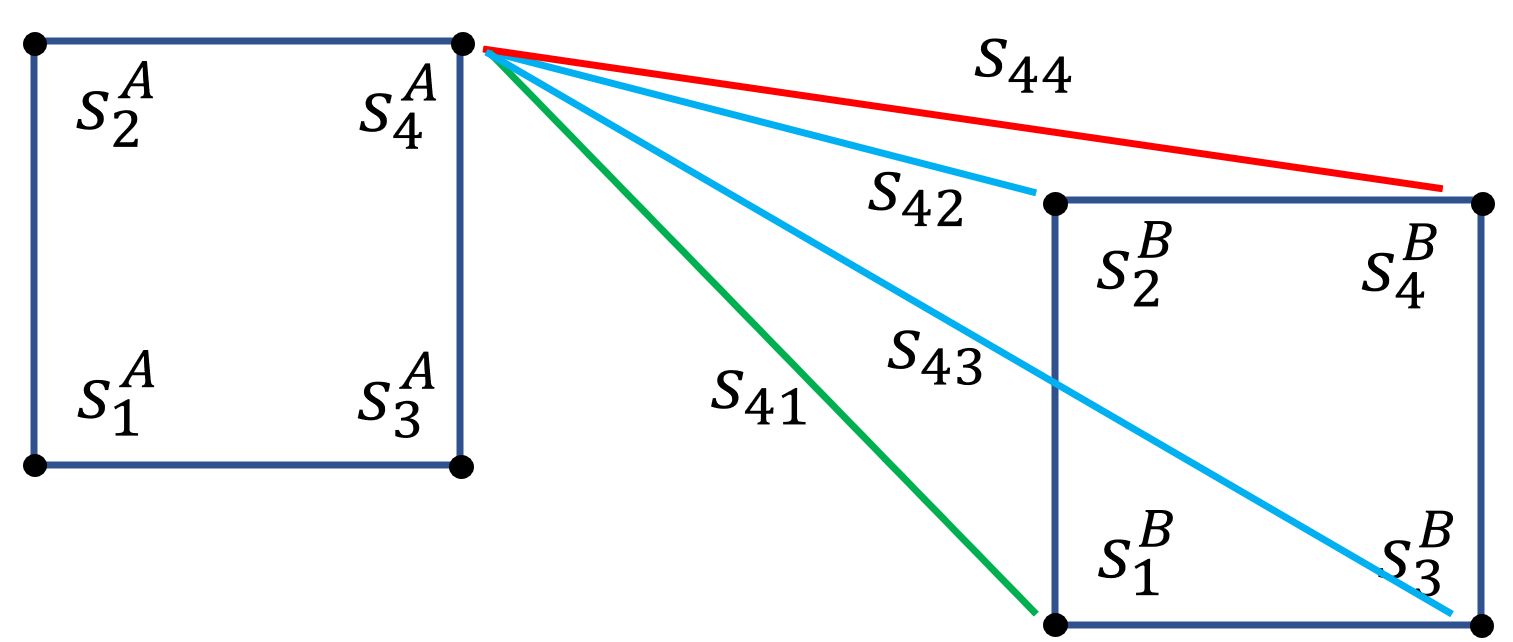}
    \caption{Selected pairs of steered states.}
    \label{fig:pairs-of-steered-warmup}
\end{figure}

We denote these four states as: 
\begin{align}
\label{eq:s}
s_{41} = s_4^A \otimes s_1^B, \quad 
s_{42} = s_4^A \otimes s_2^B, \quad 
s_{43} = s_4^A \otimes s_3^B, \quad
s_{44} = s_4^A \otimes s_4^B.
\end{align}
In the above matrix notation these are characterised by:
\begin{align}
\label{eq:s-matrix}
 s_{41}= \left[\begin{array}{ccc}
0& 0 & 0\\
0& 0 & 0\\
1& 1 & 1
    \end{array}\right] \, ,
 \quad
 s_{42}= \left[\begin{array}{ccc}
0& 0 & 0\\
0& 0 & 0\\
1& 0 & 1
    \end{array}\right]\, , 
    \nonumber \\
s_{43}= \left[\begin{array}{ccc}
0& 0 & 0\\
0& 0 & 0\\
0& 1 & 1
    \end{array}\right]\, ,
 \quad
 s_{44}= \left[\begin{array}{ccc}
0& 0 & 0\\
0& 0 & 0\\
0& 0 & 1
    \end{array}\right].
\end{align}

\noindent
\subsection{Parity Reading Measurement.}

\noindent
Now we consider the measurement that measures the parities of $XX$ and $ZZ$ in a single shot, that is, $X$ and $Z$ are \emph{parity-readable} observables. Specifically, this parity reading measurement (PRM) outputs a pair of bits: the first one reports the parity of $XX$ and the second one reports the parity of $ZZ$. 
This is expressed by  the following pair of conditions:

\begin{align}
\label{eq:prm}
    R_{XX}(\mathbf{p_s}) &\equiv p_s(00|\PRM) + p_s(01|\PRM) = p_s(00|00)+ p_s(11|00)\,, \\
    R_{ZZ}(\mathbf{p_s}) &\equiv p_s(00|\PRM) + p_s(10|\PRM) = p_s(00|11)+ p_s(11|11).
\end{align}
where we denote by   
$p_s(rq|\PRM)$ the probability of obtaining the pair of outcomes $rq$ when measuring the PRM on state $\mathbf{p_s}$. 
Notice that $R_{XX}$ ($R_{ZZ}$) have the interpretation of the probability of $XX$ ($ZZ$) being correlated (i.e., the probability of obtaining the same outcomes by Alice and Bob). In addition, note that the probabilities of the outcomes of the PRM satisfy the normalisation condition:
\begin{equation}
\label{eq:norm}
    p_s(00|\PRM)+ 
    p_s(01|\PRM)+
    p_s(10|\PRM)+
    p_s(11|\PRM)=1.
\end{equation}
We therefore have a system of three linear equations with four unknown quantities 
$p_s(rq|\PRM)$. Hence, a solution may be found with one quantity remaining an independent variable, for example the expression:
\begin{align}
\label{eq:c}
    C(\mathbf{p_s})=
    p_s(00|\PRM)-
    p_s(01|\PRM)-
    p_s(10|\PRM)+
    p_s(11|\PRM)\,.
\end{align}
Due to local tomography, the probabilities of the outcomes of any measurement for a particular state  can be written as a linear combination of state parameters, $\mathbf{p_s}$. Hence,  as $C$ is a linear combination of such probabilities, it can be computed via:
\begin{equation}
\label{eq:c2}
    C(\mathbf{p_s})= \mathcal{C}\cdot \mathbf{p_s} \,,
\end{equation}
where $\cdot$ denotes Frobenius matrix product, and 
$\mathcal{C}$ is the matrix of, at the moment, unspecified parameters describing the PRM:
\begin{align}
\label{eq:c-matrix}
    \mathcal{C}=
    \left[ \begin{array}{ccc}
c_{11}& c_{12} & c_{13}\\
c_{21}& c_{22} & c_{23}\\
c_{31}& c_{32} & c_{33}
    \end{array}\right].
\end{align}
Combining  Eqs.~\eqref{eq:prm}, \eqref{eq:norm}, \eqref{eq:c} and \eqref{eq:c2}, we obtain the formulas  for computing the probabilities of PRM outcomes,  expressed in terms of the state parameters, as well as the free parameter $\mathcal{C}$:
\begin{align}
\label{eq:prm2}
    p_s(00|\PRM)&=\frac14\left(2 R_{XX}(\mathbf{p_s}) + 2 R_{ZZ}(\mathbf{p_s}) +\mathcal{C}\cdot \mathbf{p_s} - 1 \right) \,, \nonumber \\
    p_s(01|\PRM)&=\frac14\left(2 R_{XX}(\mathbf{p_s}) - 2 R_{ZZ}(\mathbf{p_s}) -\mathcal{C}\cdot \mathbf{p_s} + 1 \right) \,, \nonumber \\
    p_s(01|\PRM)&=\frac14\left(-2 R_{XX}(\mathbf{p_s}) + 2 R_{ZZ}(\mathbf{p_s}) -\mathcal{C}\cdot \mathbf{p_s} + 1 \right) \,, \nonumber \\
    p_s(11|\PRM)&=\frac14\left(-2 R_{XX}(\mathbf{p_s}) - 2 R_{ZZ}(\mathbf{p_s}) +\mathcal{C}\cdot \mathbf{p_s} + 1 \right) \,.
\end{align}

\noindent
\subsection{Proof of nonexistence of the PR-box.}

\noindent
We shall now argue that, for any arbitrary choice of $\mathcal{C}$, the probability of at least one $\PRM$ output will necessarily be negative on some of the pairs of steered states. 
This proves that parity-readable observables cannot feature PR-box correlations, and hence cannot violate  the CHSH inequality up to its maximal algebraic bound. 

We shall first impose positivity of $p_s(qr|\PRM)$
for each of the four states in Eq.~ \eqref{eq:s}. 
Note that, for the state $s_{44}$, by definition, both $XX$ and $ZZ$ are perfectly correlated, since $X_A=X_B=1$ and $Z_A=Z_B=1$. Hence, $R_{XX}(s_{44})=R_{ZZ}(s_{44})=1$ for the state $s_{44}$. An analogous reasoning for the remaining three states yields:
\begin{align}
    R_{XX}(s_{41})&=0, \quad  R_{ZZ}(s_{41})=0, \nonumber \\
    R_{XX}(s_{42})&=0, \quad  R_{ZZ}(s_{42})=1, \nonumber \\
    R_{XX}(s_{43})&=1, \quad  R_{ZZ}(s_{43})=0, \nonumber \\
    R_{XX}(s_{44})&=1, \quad  R_{ZZ}(s_{44})=1.
\end{align}
Of course, we might alternatively compute these values from the matrix form of the states, and the definition of $R_{XX}$, $R_{ZZ}$. Inserting them into Eq.\eqref{eq:prm2} for each of the four states, yields the following conditions for the positivity of PRM probabilities:
\begin{align}
     p_{s_{41}}(qr|\PRM) &\geq 0 \quad \Leftrightarrow \quad \mathcal{C}\cdot s_{14} =1 \,, \nonumber \\
     p_{s_{42}}(qr|\PRM) &\geq 0 \quad \Leftrightarrow \quad \mathcal{C}\cdot s_{24} =-1 \,, \nonumber \\
     p_{s_{43}}(qr|\PRM) &\geq 0 \quad \Leftrightarrow \quad \mathcal{C}\cdot s_{34} =-1 \,, \nonumber \\
     p_{s_{44}}(qr|\PRM) &\geq 0 \quad \Leftrightarrow \quad \mathcal{C}\cdot s_{44} =1. 
\end{align}
We then use the matrix form of our states of Eq.~\eqref{eq:s-matrix} and the form of $\mathcal{C}$ of Eq.~\eqref{eq:c-matrix} and rewrite these conditions as: 
\begin{align}
    c_{31}+c_{32}+c_{33}=1,\quad c_{31}+c_{33}=-1, \quad c_{32}+c_{33}=-1, \quad c_{33}=1.
\end{align}
We see that this set of equalities does not have any real solutions. We therefore conclude that there does not exist a Parity Reading Measurement whose outcomes  give legitimate probabilities for the above four pairs of steered states. Since the existence of a PR-box allows for such steered states, we conclude that existence of Parity Reading Measurement excludes PR-box correlations. 

The remainder of this work explores this curious property in more detail. In particular: on the one hand, can we go beyond simply ruling out the PR-box and see how PRMs constrain the set of correlations; and, on the other hand, can we go beyond the assumptions involved in this derivation, namely, that of local tomography and that there are simply two observables per party.

\section{Generalised probabilistic theories}\label{se:GPTmain}

In order to explore the previously presented questions, we will work within the framework of 
 generalised probabilistic theories (GPTs) \cite{hardy2001quantum,barrett2007information}. This framework was developed in order to be able to describe essentially arbitrary conceivable theories of nature -- taking quantum and classical theory as just two particular points within a broad landscape of potential physical theories. The GPT framework is based on the idea that, ultimately, the way that we characterise physical devices is by the probabilities that they give rise to in experiments. From this simple observation, one can build a rich mathematical structure which any GPT must have. 
 
 GPTs have already proved a vital tool in the study of computation \cite{krumm2018quantum,barnum2018oracles,garner2018interferometric,barrett2017computational,lee2016deriving,lee2016bounds,lee2016generalised,lee2015computation,lee2017higher} and cryptography \cite{sikora2018simple,selby2018make,lami2018ultimate,barnum2011information,barnum2008nonclassicality,barrett2007information,barrett2005no} beyond quantum theory. Moreover, recently tools from convex optimisation theory have been used to gain new insight into GPTs \cite{selby2018make,sikora2018simple,fiorini2014generalized, JP17,bae2016structure,lami2018ultimate}. In particular, the generalisation of quantum theory to GPTs  is  analogous to the generalisation  from semi-definite to conic programmes. These optimisation tools will be vital for developing a complete understanding of how the structures of the GPT impact on the realisable correlations of the GPT.
 
 For simplicity of the presentation, in this manuscript we will focus on a particular class of GPTs, namely, those that satisfy the principle of Local Tomography. 
 A GPT is locally-tomographic if any state of a composite system can be uniquely determined by the information obtained from performing local measurements on its constituents (see, e.g., Ref.~\cite{hardy2001quantum} for a full formal definition). GPTs that satisfy local tomography tend to have very useful properties, and in particular they admit a useful parametrisation of their state and effect vectors, which will come in handy in various stages in this manuscript.  
 The majority of our results, such as the formulation of the hierarchy of constraints, however, do not require this principle to hold, hence we will highlight the instances where the assumption is indeed necessary.

In this paper we take a categorical approach to tomographically local GPTs. This is an intrinsically compositional approach, which allows us to describe arbitrary experimental scenarios. Moreover, the diagrammatic representation in terms of string diagrams, which comes from this approach, provides an intuitive way to reason about these complex situations. We provide a brief technical introduction to the formalism in Appendix~\ref{App:GPT}, and refer the reader  to Refs.~\cite{schmid2020structure,hardy2011reformulating,chiribella2010probabilistic,coecke_kissinger_2017,gogioso2017categorical,selby2018reconstructing} for more extensive introductions to these tools.

\subsection{Constraints on states and effects}\label{se:2.1}

We can see how the state and effect spaces constrain one another when demanding that scalars are probabilities.
For some system $V$ (see the Appendix for details on notation), the states can be thought of as vectors $s\in\Omega_V$ living inside $V$, and effects as linear functionals, $e\in \mathcal{E}_V$ living in the dual space $V^*$. Any pair of an effect and a state must satisfy:
\beq
\begin{tikzpicture}
	\begin{pgfonlayer}{nodelayer}
		\node [style=none] (0) at (-0.75, -0.5) {};
		\node [style=none] (1) at (0, -1.5) {};
		\node [style=none] (2) at (0.75, -0.5) {};
		\node [style=none] (3) at (0, -0.5) {};
		\node [style=none] (4) at (0, 0.5) {};
		\node [style=none] (5) at (-0.75, 0.5) {};
		\node [style=none] (6) at (0, 1.5) {};
		\node [style=none] (7) at (0.75, 0.5) {};
		\node [style=point] (8) at (0, -1) {$s$};
		\node [style=copoint] (9) at (0, 1) {$e$};
		\node [style={right label}] (10) at (0, -0) {$V$};
	\end{pgfonlayer}
	\begin{pgfonlayer}{edgelayer}
		\draw[qWire] (8.center) to (9.center);
	\end{pgfonlayer}
\end{tikzpicture} \in [0,1] \,.
\eeq
The geometric consequences of this for local and composite states {are presented} in App.~\ref{App:GPTNew}.

It was noted in Ref.~\cite{janotta2013generalized}, however, that even if a pair of state and effect spaces satisfy the standard constraints discussed in App.~\ref{App:GPTNew} (i.e., Eqs.~\eqref{eq7} and \eqref{eq8}), it is not straightforward that they actually define a valid GPT, at least when the No-restriction hypothesis is not assumed (i.e., when it is not required that $\Omega = \mathcal{E}^*$). An important condition that must also be checked, as shown in Theorem 9 of Ref.~\cite{janotta2013generalized}, is that the steered states are also valid states within the theory. That is, any bipartite state $s$ must satisfy:
\beq\label{eq:steest}
\InputIfFileExists{Diagrams/steerV.tikz}{}{\input{./figures/Diagrams/steerV.tikz}} \ \in \Omega_V
\quad\text{and}\quad
\InputIfFileExists{Diagrams/steerW.tikz}{}{\input{./figures/Diagrams/steerW.tikz}} \ \in \Omega_W
\eeq
for all $e_v\in\mathcal{E}_V$ and $e_w\in\mathcal{E}_W$. This constraint can be interpreted in many forms: 
\begin{compactitem}
\item as a constraint on the bipartite state space, namely, that a bipartite state must lead to valid steered states,
\item as a constraint on the local state spaces, namely, it forces them to include all of these steered states as valid local states,
\item as a constraint on the local effect spaces, namely, an effect is only allowed if it leads to valid steered states when composed with any bipartite state.
\end{compactitem}
However, we believe that the constraints of Eq.~\eqref{eq:steest} are probably best viewed not from any of these individual perspectives, but instead just as a compatibility condition between local states and effects, and bipartite states.

Similarly, we can consider bipartite effects $e$ and note that these have a similar compatibility condition together with local states and effects:
\beq
\InputIfFileExists{Diagrams/steerVeffect.tikz}{}{\input{./figures/Diagrams/steerVeffect.tikz}} \ \in \mathcal{E}_V
\quad\text{and}\quad
\InputIfFileExists{Diagrams/steerWeffect.tikz}{}{\input{./figures/Diagrams/steerWeffect.tikz}} \ \in \mathcal{E}_W
\eeq
for all $s_v\in\Omega_V$ and $s_w\in\Omega_W$.

One may be inclined to think that these constraints on bipartite states/effects, steered states, and steered effects, are sufficient to characterise a valid GPT. However, there are only the tip of the iceberg -- a whole plethora of further compatibility constraints lie underneath the surface. For example, consider a normalised bipartite state $s$ and a bipartite effect $e$. By taking two copies of each, one should be capable of wiring them as follows and obtain a valid probability:
\beq\label{eq:twotwo}
\InputIfFileExists{Diagrams/hier1.tikz}{}{\input{./figures/Diagrams/hier1.tikz}} \in [0,1]. 
\eeq
In addition, if one takes two copies of $s$ and one of $e$, one should be capable of wiring them as follows and obtain a valid bipartite state:
\beq\label{eq:twoone}
\InputIfFileExists{Diagrams/hier2.tikz}{}{\input{./figures/Diagrams/hier2.tikz}}\in\Omega_{V\otimes W}.
\eeq

Other types of compatibility constraints include diagrams like the following, which arise due to symmetry in the special case when all the local systems have the same type:
\beq\label{eq:braid1}
\InputIfFileExists{Diagrams/hiersym.tikz}{}{\input{./figures/Diagrams/hiersym.tikz}}\in[0,1],\quad %
\InputIfFileExists{Diagrams/hiersym2.tikz}{}{\input{./figures/Diagrams/hiersym2.tikz}}\in[0,1].
\eeq
One can readily see how these belong to an infinite family of constraints, each featuring the same (but arbitrary) number of normalised bipartite states $s$ and bipartite effects $e$ being connected in this ``braided'' fashion.

Even if the bipartite states consist of local systems of the same type, it is not necessary that they are symmetric under a swap operation of the local systems. Hence, a different hierarchy of braided-type constraints will arise by requiring consistent probability assignments to diagrams of the form:
\begin{align}\label{eq:braid2}
\begin{tikzpicture}
	\begin{pgfonlayer}{nodelayer}
		\node [style=none] (0) at (1, 1.5) {};
		\node [style=none] (1) at (-1, -1.5) {};
		\node [style=right label] (2) at (-1, -1.25) {$V$};
		\node [style=right label] (3) at (-3, -1.25) {$V$};
		\node [style=none] (4) at (-2, -2) {$s$};
		\node [style=none] (5) at (-0.25, -1.5) {};
		\node [style=none] (6) at (-3.75, -1.5) {};
		\node [style=none] (7) at (-2, -2.75) {};
		\node [style=none] (8) at (0.25, -1.5) {};
		\node [style=none] (9) at (2, -2) {$s$};
		\node [style=none] (10) at (2, -2.75) {};
		\node [style=none] (11) at (3.75, -1.5) {};
		\node [style=none] (12) at (7, -1.5) {};
		\node [style=none] (13) at (-1, 1.5) {};
		\node [style=right label] (14) at (-3, 1) {$V$};
		\node [style=none] (15) at (7, 1.5) {};
		\node [style=none] (16) at (-3, 1.5) {};
		\node [style=none] (17) at (2, 2) {$e$};
		\node [style=right label] (18) at (7, 1) {$V$};
		\node [style=none] (19) at (1, -1.5) {};
		\node [style=none] (20) at (3.75, 1.5) {};
		\node [style=none] (21) at (2, 2.75) {};
		\node [style=none] (22) at (0.25, 1.5) {};
		\node [style=right label] (23) at (1, -1.25) {$V$};
		\node [style=none] (24) at (-3, -1.5) {};
		\node [style=right label] (25) at (7, -1.25) {$V$};
		\node [style=right label] (26) at (1, 1) {$V$};
		\node [style=right label] (27) at (-1, 1) {$V$};
		\node [style=none] (28) at (-2, 2) {$e$};
		\node [style=none] (29) at (-0.25, 1.5) {};
		\node [style=none] (30) at (-2, 2.75) {};
		\node [style=none] (31) at (-3.75, 1.5) {};
		\node [style=none] (32) at (7.75, 1.5) {};
		\node [style=none] (33) at (6, -2) {$s$};
		\node [style=none] (34) at (7.75, -1.5) {};
		\node [style=none] (35) at (6, -2.75) {};
		\node [style=none] (36) at (4.25, 1.5) {};
		\node [style=none] (37) at (4.25, -1.5) {};
		\node [style=none] (38) at (6, 2.75) {};
		\node [style=none] (39) at (6, 2) {$e$};
		\node [style=none] (40) at (3.75, 1.5) {};
		\node [style=right label] (41) at (5, 1) {$V$};
		\node [style=none] (42) at (5, 1.5) {};
		\node [style=none] (43) at (3, -1.5) {};
		\node [style=none] (44) at (4.25, 1.5) {};
		\node [style=none] (45) at (3.75, -1.5) {};
		\node [style=right label] (46) at (3, -1.25) {$V$};
		\node [style=right label] (47) at (3, 1) {$V$};
		\node [style=none] (48) at (3, 1.5) {};
		\node [style=none] (49) at (5, -1.5) {};
		\node [style=right label] (50) at (5, -1.25) {$V$};
		\node [style=none] (51) at (4.25, -1.5) {};
	\end{pgfonlayer}
	\begin{pgfonlayer}{edgelayer}
		\draw [qWire, in=-90, out=90] (1.center) to (0.center);
		\draw (5.center) to (7.center);
		\draw (7.center) to (6.center);
		\draw (6.center) to (5.center);
		\draw (11.center) to (10.center);
		\draw (10.center) to (8.center);
		\draw (8.center) to (11.center);
		\draw [qWire, in=90, out=-90, looseness=0.50] (16.center) to (12.center);
		\draw [qWire, in=90, out=-90, looseness=0.50] (15.center) to (24.center);
		\draw [qWire, in=90, out=-90] (13.center) to (19.center);
		\draw (20.center) to (21.center);
		\draw (21.center) to (22.center);
		\draw (22.center) to (20.center);
		\draw (29.center) to (30.center);
		\draw (30.center) to (31.center);
		\draw (31.center) to (29.center);
		\draw (34.center) to (35.center);
		\draw (35.center) to (37.center);
		\draw (37.center) to (34.center);
		\draw (32.center) to (38.center);
		\draw (38.center) to (36.center);
		\draw (36.center) to (32.center);
		\draw [qWire, in=-90, out=90] (43.center) to (42.center);
		\draw [qWire, in=90, out=-90] (48.center) to (49.center);
	\end{pgfonlayer}
\end{tikzpicture}
 & \in [0,1]. 
\end{align}

In Section \ref{se:bel} we will see how to formalise these types of hierarchies, and how they can be used to constrain the potential correlations in a GPT.

\subsection{Correlations in a GPT}

{To describe correlations in a GPT we must first introduce classical systems. Here we describe a classical system by classical random variable, which can take values from a set, such as $X,Y,A,B$. We denote these classical systems by thin gray wires (to distinguish them from GPT wires). Correlations, in this formalism, are then viewed as no-signalling stochastic maps, $\mathsf{N}:X\times Y \to A\times B$, between these random variables. Diagrammatically, we denote these no-signalling boxes as:
    \beq %
\InputIfFileExists{Diagrams/nsBox.tikz}{}{\input{./figures/Diagrams/nsBox.tikz}}\,,
    \eeq
    which must satisfy the no-signalling constraints:
    \beq\label{eq:nosig}
\InputIfFileExists{Diagrams/nsBox1.tikz}{}{\input{./figures/Diagrams/nsBox1.tikz}}\ \ =\ \ %
\InputIfFileExists{Diagrams/nsBox4.tikz}{}{\input{./figures/Diagrams/nsBox4.tikz}} \qquad\text{and}\qquad %
\InputIfFileExists{Diagrams/nsBox3.tikz}{}{\input{./figures/Diagrams/nsBox3.tikz}}\ \ =\ \ %
\InputIfFileExists{Diagrams/nsBox2.tikz}{}{\input{./figures/Diagrams/nsBox2.tikz}}\,.
    \eeq
    This is equivalent to the standard view of correlations \cite{brunner2014bell} as being described by a conditional probability distribution $\mathsf{Pr}(A,B|X,Y) = \{p(ab|xy)\}_{\{a\in A,b\in B, x\in X, y\in Y\}}$, which can be seen by defining:
    \beq
    p(ab|xy) := \begin{tikzpicture}
    	\begin{pgfonlayer}{nodelayer}
    		\node [style=none] (0) at (-1, -0.5) {};
    		\node [style=none] (1) at (-1, -1.75) {};
    		\node [style=none] (2) at (-1, 0.5) {};
    		\node [style=none] (3) at (-1, 1.25) {};
    		\node [style=none] (4) at (1, 1.5) {};
    		\node [style=none] (5) at (1, 0.5) {};
    		\node [style=none] (6) at (1, -1.75) {};
    		\node [style=none] (7) at (1, -0.5) {};
    		\node [style=right label] (8) at (-1, -1) {$X$};
    		\node [style=right label] (9) at (1, -1) {$Y$};
    		\node [style=right label] (10) at (-1, 0.75) {$A$};
    		\node [style=right label] (11) at (1, 0.75) {$B$};
    		\node [style=none] (12) at (-1.5, 0.5) {};
    		\node [style=none] (13) at (-1.5, -0.5) {};
    		\node [style=none] (14) at (0, 0) {$\mathsf{N}$};
    		\node [style=none] (15) at (1.5, 0.5) {};
    		\node [style=none] (16) at (1.5, -0.5) {};
    		\node [style=copoint] (17) at (-1, 1.5) {$a$};
    		\node [style=copoint] (18) at (1, 1.5) {$b$};
    		\node [style=point] (19) at (-1, -1.75) {$x$};
    		\node [style=point] (20) at (1, -1.75) {$y$};
    	\end{pgfonlayer}
    	\begin{pgfonlayer}{edgelayer}
    		\draw [cWire, in=90, out=-90] (0.center) to (1.center);
    		\draw [cWire, in=-90, out=90] (2.center) to (3.center);
    		\draw [cWire, in=90, out=-90] (7.center) to (6.center);
    		\draw [cWire, in=-90, out=90] (5.center) to (4.center);
    		\draw (13.center) to (12.center);
    		\draw (12.center) to (15.center);
    		\draw (15.center) to (16.center);
    		\draw (16.center) to (13.center);
    	\end{pgfonlayer}
    \end{tikzpicture}\,,
    \eeq
    and checking that the no-signalling conditions of Eqs.~\eqref{eq:nosig} are equivalent to the standard no-signalling conditions for the conditional probability distribution. To do so it is useful to note that, for example:
    \beq
    \begin{tikzpicture}
	\begin{pgfonlayer}{nodelayer}
		\node [style=none] (0) at (0, -1.5) {};
		\node [style=none] (1) at (0, -0.25) {};
		\node [style=none] (2) at (0, -1.5) {};
		\node [style=none] (3) at (0, -0.25) {};
		\node [style=upground] (4) at (0, -0.0) {};
		\node [style=right label] (5) at (0, -1.25) {$X$};
	\end{pgfonlayer}
	\begin{pgfonlayer}{edgelayer}
		\draw [cWire] (3.center) to (2.center);
	\end{pgfonlayer}
\end{tikzpicture}
\quad:= \quad
\sum_{x\in X} 
\begin{tikzpicture}
	\begin{pgfonlayer}{nodelayer}
		\node [style=none] (0) at (0, -1.5) {};
		\node [style=none] (1) at (0, -0.25) {};
		\node [style=none] (2) at (0, -1.5) {};
		\node [style=none] (3) at (0, -0.25) {};
		\node [style=copoint] (4) at (0, 0.0) {$x$};
		\node [style=right label] (5) at (0, -1.25) {$X$};
	\end{pgfonlayer}
	\begin{pgfonlayer}{edgelayer}
		\draw [cWire] (3.center) to (2.center);
	\end{pgfonlayer}
\end{tikzpicture} \ .
\eeq
   
Then, in order to understand the possible correlations in a GPT, it is useful to describe measurements as transformations from a GPT to a classical system, where the choice of measurement is controlled by another classical system. These controlled measurements must satisfy the constraint:
\beq\label{eq:terminalMeasurement}
\begin{tikzpicture}
	\begin{pgfonlayer}{nodelayer}
		\node [style={right label}] (0) at (1.25, -1.5) {$V$};
		\node [style=none] (1) at (0.5, -0.5) {};
		\node [style=none] (2) at (1.25, -1.75) {};
		\node [style=none] (3) at (0.5, 0.5) {};
		\node [style=none] (4) at (-0.75, 0.5) {};
		\node [style=none] (5) at (-0.75, -0.5) {};
		\node [style=none] (6) at (0, -0) {$M$};
		\node [style=none] (7) at (-0.25, -0.5) {};
		\node [style=none] (8) at (-0.25, -1.75) {};
		\node [style=none] (9) at (-0.25, 0.5) {};
		\node [style=none] (10) at (-0.25, 1.75) {};
		\node [style={right label}] (11) at (-0.25, -1.5) {$X$};
		\node [style={right label}] (12) at (-0.25, 1.25) {$A$};
		\node [style=none] (13) at (1, -0.5) {};
		\node [style=upground] (14) at (-0.25, 2) {};
	\end{pgfonlayer}
	\begin{pgfonlayer}{edgelayer}
		\draw [qWire, in=90, out=-90, looseness=1.00] (1.center) to (2.center);
		\draw (3.center) to (4.center);
		\draw (4.center) to (5.center);
		\draw [cWire, in=90, out=-90, looseness=1.00] (7.center) to (8.center);
		\draw [cWire, in=-90, out=90, looseness=1.00] (9.center) to (10.center);
		\draw (5.center) to (13.center);
		\draw (13.center) to (3.center);
	\end{pgfonlayer}
\end{tikzpicture}\quad=\quad \begin{tikzpicture}
	\begin{pgfonlayer}{nodelayer}
		\node [style={right label}] (0) at (0.75, -0.75) {$V$};
		\node [style=none] (1) at (0.75, 0.25) {};
		\node [style=none] (2) at (0.75, -1) {};
		\node [style=none] (3) at (-0.75, 0.25) {};
		\node [style=none] (4) at (-0.75, -1) {};
		\node [style={right label}] (5) at (-0.75, -0.75) {$X$};
		\node [style=upground] (6) at (-0.75, 0.5) {};
		\node [style=upground] (7) at (0.75, 0.5) {};
	\end{pgfonlayer}
	\begin{pgfonlayer}{edgelayer}
		\draw [qWire, in=90, out=-90, looseness=1.00] (1.center) to (2.center);
		\draw [cWire, in=90, out=-90, looseness=1.00] (3.center) to (4.center);
	\end{pgfonlayer}
\end{tikzpicture}\ .
\eeq
}
Correlations that can be generated in a Bell experiment are hence of the form:
\beq\label{eq:Bell}
\InputIfFileExists{Diagrams/Bell.tikz}{}{\input{./figures/Diagrams/Bell.tikz}}
\eeq
where the local controlled measurements $M_A$ and $M_B$ (for Alice and Bob respectively) are performed on a bipartite system on state $s$, with local system types $V,W$ in the GPT.
These local measurements are controlled on the input classical variable and have an outcome recorded in the output classical variable. 

{In any GPT, such a diagram corresponds to a no-signalling stochastic map: 
\beq 
\begin{tikzpicture}
	\begin{pgfonlayer}{nodelayer}
		\node [style=none] (0) at (1, 0.75) {};
		\node [style=none] (1) at (1, -0.5) {};
		\node [style=right label] (2) at (1, -0.25) {$W$};
		\node [style=right label] (3) at (-1, -0.25) {$V$};
		\node [style=none] (4) at (0, -1) {$s$};
		\node [style=none] (5) at (1.75, -0.5) {};
		\node [style=none] (6) at (-1.75, -0.5) {};
		\node [style=none] (7) at (0, -1.75) {};
		\node [style=none] (8) at (-1, 0.75) {};
		\node [style=none] (9) at (-1, -0.5) {};
		\node [style=none] (10) at (-0.5, 0.75) {};
		\node [style=none] (11) at (-1, 1.75) {};
		\node [style=none] (12) at (-2.5, 1.75) {};
		\node [style=none] (13) at (-2.5, 0.75) {};
		\node [style=none] (14) at (-1.75, 1.25) {$M_A$};
		\node [style=none] (15) at (-1.75, -2) {};
		\node [style=none] (16) at (-1.75, -3) {};
		\node [style=none] (17) at (-1.75, 1.75) {};
		\node [style=none] (18) at (-1.75, 3) {};
		\node [style=none] (19) at (0.5, 0.75) {};
		\node [style=none] (20) at (1, 1.75) {};
		\node [style=none] (21) at (1.75, 3) {};
		\node [style=none] (22) at (1.75, 1.75) {};
		\node [style=none] (23) at (1.75, 1.25) {$M_B$};
		\node [style=none] (24) at (2.5, 0.75) {};
		\node [style=none] (25) at (2.5, 1.75) {};
		\node [style=none] (26) at (1.75, -3) {};
		\node [style=none] (27) at (1.75, -2) {};
		\node [style=right label] (28) at (-1.75, -2.75) {$X$};
		\node [style=right label] (29) at (1.75, -2.75) {$Y$};
		\node [style=right label] (30) at (-1.75, 3) {$A$};
		\node [style=right label] (31) at (1.75, 3) {$B$};
		\node [style=none] (32) at (-2.75, 2.25) {};
		\node [style=none] (33) at (-2.75, -2) {};
		\node [style=none] (34) at (2.75, -2) {};
		\node [style=none] (35) at (2.75, 2.25) {};
		\node [style=none] (36) at (-2, 0.75) {};
		\node [style=none] (37) at (2, 0.75) {};
	\end{pgfonlayer}
	\begin{pgfonlayer}{edgelayer}
		\draw [qWire, in=-90, out=90] (1.center) to (0.center);
		\draw (5.center) to (7.center);
		\draw (7.center) to (6.center);
		\draw (6.center) to (5.center);
		\draw [qWire] (8.center) to (9.center);
		\draw (13.center) to (10.center);
		\draw (10.center) to (11.center);
		\draw (11.center) to (12.center);
		\draw (12.center) to (13.center);
		\draw [cWire, in=90, out=-90] (15.center) to (16.center);
		\draw [cWire, in=-90, out=90] (17.center) to (18.center);
		\draw (24.center) to (19.center);
		\draw (19.center) to (20.center);
		\draw (20.center) to (25.center);
		\draw (25.center) to (24.center);
		\draw [cWire, in=90, out=-90] (27.center) to (26.center);
		\draw [cWire, in=-90, out=90] (22.center) to (21.center);
		\draw [thick gray dashed edge] (32.center)
			 to (35.center)
			 to (34.center)
			 to (33.center)
			 to cycle;
		\draw [cWire, in=90, out=-90] (36.center) to (15.center);
		\draw [cWire, in=90, out=-90] (37.center) to (27.center);
	\end{pgfonlayer}
\end{tikzpicture}
\quad=:\quad %
\InputIfFileExists{Diagrams/nsBox.tikz}{}{\input{./figures/Diagrams/nsBox.tikz}}
\ .
\eeq
It can then be shown that the constraint on measurements of Eq.~\ref{eq:terminalMeasurement} immediately implies the relevant no-signalling conditions, for example:
\beq
\InputIfFileExists{Diagrams/nsBox1.tikz}{}{\input{./figures/Diagrams/nsBox1.tikz}}\quad = \quad
\begin{tikzpicture}
	\begin{pgfonlayer}{nodelayer}
		\node [style=none] (0) at (1, 0.75) {};
		\node [style=none] (1) at (1, -0.5) {};
		\node [style=right label] (2) at (1, -0.25) {$W$};
		\node [style=right label] (3) at (-1, -0.25) {$V$};
		\node [style=none] (4) at (0, -1) {$s$};
		\node [style=none] (5) at (1.75, -0.5) {};
		\node [style=none] (6) at (-1.75, -0.5) {};
		\node [style=none] (7) at (0, -1.75) {};
		\node [style=none] (8) at (-1, 0.75) {};
		\node [style=none] (9) at (-1, -0.5) {};
		\node [style=none] (10) at (-0.5, 0.75) {};
		\node [style=none] (11) at (-1, 1.75) {};
		\node [style=none] (12) at (-2.5, 1.75) {};
		\node [style=none] (13) at (-2.5, 0.75) {};
		\node [style=none] (14) at (-1.75, 1.25) {$M_A$};
		\node [style=none] (15) at (-1.75, -2) {};
		\node [style=none] (16) at (-1.75, -3) {};
		\node [style=none] (17) at (-1.75, 1.75) {};
		\node [style=none] (18) at (-1.75, 3) {};
		\node [style=none] (19) at (0.5, 0.75) {};
		\node [style=none] (20) at (1, 1.75) {};
		\node [style=none] (21) at (1.75, 3) {};
		\node [style=none] (22) at (1.75, 1.75) {};
		\node [style=none] (23) at (1.75, 1.25) {$M_B$};
		\node [style=none] (24) at (2.5, 0.75) {};
		\node [style=none] (25) at (2.5, 1.75) {};
		\node [style=none] (26) at (1.75, -3) {};
		\node [style=none] (27) at (1.75, -2) {};
		\node [style=right label] (28) at (-1.75, -2.75) {$X$};
		\node [style=right label] (29) at (1.75, -2.75) {$Y$};
		\node [style=right label] (30) at (-1.75, 3) {$A$};
		\node [style=right label] (31) at (1.75, 2.5) {$B$};
		\node [style=none] (32) at (-2.75, 2.25) {};
		\node [style=none] (33) at (-2.75, -2) {};
		\node [style=none] (34) at (2.75, -2) {};
		\node [style=none] (35) at (2.75, 2.25) {};
		\node [style=none] (36) at (-2, 0.75) {};
		\node [style=none] (37) at (2, 0.75) {};
		\node [style=upground] (38) at (1.75, 3.25) {};
	\end{pgfonlayer}
	\begin{pgfonlayer}{edgelayer}
		\draw [qWire, in=-90, out=90] (1.center) to (0.center);
		\draw (5.center) to (7.center);
		\draw (7.center) to (6.center);
		\draw (6.center) to (5.center);
		\draw [qWire] (8.center) to (9.center);
		\draw (13.center) to (10.center);
		\draw (10.center) to (11.center);
		\draw (11.center) to (12.center);
		\draw (12.center) to (13.center);
		\draw [cWire, in=90, out=-90] (15.center) to (16.center);
		\draw [cWire, in=-90, out=90] (17.center) to (18.center);
		\draw (24.center) to (19.center);
		\draw (19.center) to (20.center);
		\draw (20.center) to (25.center);
		\draw (25.center) to (24.center);
		\draw [cWire, in=90, out=-90] (27.center) to (26.center);
		\draw [cWire, in=-90, out=90] (22.center) to (21.center);
		\draw [thick gray dashed edge] (32.center)
			 to (35.center)
			 to (34.center)
			 to (33.center)
			 to cycle;
		\draw [cWire, in=90, out=-90] (36.center) to (15.center);
		\draw [cWire, in=90, out=-90] (37.center) to (27.center);
	\end{pgfonlayer}
\end{tikzpicture}
\quad=\quad
\begin{tikzpicture}
	\begin{pgfonlayer}{nodelayer}
		\node [style=none] (0) at (1, 0.75) {};
		\node [style=none] (1) at (1, 0) {};
		\node [style=right label] (2) at (1, 0.25) {$W$};
		\node [style=right label] (3) at (-1, 0.25) {$V$};
		\node [style=none] (4) at (0, -0.5) {$s$};
		\node [style=none] (5) at (1.75, 0) {};
		\node [style=none] (6) at (-1.75, 0) {};
		\node [style=none] (7) at (0, -1.25) {};
		\node [style=none] (8) at (-1, 0.75) {};
		\node [style=none] (9) at (-1, 0) {};
		\node [style=none] (10) at (-0.5, 0.75) {};
		\node [style=none] (11) at (-1, 1.75) {};
		\node [style=none] (12) at (-2.5, 1.75) {};
		\node [style=none] (13) at (-2.5, 0.75) {};
		\node [style=none] (14) at (-1.75, 1.25) {$M_A$};
		\node [style=none] (15) at (-1.75, -1.5) {};
		\node [style=none] (16) at (-1.75, -3.25) {};
		\node [style=none] (17) at (-1.75, 1.75) {};
		\node [style=none] (18) at (-1.75, 3) {};
		\node [style=none] (26) at (3, -3.25) {};
		\node [style=none] (27) at (3, -2.5) {};
		\node [style=right label] (28) at (-1.75, -3) {$X$};
		\node [style=right label] (29) at (3, -3) {$Y$};
		\node [style=right label] (30) at (-1.75, 3) {$A$};
		\node [style=none] (32) at (-3, 2.25) {};
		\node [style=none] (33) at (-3, -1.5) {};
		\node [style=none] (34) at (2, -1.5) {};
		\node [style=none] (35) at (2, 2.25) {};
		\node [style=none] (36) at (-2, 0.75) {};
		\node [style=upground] (38) at (1, 1) {};
		\node [style=upground] (39) at (3, -2.25) {};
	\end{pgfonlayer}
	\begin{pgfonlayer}{edgelayer}
		\draw [qWire, in=-90, out=90] (1.center) to (0.center);
		\draw (5.center) to (7.center);
		\draw (7.center) to (6.center);
		\draw (6.center) to (5.center);
		\draw [qWire] (8.center) to (9.center);
		\draw (13.center) to (10.center);
		\draw (10.center) to (11.center);
		\draw (11.center) to (12.center);
		\draw (12.center) to (13.center);
		\draw [cWire, in=90, out=-90] (15.center) to (16.center);
		\draw [cWire, in=-90, out=90] (17.center) to (18.center);
		\draw [cWire, in=90, out=-90] (27.center) to (26.center);
		\draw [thick gray dashed edge] (32.center)
			 to (35.center)
			 to (34.center)
			 to (33.center)
			 to cycle;
		\draw [cWire, in=90, out=-90] (36.center) to (15.center);
	\end{pgfonlayer}
\end{tikzpicture}
\quad=:\quad
\InputIfFileExists{Diagrams/nsBox4.tikz}{}{\input{./figures/Diagrams/nsBox4.tikz}}\,.
\eeq
A more general version of this proof first appeared in Ref.~\cite{coecke2014terminality} and was generalised to arbitrary causal structures in Ref.~\cite{kissinger2017equivalence}.
}

Bell inequalities \cite{Bell64,brunner2014bell} are then a particular class of linear functionals from this space of stochastic maps to the reals. A linear functional corresponding to a Bell inequality, hereon denoted by $\mathcal{I}$, can be diagrammatically denoted as:
\beq
\InputIfFileExists{Diagrams/InequalityBell2.tikz}{}{\input{./figures/Diagrams/InequalityBell2.tikz}} \ \  ::\ \  %
\InputIfFileExists{Diagrams/nsBox.tikz}{}{\input{./figures/Diagrams/nsBox.tikz}}\quad \mapsto\quad %
\InputIfFileExists{Diagrams/InequalityBell3.tikz}{}{\input{./figures/Diagrams/InequalityBell3.tikz}} \in \mathds{R} \,.
\eeq
Note that $\mathcal{I}$ should not be interpreted as a process within the GPT --  $\mathcal{I}$ is simply some linear functional, and can lead to negative values.

The value of a Bell inequality on a stochastic map $\mathsf{N}$ -- realised within the GPT as per Eq.~\eqref{eq:Bell} -- is given by:
\beq
\InputIfFileExists{Diagrams/InequalityBell1.tikz}{}{\input{./figures/Diagrams/InequalityBell1.tikz}}\,.
\eeq

The maximal value of a Bell inequality $\mathcal{I}$ achievable by correlations within a given GPT $\mathcal{G}$, is therefore given by the following optimisation problem:
\beq\label{eq:optoverVtoo}
\mathcal{I}_{\max}:=\mathsf{sup}\left\{\ \ %
\InputIfFileExists{Diagrams/InequalityBell1.tikz}{}{\input{./figures/Diagrams/InequalityBell1.tikz}}\ \ \middle|\ \ V,W, M_A,M_B,s \in \mathcal{G}  \right\}\,.
\eeq
Notice that the optimisation is carried over the types of systems $V$ and $W$ present in $\mathcal{G}$, as well as over the local measurements $M_A$ and $M_B$, and bipartite states $s$. 
The solution to this optimisation problem will of course depend on the properties of the GPT being studied. However, we readily see that when maximising over states and measurements from the theory, the compatibility constraints we discussed in the previous section will play a crucial role. Indeed, if the GPT admits some bipartite effect $e$, then the above mentioned hierarchy of constraints (see Eqs.~\eqref{eq:twotwo}, \eqref{eq:twoone}, \eqref{eq:braid1}, and \eqref{eq:braid2}) will restrict the sets of states that the value of $\mathcal{I}$ is optimised over. In other words, the existence of bipartite effects $e$ within the GPT will impose a hierarchy of constraints on the correlations that such GPT may feature. In the next sections we elaborate on this fact with a concrete example.

\section{Parity reading measurement}\label{se:PRM}

In this section we will explore the constraints on the correlations that a GPT may feature, given that bipartite effects associated to a particular measurement -- which we call a \textit{Parity reading measurement} (PRM) -- exist within the GPT. 

Suppose we have a controlled measurement, $M$, for a system $V$, with a setting variable labeled by the set $\nSet:=\{0,...,n-1\}$ such that $|\nSet|=n$, and a binary outcome variable $\bSet:=\{0,1\}$ as an outcome. { This is diagrammatically denoted  by:}
\beq
\InputIfFileExists{Diagrams/ControlledMeasurement.tikz}{}{\input{./figures/Diagrams/ControlledMeasurement.tikz}}\,.
\eeq
{Recall that, as this is a measurement, it must satisfy:
\beq
\begin{tikzpicture}
	\begin{pgfonlayer}{nodelayer}
		\node [style={right label}] (0) at (0.25, -0.5) {$V$};
		\node [style=none] (1) at (0.25, -0) {};
		\node [style=none] (2) at (0.25, -0.75) {};
		\node [style=none] (3) at (-1, -0) {};
		\node [style=none] (4) at (0.75, -0) {};
		\node [style=none] (5) at (0.75, 1) {};
		\node [style=none] (6) at (-0.75, 1) {};
		\node [style=none] (7) at (0, 0.5) {$M$};
		\node [style=none] (8) at (-0.5, -0) {};
		\node [style=none] (9) at (-0.5, -0.75) {};
		\node [style=none] (10) at (0, 1) {};
		\node [style=none] (11) at (0, 1.75) {};
		\node [style={right label}] (12) at (-0.5, -0.5) {$\nSet$};
		\node [style={right label}] (13) at (0, 1.375) {$\bSet$};
		\node [style=upground] (14) at (0, 2) {};
	\end{pgfonlayer}
	\begin{pgfonlayer}{edgelayer}
		\draw [qWire] (1.center) to (2.center);
		\draw (3.center) to (4.center);
		\draw (4.center) to (5.center);
		\draw (5.center) to (6.center);
		\draw (6.center) to (3.center);
		\draw [cWire] (11.center) to (10.center);
		\draw [cWire] (8.center) to (9.center);
	\end{pgfonlayer}
\end{tikzpicture}\quad=\quad \begin{tikzpicture}
	\begin{pgfonlayer}{nodelayer}
		\node [style={right label}] (0) at (0.75, -0.75) {$V$};
		\node [style=none] (1) at (0.75, 0.25) {};
		\node [style=none] (2) at (0.75, -1) {};
		\node [style=none] (3) at (-0.75, 0.25) {};
		\node [style=none] (4) at (-0.75, -1) {};
		\node [style={right label}] (5) at (-0.75, -0.75) {$X$};
		\node [style=upground] (6) at (-0.75, 0.5) {};
		\node [style=upground] (7) at (0.75, 0.5) {};
	\end{pgfonlayer}
	\begin{pgfonlayer}{edgelayer}
		\draw [qWire, in=90, out=-90, looseness=1.00] (1.center) to (2.center);
		\draw [cWire, in=90, out=-90, looseness=1.00] (3.center) to (4.center);
	\end{pgfonlayer}
\end{tikzpicture}\,,
\eeq
which ensures that the correlations it can generate are no-signalling.
}

Then we can define a measurement $\mathcal{P}[M]$ which reads out the parity of such a measurement $M$ as follows:

\begin{definition}\textbf{Parity Reading Measurement.--}\\
 A \emph{parity reading measurement} for $M$, denoted by $\mathcal{P}[M]$, is a bipartite measurement on $V\otimes V$ with $n$ binary variables as outputs:
\beq
\InputIfFileExists{Diagrams/ParityReading.tikz}{}{\input{./figures/Diagrams/ParityReading.tikz}}\,,
\eeq
 such that tracing out all but the $i$-th outcome gives the parity of the $i$-th setting for $M$:
\beq
\forall i\in\nSet\quad,\quad %
\InputIfFileExists{Diagrams/ParityReading2.tikz}{}{\input{./figures/Diagrams/ParityReading2.tikz}}\ = \quad%
\InputIfFileExists{Diagrams/ParityReading1.tikz}{}{\input{./figures/Diagrams/ParityReading1.tikz}}\,.
\eeq
\end{definition}
We will also be interested in situations in which we have a measurement which can only read the parity of a certain subset of the setting variable ${\iSet} \subseteq \nSet$:
\begin{definition}\textbf{Partial Parity Reading Measurement.--}\\
A \emph{partial parity reading measurement} for the settings  ${\iSet} \subseteq \nSet$ of measurement $M$, denoted by $\mathcal{P}[M]_{\iSet}$ is simply a parity reading measurement for the measurement:
\beq
\begin{tikzpicture}
	\begin{pgfonlayer}{nodelayer}
		\node [style={right label}] (0) at (0.25, -0.5) {$V$};
		\node [style=none] (1) at (0.25, -0) {};
		\node [style=none] (2) at (0.25, -0.75) {};
		\node [style=none] (3) at (-1, -0) {};
		\node [style=none] (4) at (0.75, -0) {};
		\node [style=none] (5) at (0.75, 1) {};
		\node [style=none] (6) at (-0.75, 1) {};
		\node [style=none] (7) at (0, 0.5) {$M_{\iSet}$};
		\node [style=none] (8) at (-0.5, -0) {};
		\node [style=none] (9) at (-0.5, -0.75) {};
		\node [style=none] (10) at (0, 1) {};
		\node [style=none] (11) at (0, 1.75) {};
		\node [style={right label}] (12) at (-0.5, -0.5) {$\iSet$};
		\node [style={right label}] (13) at (0, 1.5) {$\bSet$};
	\end{pgfonlayer}
	\begin{pgfonlayer}{edgelayer}
		\draw [qWire] (1.center) to (2.center);
		\draw (3.center) to (4.center);
		\draw (4.center) to (5.center);
		\draw (5.center) to (6.center);
		\draw (6.center) to (3.center);
		\draw [cWire] (11.center) to (10.center);
		\draw [cWire] (8.center) to (9.center);
	\end{pgfonlayer}
\end{tikzpicture}
\ \ := \ \ \begin{tikzpicture}
	\begin{pgfonlayer}{nodelayer}
		\node [style={right label}] (0) at (0.25, -1.75) {$V$};
		\node [style=none] (1) at (0.25, -0) {};
		\node [style=none] (2) at (0.25, -2) {};
		\node [style=none] (3) at (-1, -0) {};
		\node [style=none] (4) at (0.75, -0) {};
		\node [style=none] (5) at (0.75, 1) {};
		\node [style=none] (6) at (-0.75, 1) {};
		\node [style=none] (7) at (0, 0.5) {$M$};
		\node [style=none] (8) at (-0.5, -0) {};
		\node [style=none] (9) at (-0.5, -1) {};
		\node [style=none] (10) at (0, 1) {};
		\node [style=none] (11) at (0, 1.75) {};
		\node [style={right label}] (12) at (-0.5, -0.5) {$\nSet$};
		\node [style={right label}] (13) at (0, 1.5) {$\bSet$};
		\node [style=none] (14) at (-0.5, -2) {};
		\node [style=none] (15) at (-0.5, -1) {};
		\node [style={right label}] (16) at (-0.5, -1.75) {$\iSet$};
		\node [style={black dot}] (17) at (-0.5, -1) {};
	\end{pgfonlayer}
	\begin{pgfonlayer}{edgelayer}
		\draw [qWire] (1.center) to (2.center);
		\draw (3.center) to (4.center);
		\draw (4.center) to (5.center);
		\draw (5.center) to (6.center);
		\draw (6.center) to (3.center);
		\draw [cWire] (11.center) to (10.center);
		\draw [cWire] (8.center) to (9.center);
		\draw [cWire] (15.center) to (14.center);
	\end{pgfonlayer}
\end{tikzpicture}
\eeq
where $\bullet$ is the canonical embedding of $\iSet$ into $\nSet$\footnote{That is, it maps $\iSet$ viewed as a set in its own right into $\iSet$ viewed as a subset of $\nSet$}. Using this notation we can succinctly define these partial partity reading measurements by:\beq\mathcal{P}[M]_{\iSet}:=\mathcal{P}[M_{\iSet}]\,.\eeq
\end{definition}
Note that when $\iSet=\nSet$ then we recover the notion of a PRM.

In order to see how the existence of a (partial) PRM constraints the correlations that the GPT may feature, we will fist discuss the concepts of a \textit{Fiducial Measurement} and \textit{Fiducial effects}.

Let $n$ be the affine dimension of the normalised state space, that is, $n = |V|-1$. A \emph{fiducial measurement}, $\mathcal{F}$, is a controlled measurement with $n$ settings (described by the set $\nSet$) and binary outcomes (described by the set $\bSet$):
\beq
\InputIfFileExists{Diagrams/FiducialMeasurement.tikz}{}{\input{./figures/Diagrams/FiducialMeasurement.tikz}}
\eeq
$\mathcal{F}$ is called a fiducial measurement if all of the fiducial effects can be obtained from such a measurement. Fiducial effects, in turn, form a (minimal) spanning set for the effect space of the GPT. As an example, consider the case where $n=2$: here $\mathcal{F}$ will have a binary input system, and the three fiducial effects will be given by
\beq
\left\{ %
\InputIfFileExists{Diagrams/FiducialEffect1.tikz}{}{\input{./figures/Diagrams/FiducialEffect1.tikz}}\ ,\quad %
\InputIfFileExists{Diagrams/FiducialEffect2.tikz}{}{\input{./figures/Diagrams/FiducialEffect2.tikz}}\ ,\quad  %
\InputIfFileExists{Diagrams/FiducialEffect3.tikz}{}{\input{./figures/Diagrams/FiducialEffect3.tikz}} = %
\InputIfFileExists{Diagrams/FiducialEffect5.tikz}{}{\input{./figures/Diagrams/FiducialEffect5.tikz}} = %
\InputIfFileExists{Diagrams/FiducialEffect4.tikz}{}{\input{./figures/Diagrams/FiducialEffect4.tikz}}\right\}\,,
\eeq
{where the equality for the third effect comes from the fact that $\mathcal{F}$ is a valid controlled measurement and hence satisfies:
\beq
\begin{tikzpicture}
	\begin{pgfonlayer}{nodelayer}
		\node [style={right label}] (0) at (0.25, -1.5) {$V$};
		\node [style=none] (1) at (0.25, -0.5) {};
		\node [style=none] (2) at (0.25, -1.75) {};
		\node [style=none] (3) at (0.75, 0.5) {};
		\node [style=none] (4) at (-0.75, 0.5) {};
		\node [style=none] (5) at (-1, -0.5) {};
		\node [style=none] (6) at (0, -0) {$\mathcal{F}$};
		\node [style=none] (7) at (-0.5, -0.5) {};
		\node [style=none] (8) at (-0.5, -1.75) {};
		\node [style=none] (9) at (0, 0.5) {};
		\node [style=none] (10) at (0, 1.25) {};
		\node [style={right label}] (11) at (-0.5, -1.5) {$\nSet$};
		\node [style={right label}] (12) at (0, 0.875) {$\bSet$};
		\node [style=none] (13) at (0.75, -0.5) {};
		\node [style=upground] (14) at (0, 1.5) {};
	\end{pgfonlayer}
	\begin{pgfonlayer}{edgelayer}
		\draw [qWire, in=90, out=-90, looseness=1.00] (1.center) to (2.center);
		\draw (3.center) to (4.center);
		\draw (4.center) to (5.center);
		\draw [cWire, in=90, out=-90, looseness=1.00] (7.center) to (8.center);
		\draw [cWire, in=-90, out=90, looseness=1.00] (9.center) to (10.center);
		\draw (5.center) to (13.center);
		\draw (13.center) to (3.center);
	\end{pgfonlayer}
\end{tikzpicture}\quad=\quad \begin{tikzpicture}
	\begin{pgfonlayer}{nodelayer}
		\node [style={right label}] (0) at (0.75, -0.75) {$V$};
		\node [style=none] (1) at (0.75, 0.25) {};
		\node [style=none] (2) at (0.75, -1) {};
		\node [style=none] (3) at (-0.75, 0.25) {};
		\node [style=none] (4) at (-0.75, -1) {};
		\node [style={right label}] (5) at (-0.75, -0.75) {$\nSet$};
		\node [style=upground] (6) at (-0.75, 0.5) {};
		\node [style=upground] (7) at (0.75, 0.5) {};
	\end{pgfonlayer}
	\begin{pgfonlayer}{edgelayer}
		\draw [qWire, in=90, out=-90, looseness=1.00] (1.center) to (2.center);
		\draw [cWire, in=90, out=-90, looseness=1.00] (3.center) to (4.center);
	\end{pgfonlayer}
\end{tikzpicture}\ .
\eeq
}

Coming back to the case of an arbitrary $n$, notice that the fact that fiducial effects,
\beq\left\{\begin{tikzpicture}
	\begin{pgfonlayer}{nodelayer}
		\node [style=copoint] (0) at (0, .5) {$e_j$};
		\node [style=none] (1) at (0, -1) {};
		\node [style=right label] (2) at (0, -0.5) {$V$};
	\end{pgfonlayer}
	\begin{pgfonlayer}{edgelayer}
		\draw[qWire] (1.center) to (0);
	\end{pgfonlayer}
\end{tikzpicture}
\right\}_{j=0:n}\eeq
span the corresponding vector space, means that any state $s$ can be uniquely characterised by the vector of probabilities:
\beq\label{eq:vecrepst}
\mathbf{p}_s \ := \quad \left( \begin{tikzpicture}
	\begin{pgfonlayer}{nodelayer}
		\node [style=copoint] (0) at (0, 1) {$e_0$};
		\node [style=point] (1) at (0, -1) {$s$};
		\node [style=right label] (2) at (0, -0.25) {$V$};
	\end{pgfonlayer}
	\begin{pgfonlayer}{edgelayer}
		\draw[qWire] (1.center) to (0);
	\end{pgfonlayer}
\end{tikzpicture}
 \,,\, \ldots \,,\, \begin{tikzpicture}
	\begin{pgfonlayer}{nodelayer}
		\node [style=copoint] (0) at (0, 1) {$e_{n}$};
		\node [style=point] (1) at (0, -1) {$s$};
		\node [style=right label] (2) at (0, -0.25) {$V$};
	\end{pgfonlayer}
	\begin{pgfonlayer}{edgelayer}
		\draw [qWire](1.center) to (0);
	\end{pgfonlayer}
\end{tikzpicture}
 \right)^T\,.
\eeq
For example, going back to the case where $n=2$, this vector of probabilities will be given by:
\beq\label{eq:stateAsVector}
\mathbf{p}_s \ :=\quad \left(%
\InputIfFileExists{Diagrams/FiducialProb2.tikz}{}{\input{./figures/Diagrams/FiducialProb2.tikz}}\ ,\quad %
\InputIfFileExists{Diagrams/FiducialProb3.tikz}{}{\input{./figures/Diagrams/FiducialProb3.tikz}}\ ,\quad %
\begin{tikzpicture}
	\begin{pgfonlayer}{nodelayer}
		\node [style={right label}] (0) at (0, -0.2500001) {$V$};
		\node [style=none] (1) at (0, 0.4999999) {};
		\node [style=point] (2) at (0, -0.9999999) {$s$};
		\node [style=upground] (3) at (0, 0.7500001) {};
	\end{pgfonlayer}
	\begin{pgfonlayer}{edgelayer}
		\draw [qWire] (1.center) to (2);
	\end{pgfonlayer}
\end{tikzpicture}} \right)^T.
\eeq

Now we can briefly state the case we will explore in this section: GPTs that have a  PRM for a Fiducial measurement, and where a bipartite Bell experiment is carried by Alice and Bob performing this controlled fiducial measurement in each wing.

A PRM $\mathcal{P[F]}$ for the fiducial measurement $\mathcal{F}$ will satisfy the following constraints:
\beq\label{eq:PRMfidcond}
\InputIfFileExists{Diagrams/ParFid1.tikz}{}{\input{./figures/Diagrams/ParFid1.tikz}}\quad = \quad%
\InputIfFileExists{Diagrams/ParFid2.tikz}{}{\input{./figures/Diagrams/ParFid2.tikz}} \qquad \forall i\in\nSet\,.
\eeq
It is worth mentioning that a PRM $\mathcal{P[F]}$ is not necessarily uniquely singled out by these constraints -- more than one PRM may qualify as potential candidates for the role. We denote by $\mathsf{ParMeas}[\mathcal{F}]$ the set of all PRM $\mathcal{P[F]}$ that satisfy Eq.~\eqref{eq:PRMfidcond} for the given $\mathcal{F}$. 

\bigskip

\subsection{Examples }
\label{subsec:examples}
{\bf Qubits.}
Let us conclude this discussion with an example from quantum theory. Consider the case of qubit systems, where the affine local dimension is $n=3$. A fiducial measurement corresponds to measuring the three Pauli observables $X$, $Y$, and $Z$. A PRM is given by {(a suitable post-processing of)} the Bell measurement 
\beq
\left\{\Big| \phi^+ \Big\rangle \Big\langle \phi^+ \Big|,
\Big| \phi^- \Big\rangle \Big\langle \phi^- \Big|,
\Big| \psi^+ \Big\rangle \Big\langle \psi^+ \Big|,
\Big| \psi^- \Big\rangle \Big\langle \psi^- \Big|\right\},
\eeq
with $\ket{\phi^\pm} = \frac{\ket{00} \pm \ket{11}}{\sqrt{2}}$ and $\ket{\psi^\pm} = \frac{\ket{01} \pm \ket{10}}{\sqrt{2}}$. { Denoting the four element set of outcomes of the Bell measurement as $\mathbf{B}:=\{0,1,2,3\}$, and the qubit system by $Q_2$, we can diagrammatically represent this as:
\beq
\begin{tikzpicture}\label{eq:bellmeas}
	\begin{pgfonlayer}{nodelayer}
		\node [style=none] (0) at (0, 1.5) {};
		\node [style=none] (4) at (0, 0.5) {};
		\node [style=right label] (5) at (0, 1.25) {$\mathbf{B}$};
		\node [style=none] (18) at (-0.75, 0.5) {};
		\node [style=none] (19) at (0.75, 0.5) {};
		\node [style=none] (20) at (1.25, -0.5) {};
		\node [style=none] (21) at (-1.25, -0.5) {};
		\node [style=none] (22) at (-0.75, -0.5) {};
		\node [style=none] (23) at (-0.75, -1.5) {};
		\node [style=right label] (24) at (-0.75, -1.25) {$Q_2$};
		\node [style=none] (25) at (0.75, -0.5) {};
		\node [style=none] (26) at (0.75, -1.5) {};
		\node [style=right label] (27) at (0.75, -1.25) {$Q_2$};
		\node [style=none] (28) at (0, 0) {Bell};
	\end{pgfonlayer}
	\begin{pgfonlayer}{edgelayer}
		\draw [cWire] (0.center) to (4.center);
		\draw (18.center) to (21.center);
		\draw (21.center) to (20.center);
		\draw (20.center) to (19.center);
		\draw (19.center) to (18.center);
		\draw [qWire] (22.center) to (23.center);
		\draw [qWire] (25.center) to (26.center);
	\end{pgfonlayer}
\end{tikzpicture}\,.
\eeq
To see that a post-processing is necessary {for this measurement to fit into our} definition of a PRM is easy: {the measurement of Eq.~\eqref{eq:bellmeas}} is a four outcome measurement, but, a PRM for $X$, $Y$, and $Z$ should have three binary outcomes. The {required} post-processing can be described diagrammatically as:
\beq
\begin{tikzpicture}
	\begin{pgfonlayer}{nodelayer}
		\node [style=white dot] (0) at (0, -1) {};
		\node [style=none] (1) at (-1.75, 0.25) {};
		\node [style=none] (2) at (0, 0.25) {};
		\node [style=none] (3) at (1.75, 0.25) {};
		\node [style=none] (4) at (0, -2.25) {};
		\node [style=right label] (5) at (0, -2) {$\mathbf{B}$};
		\node [style=right label] (6) at (1.75, 0) {$\mathbf{B}$};
		\node [style=right label] (7) at (0, 0) {$\mathbf{B}$};
		\node [style=right label] (8) at (-1.75, 0) {$\mathbf{B}$};
		\node [style=small box] (9) at (-1.75, 1) {$C_X$};
		\node [style=small box] (10) at (0, 1) {$C_Y$};
		\node [style=small box] (11) at (1.75, 1) {$C_Z$};
		\node [style=none] (12) at (-1.75, 2.25) {};
		\node [style=none] (13) at (0, 2.25) {};
		\node [style=none] (14) at (1.75, 2.25) {};
		\node [style=right label] (15) at (-1.75, 2) {$\bSet$};
		\node [style=right label] (16) at (0, 2) {$\bSet$};
		\node [style=right label] (17) at (1.75, 2) {$\bSet$};
	\end{pgfonlayer}
	\begin{pgfonlayer}{edgelayer}
		\draw [cWire, in=150, out=-90] (1.center) to (0);
		\draw [cWire] (0) to (2.center);
		\draw [cWire, in=-90, out=30] (0) to (3.center);
		\draw [cWire] (0) to (4.center);
		\draw [cWire] (12.center) to (9);
		\draw [cWire] (9) to (1.center);
		\draw [cWire] (13.center) to (10);
		\draw [cWire] (10) to (2.center);
		\draw [cWire] (14.center) to (11);
		\draw [cWire] (11) to (3.center);
	\end{pgfonlayer}
\end{tikzpicture}\,,
\eeq
where the white dot first makes three copies of the outcome, {and the processes} $C_X$, $C_Y$ and $C_Z$ correspond to the three different  equal bipartitions of $\mathbf{B}$. {For} example:
\beq\label{eq:thepost}
\begin{tikzpicture}
	\begin{pgfonlayer}{nodelayer}
		\node [style=none] (1) at (0, -1) {};
		\node [style=right label] (8) at (0, -1) {$\mathbf{B}$};
		\node [style=small box] (9) at (0, 0) {$C_Z$};
		\node [style=none] (12) at (0, 1) {};
		\node [style=right label] (15) at (0, 0.75) {$\bSet$};
	\end{pgfonlayer}
	\begin{pgfonlayer}{edgelayer}
		\draw [cWire] (12.center) to (9);
		\draw [cWire] (9) to (1.center);
	\end{pgfonlayer}
\end{tikzpicture}
\quad = \quad
\begin{tikzpicture}
	\begin{pgfonlayer}{nodelayer}
		\node [style=none] (1) at (0, -1.75) {};
		\node [style=right label] (8) at (0, -1.5) {$\mathbf{B}$};
		\node [style=copoint] (9) at (0, -0.75) {$0$};
		\node [style=point] (18) at (0, 0.75) {$0$};
		\node [style=none] (19) at (0, 1.75) {};
		\node [style=right label] (20) at (0, 1.5) {$\bSet$};
	\end{pgfonlayer}
	\begin{pgfonlayer}{edgelayer}
		\draw [cWire] (9) to (1.center);
		\draw [cWire] (19.center) to (18);
	\end{pgfonlayer}
\end{tikzpicture}
\ + \ 
\begin{tikzpicture}
	\begin{pgfonlayer}{nodelayer}
		\node [style=none] (1) at (0, -1.75) {};
		\node [style=right label] (8) at (0, -1.5) {$\mathbf{B}$};
		\node [style=copoint] (9) at (0, -0.75) {$1$};
		\node [style=point] (18) at (0, 0.75) {$0$};
		\node [style=none] (19) at (0, 1.75) {};
		\node [style=right label] (20) at (0, 1.5) {$\bSet$};
	\end{pgfonlayer}
	\begin{pgfonlayer}{edgelayer}
		\draw [cWire] (9) to (1.center);
		\draw [cWire] (19.center) to (18);
	\end{pgfonlayer}
\end{tikzpicture}
\ + \
\begin{tikzpicture}
	\begin{pgfonlayer}{nodelayer}
		\node [style=none] (1) at (0, -1.75) {};
		\node [style=right label] (8) at (0, -1.5) {$\mathbf{B}$};
		\node [style=copoint] (9) at (0, -0.75) {$2$};
		\node [style=point] (18) at (0, 0.75) {$1$};
		\node [style=none] (19) at (0, 1.75) {};
		\node [style=right label] (20) at (0, 1.5) {$\bSet$};
	\end{pgfonlayer}
	\begin{pgfonlayer}{edgelayer}
		\draw [cWire] (9) to (1.center);
		\draw [cWire] (19.center) to (18);
	\end{pgfonlayer}
\end{tikzpicture}
\ +\ 
\begin{tikzpicture}
	\begin{pgfonlayer}{nodelayer}
		\node [style=none] (1) at (0, -1.75) {};
		\node [style=right label] (8) at (0, -1.5) {$\mathbf{B}$};
		\node [style=copoint] (9) at (0, -0.75) {$3$};
		\node [style=point] (18) at (0, 0.75) {$1$};
		\node [style=none] (19) at (0, 1.75) {};
		\node [style=right label] (20) at (0, 1.5) {$\bSet$};
	\end{pgfonlayer}
	\begin{pgfonlayer}{edgelayer}
		\draw [cWire] (9) to (1.center);
		\draw [cWire] (19.center) to (18);
	\end{pgfonlayer}
\end{tikzpicture}\,.
\eeq
}
{To} read the parity of observable $ZZ$, 
we {hence} apply post-processing $\{0,1\}\to 0$ and $\{2,3\}\to 1$ {depicted in Eq.~\eqref{eq:thepost},} which quantum mechanically comes from 
\begin{eqnarray}
&& |\phi_+\rangle\<\phi_+| + |\phi_-\rangle\<\phi_-|=|00\rangle\<00|+
   |11\rangle\<11|,\nonumber \\ 
  && |\psi_+\rangle\<\psi_+| + |\psi_-\rangle\<\psi_-|=|01\rangle\<01|+
   |10\rangle\<10|,  \\
 && \sigma_z \otimes \sigma_z =
 |00\rangle\<00|+
   |11\rangle\<11| - (|01\rangle\<01|+
   |10\rangle\<10|).\nonumber
\end{eqnarray}
  {All three required post-processings, and the observables whose parity they read, are presented below:}
 \begin{eqnarray}
    \label{eq:Bell-postproc}
    \begin{array}{l|c}
         \text{post-processing}& \text{parity} \\
         \hline
         C_X:\quad \{\phi_+,\psi_+\}\to 0, \{\phi_-,\psi_-\}\to 1 & XX \\
         C_Y:\quad
         \{\phi_-,\psi_+\}\to 0, \{\phi_+,\psi_-\}\to 1 
         & YY \\
         C_Z:\quad
         \{\phi_+,\phi_-\}\to 0, \{\psi_+,\psi_-\}\to 1 
         & ZZ
    \end{array}
 \end{eqnarray}
 {With this we conclude the argument for why the measurement given by}
\beq
\begin{tikzpicture}
	\begin{pgfonlayer}{nodelayer}
		\node [style=white dot] (0) at (0, 0.25) {};
		\node [style=none] (1) at (-1.75, 1.5) {};
		\node [style=none] (2) at (0, 1.5) {};
		\node [style=none] (3) at (1.75, 1.5) {};
		\node [style=none] (4) at (0, -0.75) {};
		\node [style=right label] (5) at (0, -0.5) {$\mathbf{B}$};
		\node [style=right label] (6) at (1.75, 1.25) {$\mathbf{B}$};
		\node [style=right label] (7) at (0, 1.25) {$\mathbf{B}$};
		\node [style=right label] (8) at (-1.75, 1.25) {$\mathbf{B}$};
		\node [style=small box] (9) at (-1.75, 2.25) {$C_X$};
		\node [style=small box] (10) at (0, 2.25) {$C_Y$};
		\node [style=small box] (11) at (1.75, 2.25) {$C_Z$};
		\node [style=none] (12) at (-1.75, 3.5) {};
		\node [style=none] (13) at (0, 3.5) {};
		\node [style=none] (14) at (1.75, 3.5) {};
		\node [style=right label] (15) at (-1.75, 3.25) {$\bSet$};
		\node [style=right label] (16) at (0, 3.25) {$\bSet$};
		\node [style=right label] (17) at (1.75, 3.25) {$\bSet$};
		\node [style=none] (18) at (-0.75, -0.75) {};
		\node [style=none] (19) at (0.75, -0.75) {};
		\node [style=none] (20) at (1.25, -1.75) {};
		\node [style=none] (21) at (-1.25, -1.75) {};
		\node [style=none] (22) at (-0.75, -1.75) {};
		\node [style=none] (23) at (-0.75, -2.75) {};
		\node [style=right label] (24) at (-0.75, -2.5) {$Q_2$};
		\node [style=none] (25) at (0.75, -1.75) {};
		\node [style=none] (26) at (0.75, -2.75) {};
		\node [style=right label] (27) at (0.75, -2.5) {$Q_2$};
		\node [style=none] (28) at (0, -1.25) {Bell};
	\end{pgfonlayer}
	\begin{pgfonlayer}{edgelayer}
		\draw [cWire, in=150, out=-90] (1.center) to (0);
		\draw [cWire] (0) to (2.center);
		\draw [cWire, in=-90, out=30] (0) to (3.center);
		\draw [cWire] (0) to (4.center);
		\draw [cWire] (12.center) to (9);
		\draw [cWire] (9) to (1.center);
		\draw [cWire] (13.center) to (10);
		\draw [cWire] (10) to (2.center);
		\draw [cWire] (14.center) to (11);
		\draw [cWire] (11) to (3.center);
		\draw (18.center) to (21.center);
		\draw (21.center) to (20.center);
		\draw (20.center) to (19.center);
		\draw (19.center) to (18.center);
		\draw [qWire] (22.center) to (23.center);
		\draw [qWire] (25.center) to (26.center);
	\end{pgfonlayer}
\end{tikzpicture}
\eeq
is a parity reading measurement for the $X$, $Y$ and $Z$ observables.

\bigskip

\bigskip

\noindent{\bf Mirror quantum correlations.}
Such correlations have been considered in Ref.~\cite{BruknerDakic-mirror2009}. 
{To introduce them, let us first notice a feature of the Bell measurement:
\begin{compactitem}
\item the effect corresponding to $\psi_-$
appears only in the bipartitions of $\mathbf{B}$ that give rise to anticorrelations, i.e., a value of $1$ for the classical system $\bSet$,
\item the other three effects appear each only once in a bipartition that measures anticorrelations. 
\end{compactitem}
The following table summarises this feature, where we specify, for each observable whose parity we want to read, whether each effect belongs to the correlation (0) or anticorrelation (1) bipartitions: 
} 
\begin{equation}\label{eq:tableqt}
    \begin{array}{c|ccc}
        \text{effect} & \text{XX} &\text{YY} &\text{ZZ} \\
        \hline
         \phi_+ & 0 & 1 & 0  \\
         \phi_- & 1 & 0 & 0  \\
         \psi_+ & 0 & 0 & 1  \\
         \psi_- & 1 & 1 & 1  \\
    \end{array}
\end{equation}
  We then say that quantum parity reading measurement has {\it signature}
\begin{align}
    \{(+,-,+),(-,+,+),(+,+,-),(-,-,-)\}.
\end{align}
{In the case of \textit{mirror quantum mechanics} \cite{BruknerDakic-mirror2009} the table in Eq.~\eqref{eq:tableqt} does not hold anymore, and instead the following are satisfied:} 
\begin{equation}
\label{eq:tablemirror}
    \begin{array}{c|ccc}
        \text{effect} & \text{XX} &\text{YY} &\text{ZZ} \\
        \hline
         (\phi_+)^{PT} & 1 & 0 & 1  \\
         (\phi_-)^{PT} & 0 & 1 & 1  \\
         (\psi_+)^{PT} & 1 & 1 & 0  \\
         (\psi_-)^{PT} & 0 & 0 & 0  \\
    \end{array}
\end{equation}
where PT stands for partial {transposition\footnote{{In mirror quantum theory, these effects are partially transposed effects of Bell measurement}.}.}
  
Thus in mirror quantum case we have signature: 
\begin{align}
        \{(-,+,-),(+,-,-),(-,-,+),(+,+,+)\}.
\end{align}
This {case then} corresponds to the following post-processing of the measurement outcomes:
\begin{eqnarray}
    \label{eq:Mirror-postproc}
    \begin{array}{l|c}
         \text{post-processing}& \text{parity} \\
         \hline
         C_X:\quad \{0,2\}\to 0, \{1,3\}\to 1 & XX \\
         C_Y:\quad
         \{0,3\}\to 0, \{1,2\}\to 1 
         & YY \\
         C_Z:\quad
         \{2,3\}\to 0, \{0,1\}\to 1 
         & ZZ
    \end{array}
 \end{eqnarray}

\bigskip

\bigskip

\noindent
{\bf Classical theory.}
 In the quantum case, the parity reading measurement was entangled. It had to be so, because it measured parities of observables that are not jointly measurable. In classical theory we can consider three bit system, described by $X,Y,Z$ which are now jointly measurable. Then the PRM just amounts to post-process the joint measurement of all the 6 observables (three per party).

\subsection{$\mathcal{P[F]}$ and maximal violations of Bell inequalities for fiducial measurements $\mathcal{F}$}\label{se:3.1}

So, how does the existence of a $\mathcal{P[F]}$ in the GPT constrain the correlations we may observe in a Bell test? 
More precisely, what are the constraints on the correlations that a bipartite system of a certain type can produce when there exists a PRM for the fiducial measurements on that system type?
In this section we aim at optimising the value of a Bell inequality $\mathcal{I}$ when the measurements that the parties perform are given by $\mathcal{F}$ on a system of type $V$.

Notice that, in general, the cardinality of the input settings for the Bell test need not coincide with the number of settings for the fiducial measurement $\mathcal{F}$. That is, $|X|=|\nSet|=|Y|$ does not necessarily hold. In this manuscript we will work with the case where $|X|\leq|\nSet|$ and $|Y|\leq|\nSet|$, and hence only some of the settings of the controlled measurement $\mathcal{F}$ might be used for the Bell test. In such a case, then, we will focus on the constraints that a partial PRM imposes when its existence is demanded on the settings of $\mathcal{F}$ used in the Bell test. For simplicity in the discussion, in this section we will present the case where $|X|=|\nSet|=|Y|$, but the most general case follows similarly. We will return to the optimisation problems for partial PRM later on in the manuscript.  

The optimisation problem that we focus on then reads:
\beq
\mathcal{I}_{\max}:=\mathsf{sup}\left\{\ \ %
\InputIfFileExists{Diagrams/OptimisationProblem.tikz}{}{\input{./figures/Diagrams/OptimisationProblem.tikz}}\ \ \middle|\ \ s \in \mathcal{G}  \right\}\,,
\eeq
where $X,Y,A,B$ are all binary variables, and we are using the shorthand notation:
\beq
\begin{tikzpicture}
	\begin{pgfonlayer}{nodelayer}
		\node [style=right label] (0) at (-0.5, -1) {$V$};
		\node [style=none] (1) at (-0.5, -0.5) {};
		\node [style=none] (2) at (-0.5, -1.25) {};
		\node [style=none] (3) at (0.75, -0.5) {};
		\node [style=none] (4) at (-1, -0.5) {};
		\node [style=none] (5) at (-1, 0.5) {};
		\node [style=none] (6) at (0.5, 0.5) {};
		\node [style=none] (7) at (-0.25, 0) {$\mathcal{F}$};
		\node [style=none] (8) at (0.25, -0.5) {};
		\node [style=none] (9) at (0.25, -1.25) {};
		\node [style=none] (10) at (-0.25, 0.5) {};
		\node [style=none] (11) at (-0.25, 1.25) {};
		\node [style=right label] (12) at (0.25, -1) {$\nSet$};
		\node [style=right label] (13) at (-0.25, 1) {$\bSet$};
	\end{pgfonlayer}
	\begin{pgfonlayer}{edgelayer}
		\draw [qWire] (1.center) to (2.center);
		\draw (3.center) to (4.center);
		\draw (4.center) to (5.center);
		\draw (5.center) to (6.center);
		\draw (6.center) to (3.center);
		\draw [cWire] (11.center) to (10.center);
		\draw [cWire] (8.center) to (9.center);
	\end{pgfonlayer}
\end{tikzpicture}
\quad :=\quad \begin{tikzpicture}
	\begin{pgfonlayer}{nodelayer}
		\node [style=left label] (0) at (-0.5, -1.25) {$V$};
		\node [style=none] (1) at (0.25, 0) {};
		\node [style=none] (2) at (-0.5, -1.5) {};
		\node [style=none] (3) at (-1, 0) {};
		\node [style=none] (4) at (0.75, 0) {};
		\node [style=none] (5) at (0.75, 1) {};
		\node [style=none] (6) at (-0.75, 1) {};
		\node [style=none] (7) at (0, 0.5) {$\mathcal{F}$};
		\node [style=none] (8) at (-0.5, 0) {};
		\node [style=none] (9) at (0.25, -1.5) {};
		\node [style=none] (10) at (0, 1) {};
		\node [style=none] (11) at (0, 1.75) {};
		\node [style=right label] (12) at (0.25, -1.25) {$\nSet$};
		\node [style=right label] (13) at (0, 1.5) {$\bSet$};
	\end{pgfonlayer}
	\begin{pgfonlayer}{edgelayer}
		\draw [qWire, in=90, out=-90] (1.center) to (2.center);
		\draw (3.center) to (4.center);
		\draw (4.center) to (5.center);
		\draw (5.center) to (6.center);
		\draw (6.center) to (3.center);
		\draw [cWire] (11.center) to (10.center);
		\draw [cWire, in=90, out=-90] (8.center) to (9.center);
	\end{pgfonlayer}
\end{tikzpicture}
\ .
\eeq
Given a particular GPT $\mathcal{G}$ -- and, in particular, given a specification of its state and effect spaces  --  this optimisation problem reduces to a type of cone program which has been explored in recent literature \cite{sikora2018simple,selby2018make,sikora2019impossibility} regarding their relationship to GPTs. That is, there is some convex spanning cone of states $K_{V\otimes V}\subset V\otimes V$, which $s$ belongs to, and some normalisation constraint on $s$, $u_V\otimes u_V(s)=1$ so the above problem can be rewritten as:
\beq\label{eq:simpopt}
\mathcal{I}_{\max}:=\mathsf{sup}\left\{\ \ %
\InputIfFileExists{Diagrams/OptimisationProblem.tikz}{}{\input{./figures/Diagrams/OptimisationProblem.tikz}}\ \ \middle|\ \ s \in K_{V\otimes V}, u_V\otimes u_V(s)=1  \right\}.
\eeq
Here, however, we do not consider a particular GPT $\mathcal{G}$ which we optimise over. What we carry out here is an optimisation over the space of GPTs which have the relevant structure -- those which admit a PRM for the fiducial measurement. This optimisation problem is much more complex than that of Eq.~\eqref{eq:simpopt}, as we will now explain.

In Section \ref{se:2.1} we elaborated on the types of compatibility constraints between states and effects that a GPT must feature. Here, we will demand that the GPT admits the fiducial local measurement $\mathcal{F}$ and a PRM $\mathcal{P} \in \mathsf{ParMeas}[\mathcal{F}]$. By imposing the compatibility constraints motivated in Section \ref{se:2.1}, we hence restrict the possible cones of states $K[\mathcal{P}]$ that such GPT could feature. If we have a characterisation of $K[\mathcal{P}] \subset  V\otimes V$, then the optimisation problem becomes:
\beq \label{eq:opprob1}
\mathcal{I}_{\max}:=\mathsf{sup}\left\{\ \ %
\InputIfFileExists{Diagrams/OptimisationProblem.tikz}{}{\input{./figures/Diagrams/OptimisationProblem.tikz}}\ \ \middle|\,\, \mathcal{P}\in \mathsf{ParMeas}[\mathcal{F}], u_V\otimes u_V(s)=1, s \in K[\mathcal{P}]  \right\}\,.
\eeq
This turns out to be a non-linear optimisation problem, as we will show next. 

\subsection{The cone $K$ of bipartite states}

The key question here is: how to characterise the cones of states $K[\mathcal{P}]$? Here we will take the types of diagrammatic constraints motivated in Section \ref{se:2.1}, and define a systematic hierarchy of conditions that the existence of $\mathcal{P[F]}$ imposes on the cone $K$. 
This hierarchy of conditions will be specified in terms of the number of copies of the bipartite state $s$ featured in the diagram. Framing these constraints in the form of a hierarchy is useful because, as we will see, interesting results can be obtained without needing to impose all of the constraints. For example, in our case we will be interested in  possible violations of a given Bell inequality,  and, we can obtain upper bounds on this by simply working at the second level of the hierarchy. 

\bigskip

\begin{hiercond} \quad \\
Given the fiducial measurement $\mathcal{F}$ and the PRM $\mathcal{P}\in\mathsf{ParMeas}[\mathcal{F}]$, the normalised bipartite states $s \in K[\mathcal{P}]$ must satisfy: 
\begin{align}
&%
\InputIfFileExists{Diagrams/const1.tikz}{}{\input{./figures/Diagrams/const1.tikz}} \geq 0 \quad \text{,} \label{eq:const1}\\
&%
\InputIfFileExists{Diagrams/const2.tikz}{}{\input{./figures/Diagrams/const2.tikz}} \geq 0 \quad \text{,} \label{eq:singsin} \\
&%
\InputIfFileExists{Diagrams/const3.tikz}{}{\input{./figures/Diagrams/const3.tikz}}\geq 0    \quad \text{,} \label{eq:singswap}
\end{align}
where by $\geq 0$ we mean that every matrix element is non-negative. 
\end{hiercond}
Notice that the constraints that come from the deterministic effect $u$ -- in particular, the normalisation condition $u_V\otimes u_V(s)=1$ -- together with these positivity constraints, ensures that diagrams in Eqs.~\ref{eq:const1}, \ref{eq:singsin}, and \ref{eq:singswap} are stochastic maps. 

Notice moreover that the constraint of Eq.~\eqref{eq:singswap} has the same structure as that of Eq.~\eqref{eq:singsin} but applied instead to the swapped state:
\beq
\begin{tikzpicture}
	\begin{pgfonlayer}{nodelayer}
		\node [style=none] (0) at (-1, 1) {};
		\node [style=none] (1) at (1, -0.25) {};
		\node [style=left label] (2) at (-1, 0.75) {$V$};
		\node [style=right label] (3) at (1, 0.75) {$V$};
		\node [style=none] (4) at (0, -0.75) {$s$};
		\node [style=none] (5) at (1.75, -0.25) {};
		\node [style=none] (6) at (-1.75, -0.25) {};
		\node [style=none] (7) at (0, -1.5) {};
		\node [style=none] (8) at (1, 1) {};
		\node [style=none] (9) at (-1, -0.25) {};
		\node [style=none] (21) at (-1, 1) {};
	\end{pgfonlayer}
	\begin{pgfonlayer}{edgelayer}
		\draw [qWire, in=270, out=90] (1.center) to (0.center);
		\draw (5.center) to (7.center);
		\draw (7.center) to (6.center);
		\draw (6.center) to (5.center);
		\draw [qWire, in=90, out=-90] (8.center) to (9.center);
	\end{pgfonlayer}
\end{tikzpicture}
.
\eeq

We see then that there is a certain structure emerging: (i) there are two layers -- one corresponding to the state $s$ and one to the measurements $\mathcal{F}$ and $\mathcal{P}$ --, and (ii) we can vary the order in which the output wires of the state are plugged into the measurements of the second layer. This motivates the definition for the remaining levels of the hierarchy, which relies on the concept of a \emph{wiring}, which we explain next.

\begin{definition}[Wiring] \quad \\
A process which describes how a collection of input systems are connected to a collection of output systems is here referred to as a \emph{wiring}, and denoted usually by $W$.

When all the input systems are of the same type -- a case we focus on here --  wirings reduce to permutations of the systems, e.g.:
\beq
\begin{tikzpicture}
	\begin{pgfonlayer}{nodelayer}
		\node [style=none] (0) at (5, 1) {};
		\node [style=none] (1) at (5, 2.25) {};
		\node [style={right label}] (2) at (5, 1.25) {$V$};
		\node [style=none] (3) at (-3, -1) {};
		\node [style=none] (4) at (-3, -2.25) {};
		\node [style={right label}] (5) at (-3, -2) {$V$};
		\node [style={right label}] (6) at (-5, -2) {$V$};
		\node [style=none] (7) at (-5, -1) {};
		\node [style=none] (8) at (-5, -2.25) {};
		\node [style=none] (9) at (-3.25, 2.25) {};
		\node [style=none] (10) at (-3.25, 1) {};
		\node [style={right label}] (11) at (-3.25, 1.25) {$V$};
		\node [style={right label}] (12) at (-5, 1.25) {$V$};
		\node [style=none] (13) at (-5, 2.25) {};
		\node [style=none] (14) at (-5, 1) {};
		\node [style=none] (15) at (-1.25, 1) {};
		\node [style={right label}] (16) at (-1.25, 1.25) {$V$};
		\node [style=none] (17) at (-1.25, 2.25) {};
		\node [style=none] (18) at (5, -1) {};
		\node [style=none] (19) at (5, -2.25) {};
		\node [style={right label}] (20) at (5, -2) {$V$};
		\node [style={right label}] (21) at (3, -2) {$V$};
		\node [style=none] (22) at (3, -1) {};
		\node [style=none] (23) at (3, -2.25) {};
		\node [style=none] (24) at (1, -1) {};
		\node [style=none] (25) at (1, -2.25) {};
		\node [style={right label}] (26) at (1, -2) {$V$};
		\node [style={right label}] (27) at (-1, -2) {$V$};
		\node [style=none] (28) at (-1, -1) {};
		\node [style=none] (29) at (-1, -2.25) {};
		\node [style=none] (30) at (1, 2.25) {};
		\node [style=none] (31) at (1, 1) {};
		\node [style={right label}] (32) at (1, 1.25) {$V$};
		\node [style=none] (33) at (3, 1) {};
		\node [style={right label}] (34) at (3, 1.25) {$V$};
		\node [style=none] (35) at (3, 2.25) {};
		\node [style=none] (36) at (-5.5, 1) {};
		\node [style=none] (37) at (-5.5, -1) {};
		\node [style=none] (38) at (5.5, -1) {};
		\node [style=none] (39) at (5.5, 1) {};
		\node [style=none] (40) at (0, -0) {$W$};
	\end{pgfonlayer}
	\begin{pgfonlayer}{edgelayer}
		\draw [qWire, in=-90, out=90, looseness=1.00] (0.center) to (1.center);
		\draw [qWire, in=-90, out=90, looseness=1.00] (4.center) to (3.center);
		\draw [qWire] (7.center) to (8.center);
		\draw [qWire, in=-90, out=90, looseness=1.00] (10.center) to (9.center);
		\draw [qWire] (13.center) to (14.center);
		\draw [qWire, in=90, out=-90, looseness=1.00] (17.center) to (15.center);
		\draw [qWire, in=-90, out=90, looseness=1.00] (19.center) to (18.center);
		\draw [qWire] (22.center) to (23.center);
		\draw [qWire, in=-90, out=90, looseness=1.00] (25.center) to (24.center);
		\draw [qWire] (28.center) to (29.center);
		\draw [qWire, in=-90, out=90, looseness=1.00] (31.center) to (30.center);
		\draw [qWire, in=90, out=-90, looseness=1.00] (35.center) to (33.center);
		\draw (36.center) to (39.center);
		\draw (39.center) to (38.center);
		\draw (38.center) to (37.center);
		\draw (37.center) to (36.center);
	\end{pgfonlayer}
\end{tikzpicture} \ \ = \ \ \begin{tikzpicture}
	\begin{pgfonlayer}{nodelayer}
		\node [style=none] (0) at (3, 1) {};
		\node [style=none] (1) at (3, 2.25) {};
		\node [style={right label}] (2) at (3, 1.25) {$V$};
		\node [style=none] (3) at (-3, -1) {};
		\node [style=none] (4) at (-3, -2.25) {};
		\node [style={right label}] (5) at (-3, -2) {$V$};
		\node [style={right label}] (6) at (-5, -2) {$V$};
		\node [style=none] (7) at (-5, -1) {};
		\node [style=none] (8) at (-5, -2.25) {};
		\node [style=none] (9) at (-1, 2.25) {};
		\node [style=none] (10) at (-1, 1) {};
		\node [style={right label}] (11) at (-1, 1.25) {$V$};
		\node [style={right label}] (12) at (-5, 1.25) {$V$};
		\node [style=none] (13) at (-5, 2.25) {};
		\node [style=none] (14) at (-5, 1) {};
		\node [style=none] (15) at (1, 1) {};
		\node [style={right label}] (16) at (1, 1.25) {$V$};
		\node [style=none] (17) at (1, 2.25) {};
		\node [style=none] (18) at (5, -1) {};
		\node [style=none] (19) at (5, -2.25) {};
		\node [style={right label}] (20) at (5, -2) {$V$};
		\node [style={right label}] (21) at (3, -2) {$V$};
		\node [style=none] (22) at (3, -1) {};
		\node [style=none] (23) at (3, -2.25) {};
		\node [style=none] (24) at (1, -1) {};
		\node [style=none] (25) at (1, -2.25) {};
		\node [style={right label}] (26) at (1, -2) {$V$};
		\node [style={right label}] (27) at (-1, -2) {$V$};
		\node [style=none] (28) at (-1, -1) {};
		\node [style=none] (29) at (-1, -2.25) {};
		\node [style=none] (30) at (-3, 2.25) {};
		\node [style=none] (31) at (-3, 1) {};
		\node [style={right label}] (32) at (-3, 1.25) {$V$};
		\node [style=none] (33) at (5, 1) {};
		\node [style={right label}] (34) at (5, 1.25) {$V$};
		\node [style=none] (35) at (5, 2.25) {};
		\node [style=none] (36) at (-5.5, 1) {};
		\node [style=none] (37) at (-5.5, -1) {};
		\node [style=none] (38) at (5.5, -1) {};
		\node [style=none] (39) at (5.5, 1) {};
	\end{pgfonlayer}
	\begin{pgfonlayer}{edgelayer}
		\draw [qWire, in=-90, out=90, looseness=1.00] (0.center) to (1.center);
		\draw [qWire, in=-90, out=90, looseness=1.00] (4.center) to (3.center);
		\draw [qWire] (7.center) to (8.center);
		\draw [qWire, in=-90, out=90, looseness=1.00] (10.center) to (9.center);
		\draw [qWire] (13.center) to (14.center);
		\draw [qWire, in=90, out=-90, looseness=1.00] (17.center) to (15.center);
		\draw [qWire, in=-90, out=90, looseness=1.00] (19.center) to (18.center);
		\draw [qWire] (22.center) to (23.center);
		\draw [qWire, in=-90, out=90, looseness=1.00] (25.center) to (24.center);
		\draw [qWire] (28.center) to (29.center);
		\draw [qWire, in=-90, out=90, looseness=1.00] (31.center) to (30.center);
		\draw [qWire, in=90, out=-90, looseness=1.00] (35.center) to (33.center);
		\draw [thick gray dashed edge] (36.center) to (39.center);
		\draw [thick gray dashed edge] (39.center) to (38.center);
		\draw [thick gray dashed edge] (38.center) to (37.center);
		\draw [thick gray dashed edge] (37.center) to (36.center);
		\draw [qWire, in=90, out=-90, looseness=1.00] (14.center) to (7.center);
		\draw [qWire, in=90, out=-90, looseness=1.00] (10.center) to (3.center);
		\draw [qWire, in=90, out=-90, looseness=1.00] (15.center) to (28.center);
		\draw [qWire, in=90, out=-90, looseness=0.75] (31.center) to (24.center);
		\draw [qWire, in=90, out=-90, looseness=1.00] (33.center) to (22.center);
		\draw [qWire, in=90, out=-90, looseness=1.00] (0.center) to (18.center);
	\end{pgfonlayer}
\end{tikzpicture}\,.
\eeq
\end{definition}

\bigskip

\begin{hiercondarb} \quad \\
Given the fiducial measurement $\mathcal{F}$ and the PRM $\mathcal{P}\in\mathsf{ParMeas}[\mathcal{F}]$, the normalised bipartite states $s \in K[\mathcal{P}]$ must satisfy  the constraints imposed by Hierarchy Level $k'$ for all $k' < k$, as well as the following:
\begin{align}
 \begin{tikzpicture}[baseline=(current bounding box.center)]
	\begin{pgfonlayer}{nodelayer}
		\node [style=none] (12) at (4.5, 5.5) {};
		\node [style=none] (13) at (4.5, 4.25) {};
		\node [style=right label] (15) at (4.5, 5) {$\bSet$};
		\node [style=none] (17) at (4.5, 2) {};
		\node [style=none] (31) at (4.5, 3.25) {};
		\node [style=right label] (32) at (4.5, 2.25) {$V$};
		\node [style=none] (40) at (4.75, 3.75) {$\mathcal{F}$};
		\node [style=none] (45) at (3.75, 4.25) {};
		\node [style=none] (46) at (3.75, 3.25) {};
		\node [style=none] (47) at (5.75, 3.25) {};
		\node [style=none] (48) at (5.25, 4.25) {};
		\node [style=none] (52) at (6.25, 1.25) {};
		\node [style=right label] (53) at (6.25, 1.25) {$\nSet$};
		\node [style=none] (54) at (5.25, 3.25) {};
		\node [style=none] (55) at (-6.5, 0) {};
		\node [style=none] (56) at (-6.5, -1.25) {};
		\node [style=right label] (57) at (-6.5, -1) {$V$};
		\node [style=right label] (58) at (-8.5, -1) {$V$};
		\node [style=none] (59) at (-7.5, -1.75) {$s$};
		\node [style=none] (60) at (-5.75, -1.25) {};
		\node [style=none] (61) at (-9.25, -1.25) {};
		\node [style=none] (62) at (-7.5, -2.5) {};
		\node [style=none] (63) at (-8.5, 0) {};
		\node [style=none] (64) at (-8.5, -1.25) {};
		\node [style=none] (93) at (-6.75, 3.25) {};
		\node [style=none] (94) at (-6.75, 2) {};
		\node [style=right label] (95) at (-6.75, 2.25) {$V$};
		\node [style=right label] (96) at (-8.5, 2.25) {$V$};
		\node [style=none] (101) at (-8.5, 3.25) {};
		\node [style=none] (102) at (-8.5, 2) {};
		\node [style=none] (103) at (-8.5, 4.25) {};
		\node [style=none] (104) at (-8.5, 5.5) {};
		\node [style=right label] (105) at (-8.5, 5) {$\bSet$};
		\node [style=none] (106) at (-6.75, 4.25) {};
		\node [style=none] (107) at (-7.25, 3.25) {};
		\node [style=none] (108) at (-4.75, 4.25) {};
		\node [style=none] (110) at (-7.25, 4.25) {};
		\node [style=none] (111) at (-4.25, 3.25) {};
		\node [style=none] (112) at (-5.75, 3.75) {$\mathcal{P}$};
		\node [style=none] (113) at (-4.25, 3.25) {};
		\node [style=none] (114) at (-4.25, 4.25) {};
		\node [style=none] (115) at (-4.75, 2) {};
		\node [style=none] (116) at (-4.75, 5.5) {};
		\node [style=none] (117) at (-6.75, 5.5) {};
		\node [style=right label] (118) at (-6.75, 5) {$\bSet$};
		\node [style=right label] (119) at (-4.75, 2.25) {$V$};
		\node [style=right label] (120) at (-4.75, 5) {$\bSet$};
		\node [style=none] (121) at (-7.25, 4.25) {};
		\node [style=none] (122) at (-4.75, 3.25) {};
		\node [style=none] (123) at (-8.75, 3.75) {$\mathcal{F}$};
		\node [style=none] (124) at (-7.75, 4.25) {};
		\node [style=none] (125) at (-7.75, 3.25) {};
		\node [style=none] (126) at (-9.75, 3.25) {};
		\node [style=none] (127) at (-9.25, 4.25) {};
		\node [style=none] (128) at (-10.25, 1.25) {};
		\node [style=right label] (129) at (-10.25, 1.25) {$\nSet$};
		\node [style=none] (130) at (-9.25, 3.25) {};
		\node [style=none] (131) at (4.5, 0) {};
		\node [style=none] (132) at (4.5, -1.25) {};
		\node [style=right label] (133) at (4.5, -1) {$V$};
		\node [style=right label] (134) at (2.5, -1) {$V$};
		\node [style=none] (135) at (3.5, -1.75) {$s$};
		\node [style=none] (136) at (5.25, -1.25) {};
		\node [style=none] (137) at (1.75, -1.25) {};
		\node [style=none] (138) at (3.5, -2.5) {};
		\node [style=none] (139) at (2.5, 0) {};
		\node [style=none] (140) at (2.5, -1.25) {};
		\node [style=none] (141) at (0, -1.25) {$\cdots$};
		\node [style=none] (142) at (-8, -3) {};
		\node [style=none] (143) at (4, -3) {};
		\node [style=none] (144) at (-2, -3.75) {$k$};
		\node [style=none] (145) at (-2.5, 0) {};
		\node [style=none] (146) at (-2.5, -1.25) {};
		\node [style=right label] (147) at (-2.5, -1) {$V$};
		\node [style=right label] (148) at (-4.5, -1) {$V$};
		\node [style=none] (149) at (-3.5, -1.75) {$s$};
		\node [style=none] (150) at (-1.75, -1.25) {};
		\node [style=none] (151) at (-5.25, -1.25) {};
		\node [style=none] (152) at (-3.5, -2.5) {};
		\node [style=none] (153) at (-4.5, 0) {};
		\node [style=none] (154) at (-4.5, -1.25) {};
		\node [style=none] (155) at (0.75, 3.25) {};
		\node [style=none] (156) at (0.75, 2) {};
		\node [style=right label] (157) at (0.75, 2.25) {$V$};
		\node [style=none] (158) at (0.75, 4.25) {};
		\node [style=none] (159) at (0.25, 3.25) {};
		\node [style=none] (160) at (2.75, 4.25) {};
		\node [style=none] (161) at (0.25, 4.25) {};
		\node [style=none] (162) at (3.25, 3.25) {};
		\node [style=none] (163) at (1.75, 3.75) {$\mathcal{P}$};
		\node [style=none] (164) at (3.25, 3.25) {};
		\node [style=none] (165) at (3.25, 4.25) {};
		\node [style=none] (166) at (2.75, 2) {};
		\node [style=none] (167) at (2.75, 5.5) {};
		\node [style=none] (168) at (0.75, 5.5) {};
		\node [style=right label] (169) at (0.75, 5) {$\bSet$};
		\node [style=right label] (170) at (2.75, 2.25) {$V$};
		\node [style=right label] (171) at (2.75, 5) {$\bSet$};
		\node [style=none] (172) at (0.25, 4.25) {};
		\node [style=none] (173) at (2.75, 3.25) {};
		\node [style=none] (174) at (-2, 3.75) {$\cdots$};
		\node [style=none] (175) at (3.5, 6.5) {};
		\node [style=none] (176) at (-7.5, 6.5) {};
		\node [style=none] (177) at (-2, 7.25) {$k-1$};
		\node [style=none] (178) at (-9, 2) {};
		\node [style=none] (179) at (-9, 0) {};
		\node [style=none] (180) at (5, 0) {};
		\node [style=none] (181) at (5, 2) {};
		\node [style=none] (182) at (-2, 1) {$W$};
	\end{pgfonlayer}
	\begin{pgfonlayer}{edgelayer}
		\draw [cWire, in=-90, out=90] (13.center) to (12.center);
		\draw [qWire, in=-90, out=90] (17.center) to (31.center);
		\draw (45.center) to (48.center);
		\draw (48.center) to (47.center);
		\draw (47.center) to (46.center);
		\draw (46.center) to (45.center);
		\draw [cWire, in=-90, out=90] (52.center) to (54.center);
		\draw [qWire, in=-90, out=90] (56.center) to (55.center);
		\draw (60.center) to (62.center);
		\draw (62.center) to (61.center);
		\draw (61.center) to (60.center);
		\draw [qWire] (63.center) to (64.center);
		\draw [qWire, in=-90, out=90] (94.center) to (93.center);
		\draw [qWire] (101.center) to (102.center);
		\draw [cWire, in=-90, out=90] (103.center) to (104.center);
		\draw [qWire, in=90, out=-90] (122.center) to (115.center);
		\draw (107.center) to (111.center);
		\draw (121.center) to (107.center);
		\draw [cWire, in=-90, out=90] (106.center) to (117.center);
		\draw (110.center) to (114.center);
		\draw (114.center) to (113.center);
		\draw [cWire, in=-90, out=90] (108.center) to (116.center);
		\draw (127.center) to (124.center);
		\draw (124.center) to (125.center);
		\draw (125.center) to (126.center);
		\draw (126.center) to (127.center);
		\draw [cWire, in=-90, out=90] (128.center) to (130.center);
		\draw [qWire, in=-90, out=90] (132.center) to (131.center);
		\draw (136.center) to (138.center);
		\draw (138.center) to (137.center);
		\draw (137.center) to (136.center);
		\draw [qWire] (139.center) to (140.center);
		\draw [braceedge] (143.center) to (142.center);
		\draw [qWire, in=-90, out=90] (146.center) to (145.center);
		\draw (150.center) to (152.center);
		\draw (152.center) to (151.center);
		\draw (151.center) to (150.center);
		\draw [qWire] (153.center) to (154.center);
		\draw [qWire, in=-90, out=90] (156.center) to (155.center);
		\draw [qWire, in=90, out=-90] (173.center) to (166.center);
		\draw (159.center) to (162.center);
		\draw (172.center) to (159.center);
		\draw [cWire, in=-90, out=90] (158.center) to (168.center);
		\draw (161.center) to (165.center);
		\draw (165.center) to (164.center);
		\draw [cWire, in=-90, out=90] (160.center) to (167.center);
		\draw [braceedge] (176.center) to (175.center);
		\draw (178.center) to (181.center);
		\draw (181.center) to (180.center);
		\draw (180.center) to (179.center);
		\draw (179.center) to (178.center);
	\end{pgfonlayer}
\end{tikzpicture} &\geq 0
 \quad \text{,} \quad  \label{eq:hie1}\\
 \begin{tikzpicture}[baseline=(current bounding box.center)]
	\begin{pgfonlayer}{nodelayer}
		\node [style=none] (55) at (-4, 0) {};
		\node [style=none] (56) at (-4, -1.25) {};
		\node [style=right label] (57) at (-4, -1) {$V$};
		\node [style=right label] (58) at (-6, -1) {$V$};
		\node [style=none] (59) at (-5, -1.75) {$s$};
		\node [style=none] (60) at (-3.25, -1.25) {};
		\node [style=none] (61) at (-6.75, -1.25) {};
		\node [style=none] (62) at (-5, -2.5) {};
		\node [style=none] (63) at (-6, 0) {};
		\node [style=none] (64) at (-6, -1.25) {};
		\node [style=none] (93) at (-6, 3.25) {};
		\node [style=none] (94) at (-6, 2) {};
		\node [style=right label] (95) at (-6, 2.25) {$V$};
		\node [style=none] (106) at (-6, 4.25) {};
		\node [style=none] (107) at (-6.5, 3.25) {};
		\node [style=none] (108) at (-4, 4.25) {};
		\node [style=none] (110) at (-6.5, 4.25) {};
		\node [style=none] (111) at (-3.5, 3.25) {};
		\node [style=none] (112) at (-5, 3.75) {$\mathcal{P}$};
		\node [style=none] (113) at (-3.5, 3.25) {};
		\node [style=none] (114) at (-3.5, 4.25) {};
		\node [style=none] (115) at (-4, 2) {};
		\node [style=none] (116) at (-4, 5.5) {};
		\node [style=none] (117) at (-6, 5.5) {};
		\node [style=right label] (118) at (-6, 5) {$\bSet$};
		\node [style=right label] (119) at (-4, 2.25) {$V$};
		\node [style=right label] (120) at (-4, 5) {$\bSet$};
		\node [style=none] (121) at (-6.5, 4.25) {};
		\node [style=none] (122) at (-4, 3.25) {};
		\node [style=none] (131) at (2, 0) {};
		\node [style=none] (132) at (2, -1.25) {};
		\node [style=right label] (133) at (2, -1) {$V$};
		\node [style=right label] (134) at (0, -1) {$V$};
		\node [style=none] (135) at (1, -1.75) {$s$};
		\node [style=none] (136) at (2.75, -1.25) {};
		\node [style=none] (137) at (-0.75, -1.25) {};
		\node [style=none] (138) at (1, -2.5) {};
		\node [style=none] (139) at (0, 0) {};
		\node [style=none] (140) at (0, -1.25) {};
		\node [style=none] (141) at (-2, -1.25) {$\cdots$};
		\node [style=none] (142) at (-5.5, -3) {};
		\node [style=none] (143) at (1.5, -3) {};
		\node [style=none] (144) at (-2, -3.75) {$k$};
		\node [style=none] (155) at (0, 3.25) {};
		\node [style=none] (156) at (0, 2) {};
		\node [style=right label] (157) at (0, 2.25) {$V$};
		\node [style=none] (158) at (0, 4.25) {};
		\node [style=none] (159) at (-0.5, 3.25) {};
		\node [style=none] (160) at (2, 4.25) {};
		\node [style=none] (161) at (-0.5, 4.25) {};
		\node [style=none] (162) at (2.5, 3.25) {};
		\node [style=none] (163) at (1, 3.75) {$\mathcal{P}$};
		\node [style=none] (164) at (2.5, 3.25) {};
		\node [style=none] (165) at (2.5, 4.25) {};
		\node [style=none] (166) at (2, 2) {};
		\node [style=none] (167) at (2, 5.5) {};
		\node [style=none] (168) at (0, 5.5) {};
		\node [style=right label] (169) at (0, 5) {$\bSet$};
		\node [style=right label] (170) at (2, 2.25) {$V$};
		\node [style=right label] (171) at (2, 5) {$\bSet$};
		\node [style=none] (172) at (-0.5, 4.25) {};
		\node [style=none] (173) at (2, 3.25) {};
		\node [style=none] (174) at (-2, 3.75) {$\cdots$};
		\node [style=none] (175) at (2.25, 6.5) {};
		\node [style=none] (176) at (-6.25, 6.5) {};
		\node [style=none] (177) at (-2, 7.25) {$k$};
		\node [style=none] (178) at (-6.5, 2) {};
		\node [style=none] (179) at (-6.5, 0) {};
		\node [style=none] (180) at (2.5, 0) {};
		\node [style=none] (181) at (2.5, 2) {};
		\node [style=none] (182) at (-2, 1) {$W'$};
	\end{pgfonlayer}
	\begin{pgfonlayer}{edgelayer}
		\draw [qWire, in=-90, out=90] (56.center) to (55.center);
		\draw (60.center) to (62.center);
		\draw (62.center) to (61.center);
		\draw (61.center) to (60.center);
		\draw [qWire] (63.center) to (64.center);
		\draw [qWire, in=-90, out=90] (94.center) to (93.center);
		\draw [qWire, in=90, out=-90] (122.center) to (115.center);
		\draw (107.center) to (111.center);
		\draw (121.center) to (107.center);
		\draw [cWire, in=-90, out=90] (106.center) to (117.center);
		\draw (110.center) to (114.center);
		\draw (114.center) to (113.center);
		\draw [cWire, in=-90, out=90] (108.center) to (116.center);
		\draw [qWire, in=-90, out=90] (132.center) to (131.center);
		\draw (136.center) to (138.center);
		\draw (138.center) to (137.center);
		\draw (137.center) to (136.center);
		\draw [qWire] (139.center) to (140.center);
		\draw [braceedge] (143.center) to (142.center);
		\draw [qWire, in=-90, out=90] (156.center) to (155.center);
		\draw [qWire, in=90, out=-90] (173.center) to (166.center);
		\draw (159.center) to (162.center);
		\draw (172.center) to (159.center);
		\draw [cWire, in=-90, out=90] (158.center) to (168.center);
		\draw (161.center) to (165.center);
		\draw (165.center) to (164.center);
		\draw [cWire, in=-90, out=90] (160.center) to (167.center);
		\draw [braceedge] (176.center) to (175.center);
		\draw (178.center) to (181.center);
		\draw (181.center) to (180.center);
		\draw (180.center) to (179.center);
		\draw (179.center) to (178.center);
	\end{pgfonlayer}
\end{tikzpicture} &\geq 0
 \quad \text{,} \label{eq:hie2} 
\end{align}
for all distinct totally connected wirings $W$ and $W'$. 

Here, two wirings are \emph{distinct} if they do not give the same diagram under some permutation of the external wires. A wiring is \emph{totally connected} if the diagram it gives rise to does not factorise.
\end{hiercondarb}
 First, notice that if the constraints of Level k+1 are satisfied, then the constraints of Level k are also by definition satisfied. Moreover,
the constraint that the wirings are distinct ensures that there is no redundancy within a particular level in the hierarchy. In addition, the condition that the wirings are totally connected ensures that the constraints they impose do not reduce to constraints at lower levels in the hierarchy.  We discuss the convergence of this hierarchy in App.~\ref{App:Convergence}. 
 Enumerating and finding a simple description of the distinct totally connected wirings is left as an interesting open problem. 

We see that the constraints that each level $k$ imposes are then of two types: (i) one where the process in the second layer is the product of $k$ copies of $\mathcal{P}$ -- Eq.~\eqref{eq:hie2} --, and (ii) one where the process in the second layer is the product of $k-1$ copies of $\mathcal{P}$ and two copies of $\mathcal{F}$ -- Eq.~\eqref{eq:hie1}. Note that if we had more copies of $\mathcal{F}$ (and so fewer of $\mathcal{P}$) then it would be impossible to have a totally connected wiring, hence we need consider at most two copies of $\mathcal{F}$.

\bigskip 

One particular example of a of constraint imposed by the hierarchy is
\begin{align}
    &%
\InputIfFileExists{Diagrams/const6.tikz}{}{\input{./figures/Diagrams/const6.tikz}} \geq 0\,. \label{eq:steestatmeas}
\end{align}
This condition, imposed first in Level 2, can be given the following interpretation: the PRM $\mathcal{P}$ must give valid probabilities on products of two steered states, each constructed from $s$. 
\bigskip

\section{The optimisation problems}

The hierarchy of constraints presented in the previous subsection ultimately defines some convex cone. To see this, suppose $\sigma_1$ and $\sigma_2$ satisfy the compatibility constraints of Eqs.~\eqref{eq:hie1} and \eqref{eq:hie2} for all $k$. Then, $r_1\sigma_1 + r_2\sigma_2$ will satisfy the constraints for all $r_1,r_2 \in \mathds{R}^+$.

The optimisation problem of Eq.~\eqref{eq:opprob1} is therefore carried out over $\Sigma$, the union of the cones $K[\mathcal{P}]$:
\beq
s \in \bigcup_{\mathcal{P}\in \mathsf{ParMeas}[\mathcal{F}]}K[\mathcal{P}]\quad =:\quad \Sigma.
\eeq

This allows us to cast the optimization problem in a deceptively simple form:
\begin{prob} \label{op:orig}
\beq
\mathcal{I}_{\max}:=\mathsf{sup}\left\{\ \ %
\InputIfFileExists{Diagrams/OptimisationProblem.tikz}{}{\input{./figures/Diagrams/OptimisationProblem.tikz}}\ \ \middle|\  s \in \Sigma, u_V\otimes u_V(s)=1\right\}.
\eeq
\end{prob}
While this form of the optimisation may appear simple, determining membership of the set $\Sigma$ is computationally extremely difficult. Indeed, $\Sigma$ is defined as the union of a (potentially infinite) set of cones, each of which is defined by an infinite hierarchy of constraints. In the remaining of the paper we will see how to relax these constraints, to make the optimisation problem computationally tractable, and by so compute upper bounds to $\mathcal{I}_{\max}$. Note that, since the objective function is linear, we can make this a convex optimisation problem by optimising over the convex closure of $\Sigma$.

{
In addition, one may wish to optimise the value of a Bell inequality where the cardinality of the input variables in the Bell test does not coincide with the number of settings in the fiducial measurement, i.e., $|X| \leq |\nSet|$ and/or $|Y| \leq |\nSet|$. In this case, only a subset $\iSet \subset \nSet$ of the control settings are of interest, and the relevant constraint is the existence of a partial PRM for $\iSet$. In this case, the Optimisation problem becomes: 
\begin{prob'}{op:orig} \label{op:origp}
\beq
\mathcal{I}_{\max}:=\mathsf{sup}\left\{\ \ %
\InputIfFileExists{Diagrams/OptPartial.tikz}{}{\input{./figures/Diagrams/OptPartial.tikz}}\ \ \middle|\  s \in \Sigma_{\iSet}, u_V\otimes u_V(s)=1\right\},
\eeq
where $\Sigma_{\iSet}$ is the set of potential states compatible with the existence of a partial PRM $\mathcal{P[F]}_{\iSet}$ for the settings  $\iSet \subseteq \nSet$.
\end{prob'}
}

\subsection{A relaxation to Optimisation Problem \ref{op:orig}}\label{se:bel}

In this section we will specify a particular subset of constraints imposed by the hierarchy that defines $K[\mathcal{P}]$. We will focus on a particular set of minimum requirements to demand to the GPT, which are colloquially stated as: 
\begin{compactitem}
\item Valid states under local fiducial measurements -- Eq.~\eqref{eq:const1},
\item $\mathcal{P[F]}$ is a valid measurement on generic composite states -- Eq.~\eqref{eq:singsin},
\item $\mathcal{P[F]}$ is a valid measurement on products of steered states -- Eq.~\eqref{eq:steestatmeas}, 
\item $\mathcal{P[F]}$ is a PRM -- Eq.~\eqref{eq:PRMfidcond}.
\end{compactitem}
Implementing these conditions is already computationally demanding, since we are optimising over the possible PRM $\mathcal{P[F]}$ and the possible bipartite states $s$ of the GPT. 

The optimisation problem to solve therefore reads as follows: 
\begin{prob} \label{op:rel} \quad \\
\begin{align}
    \mathcal{I}^\mathrm{R}_{\max} \quad &= \quad \stackrel[\{s, \mathcal{P}\}]{}{\mathsf{sup}}\,\, %
\InputIfFileExists{Diagrams/OptimisationProblem.tikz}{}{\input{./figures/Diagrams/OptimisationProblem.tikz}} \,\,,\\ \nonumber \\
    &\textrm{s.t.} \begin{cases} \nonumber 
    \quad 
    \begin{tikzpicture}
	\begin{pgfonlayer}{nodelayer}
		\node [style=none] (0) at (1, 0.75) {};
		\node [style=none] (1) at (1, 0) {};
		\node [style=right label] (2) at (1, 0.25) {$V$};
		\node [style=right label] (3) at (-1, 0.25) {$V$};
		\node [style=none] (4) at (0, -0.5) {$s$};
		\node [style=none] (5) at (1.75, 0) {};
		\node [style=none] (6) at (-1.75, 0) {};
		\node [style=none] (7) at (0, -1.25) {};		
		\node [style=none] (8) at (-1, 0.75) {};
		\node [style=none] (9) at (-1, 0) {};
		\node [style=upground] (38) at (1, 1) {};
		\node [style=upground] (39) at (-1, 1) {};
	\end{pgfonlayer}
	\begin{pgfonlayer}{edgelayer}
		\draw [qWire, in=-90, out=90] (1.center) to (0.center);
		\draw (5.center) to (7.center);
		\draw (7.center) to (6.center);
		\draw (6.center) to (5.center);
		\draw [qWire] (8.center) to (9.center);
	\end{pgfonlayer}
\end{tikzpicture} = 1  \hfill  [normalisation] \,,\\
    \quad %
\InputIfFileExists{Diagrams/const1.tikz}{}{\input{./figures/Diagrams/const1.tikz}} \geq 0\,,
    \quad %
\InputIfFileExists{Diagrams/const2New.tikz}{}{\input{./figures/Diagrams/const2New.tikz}} \geq 0 \,,\quad
    \quad %
\InputIfFileExists{Diagrams/const6New.tikz}{}{\input{./figures/Diagrams/const6New.tikz}} \geq 0 \,,\\
    \quad %
\InputIfFileExists{Diagrams/ParFid1.tikz}{}{\input{./figures/Diagrams/ParFid1.tikz}} = %
\InputIfFileExists{Diagrams/ParFid2.tikz}{}{\input{./figures/Diagrams/ParFid2.tikz}} \qquad \forall i\in\nSet\,.
    \end{cases}
\end{align}
\end{prob}
It is readily seen how the Optimisation Problem \ref{op:rel} is a relaxation of the Optimisation Problem \ref{op:orig} -- a solution $\mathcal{I}^\mathrm{R}_{\max}$ to the former will yield an upper bound to the solution $\mathcal{I}_{\max}$ of the latter. 

\bigskip

{To be able to implement this sort of optimisation problem on a computer, we must switch from the high level diagrammatic description of the processes in a GPT, to a lower level tensorial representation. The method for doing this is presented in App.~\ref{sec:tensor},  together with a spelled-out example for a particular scenario.  }

The constraints that appear in Optimisation Problem \ref{op:rel} can {then} be recast by means of the tensor representation as constraints on real vectors. This lower level form of Eqs.~\eqref{eq:const1}, \eqref{eq:singsin}, \eqref{eq:steestatmeas}, and \eqref{eq:PRMfidcond}, will be used when coding the scripts to carry out the numerical calculations of the next section. 

\bigskip

\noindent \textbf{Example 1.}\\
Consider the case where $n=2$. Here, as we discussed in Section \ref{se:PRM}, a normalised state $s$ can be fully parametrised as in Eq.~\eqref{eq:stateAsVector} by the vector of probabilities:
\begin{align}
    \mathbf{p}_s = \left( p_s(0|0)\,,\, p_s(0|1)\,,\, 1 \right)^T\,,
\end{align}
where $p_s(a|x)$ is the probability that outcome $a$ is obtained when the fiducial measurement $x$ is performed on a system on state $s$. In a locally tomographic GPT, a bipartite system can be parametrised as follows:
\begin{align}
\label{eq:LTparam-ex1}
    \mathbf{p}_s = \left( p_s(00|00)\,,\, p_s(00|10)\,,\, p^\mathrm{(2)}_s(0|0) \,,\, p_s(00|01)\,,\, p_s(00|11)\,,\, p^\mathrm{(2)}_s(0|1) \,,\, p^\mathrm{(1)}_s(0|0)\,,\, p^\mathrm{(1)}_s(0|1)\,,\, 1 \right)^T\,,
\end{align}
where $p^\mathrm{(j)}_s(a|x)$ denotes the marginal {conditional} probability of subsystem $j$, {and $p_s(ab|xy)$ denotes the joint conditional probabilities. Note that these parameters are also precisely those required to characterise a no-signalling box with binary inputs and outputs. This is not a {coincidence -- indeed,} this is precisely the no-signalling box that we will obtain when we measure this state with the fiducial measurement $\mathcal{F}$ on both systems. Hence, this parameterisation of the bipartite state and the form of $\mathcal{F}$ ensure that the observed correlations are no-signalling.}

{Using the tensorial notation described in App.~\ref{sec:tensor} and, in particular, the above parameterisation of the composite state (Eq.~\eqref{eq:LTparam-ex1}),} the constraints in Optimisation Problem \ref{op:rel} can be recast as follows. 

The first one, i.e., Eq.~\eqref{eq:const1}, reads: 
\begin{align}
    p_s(ab|xy) \geq 0 \,\, \forall \,  a,b,x,y \in \{0,1\} \,, \quad 
    \sum_{a,b=0:1} p_s(ab|xy) = 1 \,\, \forall \, x,y \in \{0,1\} \,,
\end{align}
 
\begin{align}
    \sum_{a=0:1} p_s(ab|xy) = p_s(b|y)\,\, \quad 
    \sum_{b=0:1} p_s(ab|xy) = p_s(a|y)\,\,
    \quad
    \forall \, a,b,x,y \in \{0,1\} \,.
\end{align}

That is, the outcome statistics of fiducial measurements on a state $s$ are a well-defined no-signalling normalised conditional probability distribution. 

The second constraint, i.e., Eq.~\eqref{eq:singsin}, reads
\begin{align}
    \sum_{vw} \mathcal{P}^{qr}_{vw} \, s^{vw} \geq 0 \quad \forall q,r \in \{0,1\}\,,
\end{align}
{where $v$ and $w$ are the indices associated to the two GPT vector spaces $V$, and $q$ and $r$ are the indices associated to the classical outcomes of the parity reading measurement, that is, each corresponds to one of the parities which is being read.}
Since $s^{vw}$ is represented by a $9$-dimensional probability vector $\mathbf{p}_s$, $\mathcal{P}^{qr}_{vw}$ may be represented by a $4 \times 9$ matrix, {$[\mathcal{P}]$}. 

The third constraint, i.e., Eq.~\eqref{eq:steestatmeas}, reads: 
\begin{align}
    \sum_{v_1,w_1,v_2,w_2} \mathcal{F}_{xv_1}^{a} \, \mathcal{P}^{rq}_{w_1 v_2} \, \mathcal{F}_{yw_2}^{b} \, s^{v_1w_2} \, s^{v_2w_2} \geq 0 \quad \forall x,y,q,r,a,b \in \{0,1\} \,,
\end{align}
{where the indices $x$ and $y$ are the indices associated to the measurement settings of the two fiducial measurements, and the indices $a$ and $b$ to their outcomes.}
Notice that the tensors $\mathcal{F}_{xv_1}^{a}$ and $\mathcal{F}_{yw_2}^{b}$  correspond to the definition of the fiducial effects for system of type $V$. {Specifically, we can write that:
\beq
e_{a|x} = (\mathcal{F}^a_{x0},\mathcal{F}^a_{x1},\mathcal{F}^a_{x2})\quad\text{and}\quad e_{b|y} = (\mathcal{F}^b_{y0},\mathcal{F}^b_{y1},\mathcal{F}^b_{y2}) \ .
\eeq
}
Hence, this equation can be further written as: 
\begin{align}
    [\mathcal{P}] \circ \left( (e_{a|x} \otimes \mathds{1}_V) \circ s \right) \otimes \left( (\mathds{1}_V \otimes e_{b|y}) \circ s \right) \geq 0 \quad \forall x,y,a,b \in \{0,1\} \,,
\end{align}
where by $\geq 0$ we mean that every element of the matrix must be $\geq 0$. This equivalent form of Eq.~\eqref{eq:steestatmeas} makes it clear to see that it indeed imposes that $\mathcal{P[F]}$ is a valid measurement on products of steered states, which are steered by fiducial measurements. 
We can denote the {(subnormalised)} steered states explicitly by \begin{align}
\label{eq:steered-ex1}
    s^{(1)}_{b|y}:=(\mathds{1}_V \otimes e_{b|y}) \circ s \\
    s^{(2)}_{a|x}:=(e_{a|x} \otimes \mathds{1}_V) \circ s. 
\end{align}
Then, the condition \eqref{eq:steestatmeas}
can be finally written as 
\begin{align}
    [\mathcal{P}] \circ  s^{(1)}_{b|x} \otimes
    s^{(2)}_{a|x} \geq 0
    \quad \forall x,y,a,b \in \{0,1\} \,,
\end{align}
{that is, the parity reading measurement must give valid probabilities on products of steered states.}

The last constraint, given by Eq.~\eqref{eq:PRMfidcond}, can be recast in the $n=2$ case as follows: 
\begin{align}
\label{eq:PRMfidcond-2}
    \sum_{q=0:1}  \mathcal{P}_{vw}^{qr} = \mathcal{F}^{0}_{1v} \mathcal{F}^{r}_{1w} + \mathcal{F}^{1}_{1v} \mathcal{F}^{1 \oplus r}_{1w} \quad \forall \, r\in\{0,1\} \,, \\
    \sum_{r=0:1}  \mathcal{P}_{vw}^{qr} = \mathcal{F}^{0}_{0v} \mathcal{F}^{q}_{0w} + \mathcal{F}^{1}_{0v} \mathcal{F}^{1 \oplus q}_{0w} \quad \forall \, q\in\{0,1\}\,, 
\end{align}
where $\oplus$ denotes sum mod 2. These tensorial equations indeed correspond to equality constraints between $9$-dimensional covectors:
\begin{align}
    &u_{\boldsymbol{\beta}} \otimes \vec{r} \circ [\mathcal{P}] = e_{0|1} \otimes e_{r|1} + ( u_V - e_{0|1}) \otimes (u_V - e_{r|1}) \quad \forall \, r\in\{0,1\} \,, \\
    & \vec{q} \otimes u_{\boldsymbol{\beta}} \circ [\mathcal{P}] = e_{0|0} \otimes e_{q|0} + ( u_V - e_{0|0}) \otimes (u_V - e_{q|0}) \quad \forall \, q\in\{0,1\} \,.
\end{align}
{which can be straightforwardly verified by noting that:
\beq
u_{\bSet} = (1,1)\ , \quad \vec{0} = (1,0)\ , \quad \vec{1} = (0,1)\,,
\eeq
and that 
\beq
u_V=(\mathcal{F}_{x0}^{0}+\mathcal{F}_{x0}^{1},\ \mathcal{F}_{x1}^{0}+\mathcal{F}_{x1}^{1},\ \mathcal{F}_{x2}^{0}+\mathcal{F}_{x2}^{1})\quad \forall\ x\in\{0,1\}. 
\eeq
}
\hfill $\blacksquare$

\bigskip

Optimisation Problem \ref{op:rel}, despite being a relaxation of Optimisation Problem \ref{op:orig} ,  still shares a common feature with the latter: they are both nonlinear optimisation problems. Indeed, we can see clearly in the formulation of Optimisation Problem \ref{op:rel} how the constraints feature products of the variables being optimised over. Solutions to such polynomial optimisation problems may be approximated by standard techniques in the literature. Here, we will consider the hierarchy of semidefinite relaxations to polynomial optimisation problems given by Lasserre \cite{lasserre2001global}. Each level of such hierarchy will give an upper bound to the solution $\mathcal{I}^\mathrm{R}_{\max}$ of Optimisation Problem \ref{op:rel}. 

\bigskip

Optimisation Problem \ref{op:rel} is formulated for the situations where the cardinality of the input variables in the Bell test match the number of settings in the fiducial measurement, i.e., $|X|=|Y|=|\boldsymbol{\eta}|$. However, as we mentioned in Section \ref{se:3.1}, this is not always necessarily the case. We will therefore next reformulate Optimisation Problem \ref{op:rel} to encompass the case of partial PRMs. 
This adjusted version of the optimisation problem will come in handy when exploring quantum correlations, {{since for example} it allows the study of Bell inequalities with two measurement settings per wing (see, e.g., the CHSH scenario in which $|X|=|Y|=2$) on qubits (whose affine dimension is $3$ rather than $2$).}  
{In addition, importantly, this adjusted version of the optimisation problem might allow us to make device-independent studies of the results\footnote{ As mentioned in Sec.~\ref{se:3}, in such a black-box approach to describing the states of the systems we only want to use the information provided by the correlations read out in the experiment. Hence, we cannot make any assumptions on what the dimension of the underlying physical systems is. }, since do not require full knowledge of the dimension of the local systems to impose the constraint of existence of partial PRMs.} Namely, we hope that the constraints for correlations 
obeyed by set of local observables imposed by existence of PRM persists, regardless of the dimension of the system.  

\newpage

\begin{prob'}{op:rel} \label{op:relQuant} \quad \\
Let $\iSet \subset \nSet$. 
\begin{align}
   \mathcal{I}^\mathrm{RP}_{\max} \quad &= \quad \stackrel[\{s, \mathcal{P_\iSet}\}]{}{\mathsf{sup}}\,\, %
\InputIfFileExists{Diagrams/OptQuant1.tikz}{}{\input{./figures/Diagrams/OptQuant1.tikz}} \,\,,\\ \nonumber \\
    &\textrm{s.t.} \begin{cases} \nonumber
        \quad 
    \begin{tikzpicture}
	\begin{pgfonlayer}{nodelayer}
		\node [style=none] (0) at (1, 0.75) {};
		\node [style=none] (1) at (1, 0) {};
		\node [style=right label] (2) at (1, 0.25) {$V$};
		\node [style=right label] (3) at (-1, 0.25) {$V$};
		\node [style=none] (4) at (0, -0.5) {$s$};
		\node [style=none] (5) at (1.75, 0) {};
		\node [style=none] (6) at (-1.75, 0) {};
		\node [style=none] (7) at (0, -1.25) {};		
		\node [style=none] (8) at (-1, 0.75) {};
		\node [style=none] (9) at (-1, 0) {};
		\node [style=upground] (38) at (1, 1) {};
		\node [style=upground] (39) at (-1, 1) {};
	\end{pgfonlayer}
	\begin{pgfonlayer}{edgelayer}
		\draw [qWire, in=-90, out=90] (1.center) to (0.center);
		\draw (5.center) to (7.center);
		\draw (7.center) to (6.center);
		\draw (6.center) to (5.center);
		\draw [qWire] (8.center) to (9.center);
	\end{pgfonlayer}
\end{tikzpicture} = 1\,, \hfill  [normalisation]  \\
    \quad %
\InputIfFileExists{Diagrams/const1.tikz}{}{\input{./figures/Diagrams/const1.tikz}} \geq 0\,,\quad
    \quad %
\InputIfFileExists{Diagrams/OptQuant2.tikz}{}{\input{./figures/Diagrams/OptQuant2.tikz}} \geq 0 \,,\quad
    \quad %
\InputIfFileExists{Diagrams/OptQuant3.tikz}{}{\input{./figures/Diagrams/OptQuant3.tikz}} \geq 0 \,,\\
    \quad %
\InputIfFileExists{Diagrams/OptQuant4.tikz}{}{\input{./figures/Diagrams/OptQuant4.tikz}} = %
\InputIfFileExists{Diagrams/OptQuant5.tikz}{}{\input{./figures/Diagrams/OptQuant5.tikz}} \qquad \forall i\in\iSet\subseteq\nSet\,.
    \end{cases}
\end{align}
\end{prob'}
{Notice finally that Optimisation Problem \ref{op:relQuant} is indeed a relaxation of Optimisation Problem \ref{op:origp}, in the same way that Optimisation Problem \ref{op:rel} is a relaxation of Optimisation Problem \ref{op:orig}.}

\bigskip

\noindent \textbf{Example 2.}\\
Consider the case where $n=3$. Here, as we discussed in Section \ref{se:PRM}, a normalised state $s$ can be fully parameterised as in Eq.~\eqref{eq:stateAsVector} by the vector of probabilities:
\begin{align}
    \mathbf{p}_s = \left( p_s(0|0)\,,\, p_s(0|1)\,,\,p_s(0|2)\,,\, 1 \right)^T\,,
\end{align}
where $p_s(a|x)$ is the probability that outcome $a$ is obtained when the fiducial measurement $x$ is performed on a system on state $s$. In a locally tomographic GPT, a bipartite system can be parameterised as follows:
\begin{align} \nonumber
    \mathbf{p}_s = \Big( &p_s(00|00)\,,\, p_s(00|10)\,,\,p_s(00|20)\,,\, p^\mathrm{(2)}_s(0|0) \,,\, p_s(00|01)\,,\, p_s(00|11)\,,\,p_s(00|21)\,,\, p^\mathrm{(2)}_s(0|1) \,,\\ &p_s(00|02)\,,\, p_s(00|12)\,,\,p_s(00|22)\,,\, p^\mathrm{(2)}_s(0|2) \,,\, p^\mathrm{(1)}_s(0|0)\,,\, p^\mathrm{(1)}_s(0|1)\,,\,p^\mathrm{(1)}_s(0|2)\,,\, 1 \Big)^T\,,
\end{align}
where $p^\mathrm{(j)}_s(a|x)$ denotes the marginal probability of subsystem $j$. 

We will furthermore consider the case where $\iSet=\{0,1\}\subset \{0,1,2\}=\nSet$.
With this choice, {using the tensorial notation of App.~\ref{sec:tensor}}, the constraints in Optimisation Problem \ref{op:relQuant} can be recast as follows. The first one, i.e., Eq.~\eqref{eq:const1},  reads: {
\begin{align}
    &p_s(ab|xy) \geq 0 \,\, \forall \,  a,b \in \{0,1\},\ x,y \in \{0,1,2\} \,, \quad 
    \sum_{a,b=0:1} p_s(ab|xy) = 1 \,\, \forall \, x,y \in \{0,1,2\} \,,\\
    &\sum_{a=0:1} p_s(ab|xy) = p_s(b|y)\,\, \quad 
    \sum_{b=0:1} p_s(ab|xy) = p_s(a|y)\,\,
    \quad
    \forall \, a,b \in \{0,1\},x,y \in \{0,1,2\} \,.
\end{align}}
That is, the outcome statistics of fiducial measurements on a state $s$ are a well-defined no-signalling normalised conditional probability distribution. 

The second constraint, i.e., Eq.~\eqref{eq:singsin}, reads
\begin{align}
    \sum_{vw} {\mathcal{P}_{\iSet}}^{qr}_{vw} \, s^{vw} \geq 0 \quad \forall q,r \in \{0,1\}\,.
\end{align}
Since $s^{ij}$ is represented by a $16$-dimensional probability vector $\mathbf{p}_s$, ${\mathcal{P}_{\iSet}}^{qr}_{vw}$ may be represented by a $4 \times 16$ matrix. Notice that the fact that ${\mathcal{P}_{\iSet}}$ is a partial PRM is captured by the fact that it only has two output systems -- hence its matrix representation has four rows.

The third constraint, i.e., Eq.~\eqref{eq:steestatmeas}, reads: 
\begin{align}
    \sum_{v_1,w_1,v_2,w_2} \mathcal{F}_{xv_1}^{a} \, {\mathcal{P}_{\iSet}}^{rq}_{w_1 v_2} \, \mathcal{F}_{yw_2}^{b} \, s^{v_1w_2} \, s^{v_2w_2} \geq 0 \quad \forall a,b,q,r \in \{0,1\},\ x,y\in\{0,1,2\} \,.
\end{align}
Notice that the tensors $\mathcal{F}_{xv_1}^{a}$ and $\mathcal{F}_{yw_2}^{b}$ actually correspond to the definition of the fiducial effects for system of type $V$. If we represent ${\mathcal{P}_{\iSet}}^{qr}_{vw}$ by a $4 \times 16$ matrix $[\mathcal{P}_{\iSet}]$,  hence, this equation can be further written as: 
\begin{align}
    [\mathcal{P}_{\iSet}] \circ \left( (e_{a|x} \otimes \mathds{1}_V) \circ s \right) \otimes \left( (\mathds{1}_V \otimes e_{b|y}) \circ s \right) \geq 0 \quad \forall a,b \in \{0,1\},\ x,y \in \{0,1,2\} \,,
\end{align}
where by $\geq 0$ we mean that every element of the matrix must be $\geq 0$. This equivalent form of Eq.~\eqref{eq:steestatmeas} makes it clear to see that it indeed imposes that $\mathcal{P}_{\iSet}$ is a valid measurement on products of steered states, which are steered by fiducial measurements. 

The last constraint, given by Eq.~\eqref{eq:PRMfidcond}, can be recast in the $n=3$ case as follows: 
\begin{align}
    \sum_{q=0:1}  {\mathcal{P}_{\iSet}}_{vw}^{qr} = \mathcal{F}^{0}_{1v} \mathcal{F}^{r}_{1w} + \mathcal{F}^{1}_{1v} \mathcal{F}^{1 \oplus r}_{1w} \quad \forall \, r\in\{0,1\} \,, \\
    \sum_{r=0:1}  \mathcal{P_{\iSet}}_{vw}^{qr} = \mathcal{F}^{0}_{0v} \mathcal{F}^{q}_{0w} + \mathcal{F}^{1}_{0v} \mathcal{F}^{1 \oplus q}_{0w} \quad \forall \, q\in\{0,1\}\,, 
\end{align}
where $\oplus$ denotes sum mod 2. These tensorial equations indeed correspond to equality constraints between $16$-dimensional covectors:
\begin{align}
    &u_{\boldsymbol{\beta}} \otimes \vec{r} \circ [\mathcal{P}] = e_{0|1} \otimes e_{r|1} + ( u_V - e_{0|1}) \otimes (u_V - e_{r|1}) \quad \forall \, r\in\{0,1\} \,, \\
    & \vec{q} \otimes u_{\boldsymbol{\beta}} \circ [\mathcal{P}] = e_{0|0} \otimes e_{q|0} + ( u_V - e_{0|0}) \otimes (u_V - e_{q|0}) \quad \forall \, q\in\{0,1\} \,.
\end{align}
\hfill $\blacksquare$

\bigskip

\section{ Approximating $\mathcal{I}_{\max}$ for various Bell inequalities }

In Ref.~\cite{czekaj2018bell} it was shown that existence of the Bell measurement for two systems, each with three observables (i.e., for $|\nSet|=3$) imposes {that there are} no {post-quantum} correlations. 
 In our work we want to pose {the} more general problem of whether parity-readable observables can lead to {post-quantum} correlations. 
In order to tackle this question, we defined a hierarchy of constraints that the existence of parity-readable observables imposes on the states and effects of the underlying GPT, and therefore on the sets of correlations that the GPT allows. We explored the boundary of the allowed correlations by tackling the Optimization Problems defined in the previous section. The results of these numerical explorations led us to further formulate a conjecture, re-stated below: 

\setcounter{conj}{0}

\begin{conj}\label{conject}
 
Suppose we have a bipartite system in a GPT that satisfies local tomography and no-signaling. 
The, the observables for which there exists a parity reading measurement 
cannot be used to violate any Bell inequality more than quantum mechanics does.

\noindent [Old below:]\\
Suppose that a bipartite system satisfies local tomography and no-signaling.
Moreover, suppose that product of states is a valid state. Then the observables for which there exists parity reading measurement 
do not violate any Bell inequality more than quantum mechanics does. 
\end{conj}
Formally, the conjecture, if true, means that for any Bell inequality the Optimization Problem \ref{op:origp} returns at most the quantum  bound. {Note, however,} that we do not always expect {OP\ref{op:origp}} to actually reach the quantum bound, as the maximal quantum value is not necessarily achieved by observables which are parity-readable within quantum theory.

In this section we present numerical computations towards upper-bounding the solution to OP\ref{op:rel}. 
Indeed, our numerical explorations focus on the relaxed problems OP\ref{op:rel} and  OP\ref{op:relQuant}, depending on the cardinalities of $\bSet$ and $\nSet$. 
As mentioned in the previous section, we will approximate the solution to OP\ref{op:relQuant} by means of a hierarchy of semidefinite relaxations formulated by  Lasserre \cite{lasserre2001global}, using mostly the Lasserre hierarchy levels 1+AB and 2. A brief explanation of what these two levels mean is presented below, and refer the reader to Ref.~\cite{lasserre2001global} for a thorough exposition.  Throughout the next subsection we also discuss the relation between the numerical results and the conjecture we formulated.  

Finally, in this section we provide an analytical proof that GPTs which violate local tomography admit PR-box correlations under the constraints of OP\ref{op:rel} -- that is, OP\ref{op:rel} may yield a value of $\frac{1}{2}$ for the CHSH inequality (in the notation of Eq.~\eqref{eq:chsh}) within non-tomographically local GPTs.  This highlights the relevance and impact of the assumption of local tomography. We also discuss how the violation of local tomography by a GPT may impact whether its correlations satisfy or not the conjecture. In particular, we discuss how the result we show does not necessarily imply that non-tomographically local GPTs violate Conjecture \ref{conject}, since the actual optimisation problem to be solved -- OP\ref{op:origp} -- imposes additional constraints to those appearing in OP\ref{op:rel}.

 Before moving on to presenting the numerical results, let us briefly comment on the so-called Lasserre hierarchy. Each Lasserre hierarchy level is related to the semidefiniteness of a matrix (whose definition we will not give here), and the rows and columns of this matrix have particular labels depending on what level we are focusing on. Let $\Upsilon$ be the set that contains the variables we are optimising over plus the element $1$. In the first level of the Lasserre hierarchy, the matrix under study has row and columns labelled by the elements of $\Upsilon$. In the second level of the hierarchy, however, the matrix under study is of much larger size, and its rows and columns are labelled by the elements of $\Upsilon \times \Upsilon$, where $\times$ denotes the Cartesian product of sets. The so-called 1+AB level lies in between the first and the second -- the matrix corresponding to 1+AB is a sub-matrix of that of level 2, and  the matrix corresponding to level 1 is a sub-matrix of that of 1+AB. In particular, the rows and columns of the matrix corresponding to level 1+AB are labelled by the elements of the set $\Upsilon \times \Upsilon \setminus \{{(\upsilon, \upsilon)} | \upsilon \in \Upsilon\}$.

The numerical computations from Sec.~\ref{se:chsh} were performed with Python 3.7. The SDP relaxation of polynomial programming was calculated with the package Ncpol2sdpa \cite{npcol2sdpa_ref}. The SDP problem was solved using SDPA \cite{sdpa_ref}. 
All other numerical computations were carried out by a sparsity-adapted SDP relaxation of the polynomial optimization problem modeled with the TSSOS \cite{wang2021tssos} algorithm. For more details on the modeling syntax, we refer the interested reader to the tutorial from \cite[Appendix~B.2]{sparsebook} and the online website \url{https://github.com/wangjie212/TSSOS}. Each SDP problem was solved using Mosek \cite{andersen2000mosek}.

\subsection{CHSH inequality}\label{se:chsh}
In a bipartite Bell scenario featuring two dichotomic measurements per party, the most studied inequality is the  Clauser, Horne,  Shimony, Holt (CHSH) {inequality} { \cite{CHSH}, which, using the notation of 
Eq.~\eqref{eq:LTparam-ex1}, reads: }
\begin{align}
\label{eq:chsh}
    I_{\text{CHSH}}(\mathbf{p}_s)=-p_s^{(1)}(0|0)-p_s^{(2)}(0|0)+p_s(00|00)+p_s(00|10)+p_s(00|01)-p_s(00|11)\, .
\end{align}
This inequality is bounded from above, and the corresponding classical, quantum, and non-signalling bounds are:
\begin{align}
    \beta_{\text{CHSH}}^\text{C}=0\,,\quad \beta_{\text{CHSH}}^\text{Q}=\frac{\sqrt{2}-1}{2}\sim 0.2071\,, \text{ {and}}\quad \beta_{\text{CHSH}}^\text{NS}=\frac{1}{2}\,.
\end{align}
{Note that} in this section we {shall make a slight abuse of notation, and} denote  
Alice and Bob's observables  by $X,Y,Z$, {which is not to be confused with the use of
$X$ and $Y$ to denote the} sets of inputs.

{The numerical results presented in this subsection are summarized in Fig.~\ref{fig:table-Bell}.}

\bigskip

\noindent{\bf Case $|\iSet|=|\nSet|=2$.} This is the simplest possible problem, where there are only two observables per {party} (that is, Alice has two observables $X,Z$, and the same for Bob) and the PRM measures {the} two parities {$XX$ and $ZZ$}. 
We {approximated the solution of OP\ref{op:rel} applied to the CHSH inequality, by applying a Lasserre SDP relaxation with hierarchy level `a bit lower than 1+AB'. We will specify shortly what this means, but will first elaborate on the specific parameterisation we chose for OP\ref{op:rel}.}
The state is described by 8 parameters coming from local tomography {(see Eq.~\eqref{eq:LTparam-ex1})} which we recall here in a more compact, matrix notation:
\begin{equation}
    \label{eq:ps-matrixform}
    \mathbf{p_s}=\left[ \begin{array}{ccc}
p_s(00|00)&p_s(00|01)   & p_s^{\mathrm{(1)}}(0|0)   \\
p_s(00|10)& p_s(00|11)  &  p_s^\mathrm{(1)}(0|1)  \\
p_s^\mathrm{(2)}(0|0)& p_s^\mathrm{(2)}(0|1)  & 1  
    \end{array}
    \right].
\end{equation}
The (unnormalized) Alice states steered by Bob
given by Eq.~\eqref{eq:steered-ex1} {are expressed as:}
\begin{align}
s^\mathrm{(1)}_{0|0} &= (p_s(00|00), p_s(00|10), p^\mathrm{(2)}(0|0))^T \,, \nonumber \\
s^\mathrm{(1)}_{1|0} &= (p^\mathrm{(1)}(0|0) - p_s(00|00), p^\mathrm{(1)}(0|1) - p_s(00|10), 1 - p^\mathrm{(2)}(0|0))^T \,,
\nonumber \\
s^\mathrm{(1)}_{0|1} &= (p_s(00|01), p_s(00|11),p^\mathrm{(2)}(0|1))^T \,, \text{ and}
  \nonumber \\
s^\mathrm{(1)}_{1|1} &= (p^\mathrm{(1)}(0|0) - p_s(00|01), p^\mathrm{(1)}(0|1) - p_s(00|11), 1 - p^\mathrm{(2)}(0|1))^T \,. \label{eq:st}
\end{align}
{ Bob's states steered by Alice have the same form as in Eq.~\eqref{eq:st} but exchanging $XZ\leftrightarrow ZX$  and $\mathrm{(1)}\leftrightarrow \mathrm{(2)}$. }
Now, the constraint that PRM measures parity 
{(given by Eq.~\eqref{eq:PRMfidcond-2})}
reads 
\begin{align}
    \mathcal{P}^{00}+  \mathcal{P}^{01}
    =\left[ \begin{array}{ccc}
          2 & 0 & -1  \\
          0 & 0 & 0 \\
         -1 & 0 & 1 
    \end{array} \right]\,,
    \nonumber \\
    \mathcal{P}^{10}+  \mathcal{P}^{11}
    =\left[ \begin{array}{ccc}
          0 & 0 &  0  \\
          0 & 2 & -1 \\
          0 & -1 & 1 
    \end{array} \right]\,.
\end{align}
{This can be obtained from Eq.~\eqref{eq:PRMfidcond-2} as follows. First we have,} for example,
\begin{equation}
    (\mathcal{P}^{00}+  \mathcal{P}^{01}) \cdot \mathbf{p}_s=
    p_s(00|00)+p_s(11|00)\,,
\end{equation}
{where $\cdot$ represents the Frobenius inner product of the two matrices.}
Then using
\begin{align}
p_s(00|00)+p_s(01|00) &= p^{(1)}(0|0)\,, \nonumber \\ p_s(11|00)+p_s(01|00) &= p^{(2)}(0|0) \,, \nonumber \\
 p_s(00|00)+  p_s(01|00)+ p_s(10|00)+ p_s(11|00) &= 1\,,
\end{align}
we get 
\begin{equation}
    (\mathcal{P}^{00}+  \mathcal{P}^{01}) \cdot \mathbf{p}_s=
    1+2 p_s(00|00) - p^{(1)}(0|0)-p^{(2)}(0|0)\,,
\end{equation}
which leads to the form above. 
Preserving probability by PRM reads as 
\begin{align}
\label{eq:id}
    \mathcal{P}^{00}+  \mathcal{P}^{01} +
    \mathcal{P}^{10}+  \mathcal{P}^{11}=
    \left[ \begin{array}{ccc}
0&0&0\\
0&0&0\\
0&0&1
    \end{array}
    \right]=\Id\,.
\end{align}
Thus, {the condition that $\mathcal{P}$ is a PRM is captured by} the following free parameters: 
\begin{align}
    \label{eq:free}    \mathcal{P}^{00}-  \mathcal{P}^{01} -
    \mathcal{P}^{10}+  \mathcal{P}^{11}
    =\left[ \begin{array}{ccc}
c_{11}& c_{12} & c_{13}\\
c_{21}& c_{22} & c_{23}\\
c_{31}& c_{32} & c_{33}
    \end{array}
    \right]\equiv \cal C.
\end{align}

{Finally, the constraints that we still need to impose are the positivity} of PRM effects 
both on the state, as well as on tensor products on all pairs of 
steered states that can be obtained from it. 
{We see then that OP\ref{op:rel} requires us to optimise over the free parameters (state $s$ given by Eq.~\eqref{eq:ps-matrixform}
 and $\cal C$), under the positivity constraints of the previous sentence. }

{Now, to approximate the solution to OP\ref{op:rel}, we apply a particular level of the Lasserre hierarchy, which is slightly lower than the previously described 1+AB, and which we will denote by 1+AB${}^*$. Let $\Upsilon_p$ denote the set of free parameters given by given by Eq.~\eqref{eq:ps-matrixform}, and  $\Upsilon_{\mathcal{C}}$ that given by the free parameters in $\cal C$. Here, $\Upsilon = \Upsilon_p \bigcup \Upsilon_{\mathcal{C}} \bigcup \{1\}$. However, the matrix under study in the level we consider here has rows and columns labeled by the elements of $\left(\Upsilon \times \Upsilon\right) \setminus \left( \Upsilon_p \times \Upsilon_p \right) \setminus \left( \Upsilon_{\mathcal{C}} \times \Upsilon_{\mathcal{C}} \right)  $ -- that is, it is a submatrix of that considered in level 1+AB.}

{The upper bound to OP\ref{op:rel} given by the 1+AB${}^*$ level of the Lasserre hierarchy, gives a value of $\sim 0.2071$, which  agrees up to numerical precision with  $\mathcal{I}^\mathrm{R}_{\max} = \frac{\sqrt{2}-1}{2}$. This equality follows from recalling that the Tsirelson's bound value can be achieved within Quantum theory by a Bell measurement, and hence yields a lower bound to $\mathcal{I}^\mathrm{R}_{\max}$. In other words, here we recover Tsirelson's bound for the CHSH inequality.

 \bigskip
 
 \noindent{\bf Case $|\iSet|=2$, $|\nSet|=3$.}
 {Here} we still assume that the PRM measures just two parities (i.e., those of $XX$ and $ZZ$), but {now} Alice and Bob have one more additional observable (i.e., Y). This is an important case, as it allows for the possibility that the constraints {imposed} by the existence of a PRM are sensitive to the dimension of the local systems.  Notice that if the constrains stemming from the existence of a PRM turn out to be independent of the dimensions of the local systems, one can then take a device-independent approach to the problem and only rely on the black-box statistics to make assessments on the possible violations of the Bell inequality.

 {Our numerical results show that in this case Tsirelson's bound is also not violated. The results are computed exactly as in the previous case with $|\nSet|=2$: we upperbound the value of $\mathcal{I}^\mathrm{RP}_{\max}$ via the 1+AB${}^*$ level of the Lasserre hierarchy, which  agrees up to numerical precision with Tsierlson's bound for the CHSH inequality as the solution to OP\ref{op:relQuant}. }
 
 \bigskip
 
\noindent{\bf Case $|\iSet|=3$, $|\nSet|=3$.}
Here, both parties have three observables, and the PRM measures the parity of all three of them. In this case, there is no need to run numerics to approximate the solution to $\mathcal{I}^\mathrm{R}_{\max}$.  On the one hand, notice that the optimisation to be carried out is the same as that for the case with  $|\iSet|=2$ and $|\nSet|=3$, with some additional constraints given by the requirement that the PRM reads out the parity of the extra pair of fiducial measurements (since now $|\iSet|=3$). Hence, $\mathcal{I}^\mathrm{R}_{\max}[|\iSet|=3\,,\,|\nSet|=3] \leq \mathcal{I}^\mathrm{RP}_{\max}[|\iSet|=2\,,\,|\nSet|=3]$. Given the results from above it follows that, up to numerical precision, $\mathcal{I}^\mathrm{R}_{\max}[|\iSet|=3\,,\,|\nSet|=3] \leq \frac{\sqrt{2}-1}{2}$, the latter being Tsirelson's bound for CHSH. 

On the other hand, quantum theory is an example of a generalised probabilistic theory that yields the value $CHSH = \frac{\sqrt{2}-1}{2}$ while complying with the constraints from OP\ref{op:orig}. Indeed, taking the three fiducial measurements per side to be the Pauli observables, one can construct a PRM that does the trick. Hence, $\mathcal{I}^\mathrm{R}_{\max}[|\iSet|=3\,,\,|\nSet|=3] \geq \frac{\sqrt{2}-1}{2}$. From this follows that, up to numerical precision, $\mathcal{I}^\mathrm{R}_{\max}[|\iSet|=3\,,\,|\nSet|=3] \sim \frac{\sqrt{2}-1}{2}$.

\subsection{AMP inequalities}
In a bipartite Bell scenario featuring two dichotomic measurements per party, a relevant family of inequalities was defined by Ac\'in, Massar, and Pironio (AMP) \cite{AMP-inequality-2012}. {These} correspond to tilted CHSH inequalities, and {have been found} to be useful for randomness `generation' \cite{AMP-inequality-2012}. In the traditional language, the {value assigned to the} linear functional associated to the inequality reads:
\begin{align}
I^{\mathrm{AMP}}_{\alpha,\gamma} = \langle \gamma \, A_0 + \alpha \, {A_0B_0} + \, \alpha {A_0B_1} + {A_1B_0} - {A_1B_1} \rangle \,,
\end{align}
where the parameters $\alpha$ and $\gamma$ satisfy: $\alpha \geq 1$, $\gamma \geq 0$, and $\gamma < 2$.
{These inequalities are} bounded from above, and {their} corresponding classical, quantum, and non-signalling bounds {when $\alpha \gamma \leq 2$} are:
\begin{align}
\beta_{\text{AMP}}^{\text{C}}(\alpha,\gamma) &= 2\alpha + \gamma,\\
\beta_{\text{AMP}}^{\text{Q}}(\alpha,\gamma) &= 2\sqrt{(1+\alpha^2)\left(1+\frac{\gamma^2}{4}\right)}\,, \text{ and}\\
\beta_{\text{AMP}}^{\text{NS}}(\alpha,\gamma) &= 2+2\alpha\,.
\end{align}

In {our notation, that is, in} terms of the probabilities,  the AMP inequialities are equivalently captured by the following linear functional: 
\begin{align} \nonumber
I_{\alpha,\gamma} = &-(2\alpha+\gamma) \, p_s^{(1)}(0|0) - (1+\alpha) \, p_s^{(2)}(0|0) + 2\alpha \, p_s(00|00)+  2\alpha \, p_s(00|01)   \\
&+ (1-\alpha) \, p_s^{(2)}(0|1) +  2 \, p_s(00|10)  -2 \, p_s(00|11)\,,
\end{align}
 whose corresponding classical, quantum, and non-signalling bounds when $\alpha \gamma \leq 2$ are: 
\begin{align}
\beta_{\alpha,\gamma}^{\text{C}}(\alpha,\gamma) &= 0\,,\quad \beta_{\alpha,\gamma}^{\text{NS}}(\alpha,\gamma) = 1-\frac{\gamma}{2}\,,\text{ and} \\
\beta_{\alpha,\gamma}^{\text{Q}}(\alpha,\gamma) &= \sqrt{(1+\alpha^2)\left(1+\frac{\gamma^2}{4}\right)} - \frac{2\alpha + \gamma}{2}\,, \quad 
\end{align}

 To explore the case of these inequalities in this section,  we have only considered the case of $|\iSet|=|\nSet|=2$.

The values of $\alpha$ and $\gamma$ that we considered are quite varied. On the one hand, we took $\alpha$ from the set
$\{1,3,5,7,9,11\}$ and then, for each such $\alpha$, considered six equally-spaced values for $\gamma$ (see Sec.~\ref{sec:AMP-table} in the
Appendix). On the other hand, we wanted to explore the transition between $\alpha=1$ and $\alpha=3$ more deeply, hence we explored the linear functionals $I_{\alpha,\gamma}$ also for $\alpha$ taken from the set $\{1.01,\, 1.05,\, 1.1,\, 1.2,\, 1.5,\, 2,\, 2.2,\, 2.4,\, 2.6,\, 2.8,\, 2.9,\, 2.95\}$ whilst keeping $\gamma=0$. The motivation for this will hopefully become clear later on. 

 For each of the linear functionals $I_{\alpha,\gamma}$ defined by the above-mentioned values of $\alpha$ and $\gamma$, we asked what the value of $\mathcal{I}_{\max}$ -- the solution to the optimisation problem OP\ref{op:orig} -- is. Here we computed an upper bound to $\mathcal{I}_{\max}$ for each inequality, by applying two relaxations to OP\ref{op:orig}:
\begin{compactitem}
\item First, instead of demanding that $s \in \Sigma$, we only request that the state $s$ belongs to the cone $K[\mathcal{P}]$ that satisfies the second level of our hierarchy (see Eqs.~\eqref{eq:const1}, \eqref{eq:singsin}, \eqref{eq:singswap}, \eqref{eq:hie1}, and \eqref{eq:hie2}). 
\item Second, by solving the associated level 3 of the Lasserre hierarchy, we upper bound the solution to the relaxation to OP\ref{op:orig} defined in the previous item.
\end{compactitem}

 Our numerical calculations show that, in the cases where $\alpha \geq 3$ the upper bound for $\mathcal{I}_{\max}$ is smaller than the inequality's Tsirelson's bound (see Fig.~\ref{fig:AMP-Table}). Indeed, up to numerical precision $\mathcal{I}_{\max} \leq 0$, where $0$ is the classical bound of the inequality. For the case where $\alpha=1$, the inequality becomes the CHSH inequality plus a extra term corresponding to the single-party observable $A_0$. Beyond the case $\gamma = 0$ (which corresponds to the traditional CHSH inequality), other values of $\gamma$ give an upper bound to $\mathcal{I}_{\max}$ that is larger than yet close to the inequality's Tsirelson's bound (see Fig.~\ref{fig:AMP-Table}). Finally, for the values of $\alpha \in \{1,\, 1.01,\, 1.05,\, 1.1,\, 1.2,\, 1.5,\, 2,\, 2.2,\, 2.4,\, 2.6,\, 2.8,\, 2.9,\, 2.95,\, 3,\, 5,\, 7,\, 9,\, 11\}$ and $\gamma=0$ one observes that the upper bound to $\mathcal{I}_{\max}$ drops to 0 when $\alpha$ goes from $1$ to $3$. 
Reading into the data of Fig.~\ref{fig:AMP-Table}, one can notice some additional interesting behavior. For instance, in a few cases the upper bound to $\mathcal{I}_{\max}$ is equal to Tsirelson's bound, at least, up to the numerical precision. In particular, this happens for the cases where $\alpha\gamma=2$ explored in this manuscript. The numerical results presented in this subsection are further summarized in Fig.~\ref{fig:table-Bell}.

The cases in which the PRM bounds the value of the AMP inequality to be smaller than the maximal quantum value 
(in contrast to CHSH in which the exact quantum bound was obtained) are likely to be cases in which the quantum bound is achieved for quantum observables for which there does not exist a PRM. This suggests that the observables {which allow for a} PRM 
{may feature some particular properties regarding them being maximally complementary.}
Understanding the scope of parity-readable observables within quantum theory, and the correlations which they can realise, is therefore an important topic for future work.

 Now, what does this all mean for the purpose of our conjecture? Well, no conclusive statement can be drawn from the numerics run for $\alpha=1\,,\,\gamma\neq0$. However all other cases are consistent with (and hence support) Conjecture \ref{conject}.

\begin{figure}
\vskip 0.5cm
\begin{subfigure}[b]{0.75\textwidth}
  \centering
  \includegraphics[width=.9\linewidth]{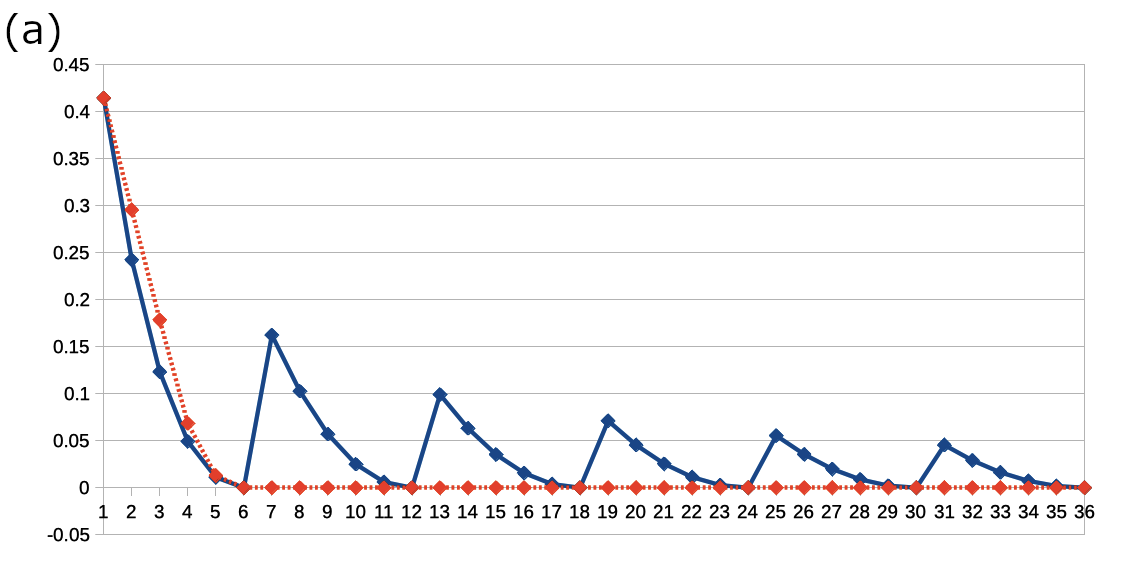}
\end{subfigure}\hspace{5mm}
\centering
\begin{subfigure}[b]{0.75\textwidth}
  \centering
  \includegraphics[width=.9\linewidth]{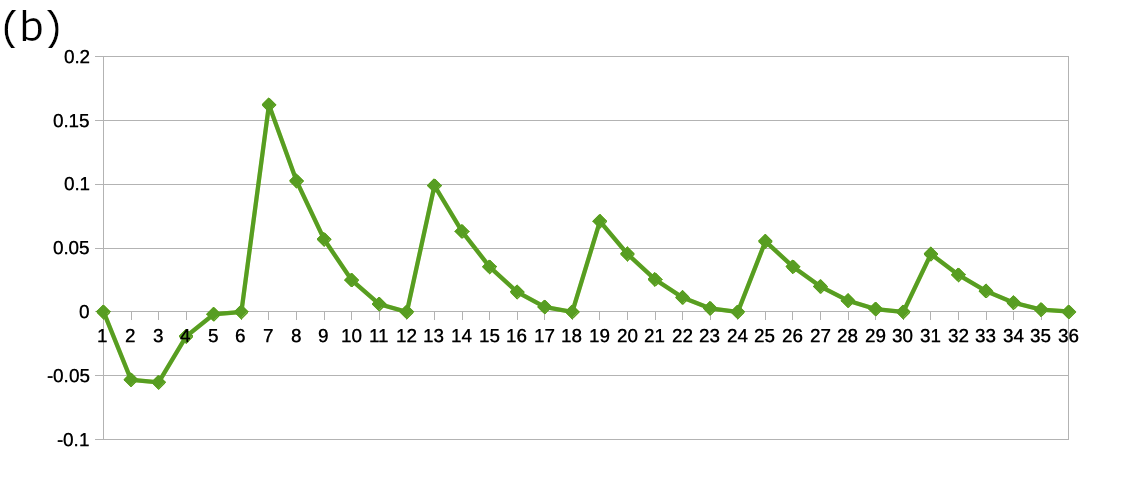}
\end{subfigure}\hspace{5mm}
\begin{subfigure}[b]{0.75\textwidth}
  \centering
  \includegraphics[width=.9\linewidth]{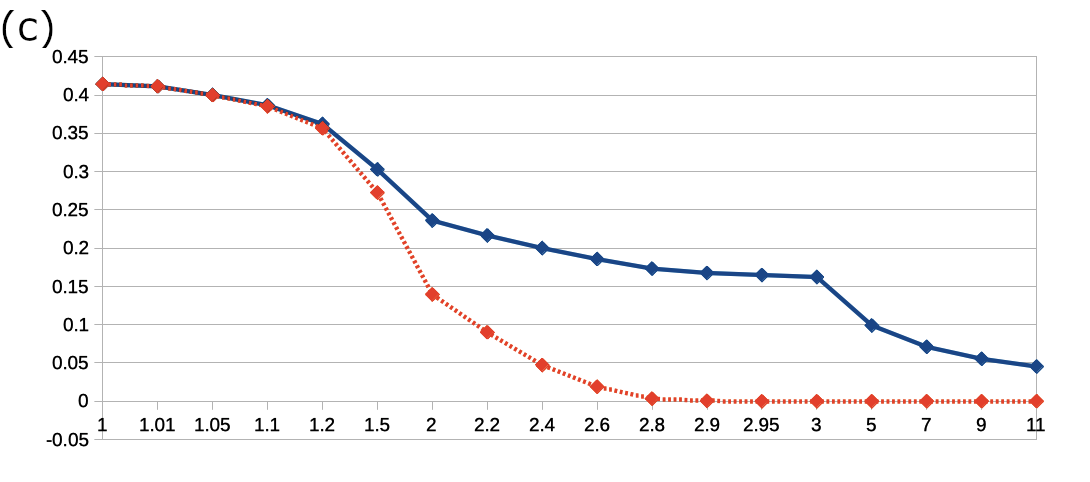}
\end{subfigure}
\begin{subfigure}[b]{0.75\textwidth}
  \centering
  \includegraphics[width=.9\linewidth]{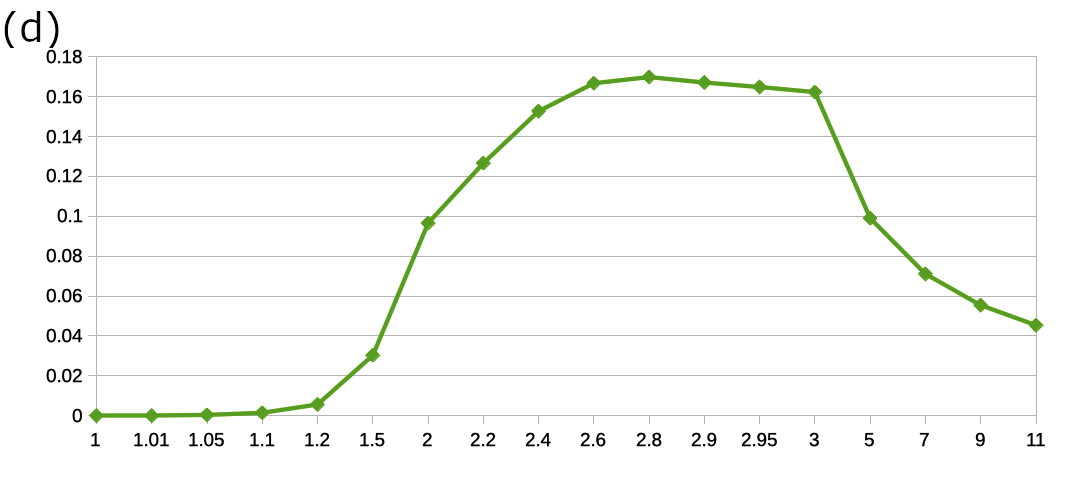}
\end{subfigure}
\caption{  Numerical results for the AMP inequalities. 
(a) Each integer value, $x$ in the horizontal axis corresponds to a different choice of pair of parameters $(\alpha,\gamma)$, i.e., a different AMP inequality (see Table in Eq.~\eqref{table:alphagamma}). The vertical axis plots, for each inequality, both the numerical approximation to OP\ref{op:orig} (dashed red line) and the Tsirelson's bound of the inequality (solid blue line). 
(b) The horizontal axis is the same as for the case (a). The vertical axis plots the difference between the numerical approximation to  OP\ref{op:orig} and the Tsirelson's bound of the inequality. 
(c) The value of the horizontal axis corresponds to the value of $\alpha$. The value of $\gamma$ is always 0. The vertical axis plots, for each inequality, both the numerical approximation to OP\ref{op:orig} (dashed red line) and the Tsirelson's bound of the inequality (solid blue line). 
(d) The horizontal axis is the same as for the case (c). The vertical axis plots the difference between the numerical approximation to  OP\ref{op:orig} and the Tsirelson's bound of the inequality. 
In (b) and (c), witnessing a nonnegative value in the plot means that $\mathcal{I}^\mathrm{R}_{\max}$ lies below Tsirelson's bound for that particular inequality. 
In all these figures, we approximate OP\ref{op:orig} by first relaxing OP\ref{op:orig} and then using the third level of the Lasserre hierarchy -- see main text for details.}
\label{fig:AMP-Table}
\end{figure}

\subsection{AQ inequality}
In Ref.~\cite{Navascues2015-AQ}
an inequality was provided, which 
is violated by so called  ``almost quantum'' correlations \cite{Navascues2015-AQ}, 
but is not  violated by any quantumly realisable correlations. {Here we refer to this inequality as \textit{AQ inequality}.}
In our notation it is given by 
\begin{align}\nonumber
    I_{\text{AQ}}(\mathbf{p}_s)=
    &\frac{30}{31} {p^{(1)}(0|0)}
    -\frac{167}{9} {p^{(1)}(0|1)}
    +\frac{30}{31} {p^{(2)}(0|0)}
    -\frac{74}{11} p(00|00)
    +\frac{174}{11} p(00|10) \\
    &-\frac{167}{9} {p^{(2)}(0|1)}
    +\frac{174}{11} p(00|01)
    +\frac{244}{23} p(00|11)\,.
\end{align}
This inequality is bounded from above, and the corresponding classical, quantum, almost-quantum, and non-signalling bounds are:
\begin{align}
    &\beta_{\text{AQ}}^\text{C}=\frac{30}{31}\sim 0.9677\,,\quad {\beta_{\text{AQ}}^\text{Q} < 1}\,,\\
    &\beta_{\text{AQ}}^\text{AQ}=1.0232\,, \text{ and}\quad
    \beta_{\text{AQ}}^\text{NS}=3.5347\,.
\end{align}

Let us first consider the case  $|\iSet|=2$, $|\nSet|=2$. Similarly to the case for the AMP inequalities, we upper-bound $\mathcal{I}_{\max}$ by the solution to  OP\ref{op:orig} provided by the second level of the PRM-hierarchy of constraints presented in this paper. This solution is moreover estimated (i.e., upper bounded) by using the third level of the Lasserre hierarchy for polynomial optimisation problems. In this case, we obtain $\mathcal{I}_{\max} \leq 1.387818418422242$. This upper bound to $\mathcal{I}_{\max}$ is quite larger than the quantum bound, and hence not much can be concluded. Going to higher levels in the PRM-hierarchy might be the most promising step to take, however our current computational capabilities cannot handle the number of constraints and hence we defer this option for future work.

Next, we considered the case of  $|\iSet|=3$, $|\nSet|=3$. Approximating the solution of OP\ref{op:rel} via the 1+AB level of the Lasserre hierarchy gives $\mathcal{I}^\mathrm{R}_{\max} < 1.7$. This number is substantially larger than the quantum bound for the inequality, and hence we are in a similar situation to the case presented before. 

In this case, however, one can further explore the specific cases where some extra properties are required of the PRM being optimised over. This is similar to what we discussed in Sec.~\ref{subsec:examples}. So let us remind ourselves first of what these two specific types of PRM we focus on are. Notice that since $|\iSet|=3$ and $|\nSet|=3$ the outcome of a PRM is of the form $(\pm,\pm,\pm|1,2,3)$, where $\pm_j$ tells whether the pair of fiducial measurements  $(j,j)$ is correlated ($+$) or anti-correlated ($-$). A PRM is of `quantum' type if it assigns non-zero probability only to the outcomes 
\begin{align}
\{(-,-,-|1,2,3),(-,+,+|1,2,3),(+,-,+|1,2,3),(+,+,-|1,2,3)\}\,, \quad \text{(`quantum')}\,.
\end{align}
The motivation behind the name is that a PRM of Pauli measurements satisfies this constraints. In turn, a PRM is of `mirror quantum' type if it assigns non-zero probability only to the outcomes 
\begin{align}
\{(-,-,+|1,2,3),(-,+,-|1,2,3),(+,-,-|1,2,3),(+,+,+|1,2,3)\}\,, \quad\text{(`mirror quantum')}\,.
\end{align}
The motivation behind the name is that it is exactly the complement of the `quantum'-type PRM. 

With this in mind, we approximated the solution of OP\ref{op:rel} via the second level of the Lasserre hierarchy. The results we obtained are: 
\begin{align}
\mathcal{I}^\mathrm{R}_{\max} &< 0.8782940363666021 \,, \quad \text{(`quantum')}\,,\\
\mathcal{I}^\mathrm{R}_{\max} &< 1.3953137950470862 \,, \quad \text{(`mirror quantum')}\,.
\end{align}

We see that restricting the optimisation to PRMs that are of `quantum' type give quite a strong constraint on the possible value of $\mathcal{I}_{\max}$, which here happens to be below the quantum bound (even below the classical bound) of the inequality. We believe that demanding that a PRM of `quantum' type exists somehow forces the fiducial measurements to display some complementarity properties, and hence are not ideal for maximising the value of the linear functional $I_{\text{AQ}}$. 

The numerical results presented in this subsection are summarized in Fig.~\ref{fig:table-Bell}.

\begin{figure}
	\centering
	\includegraphics[width=\linewidth]{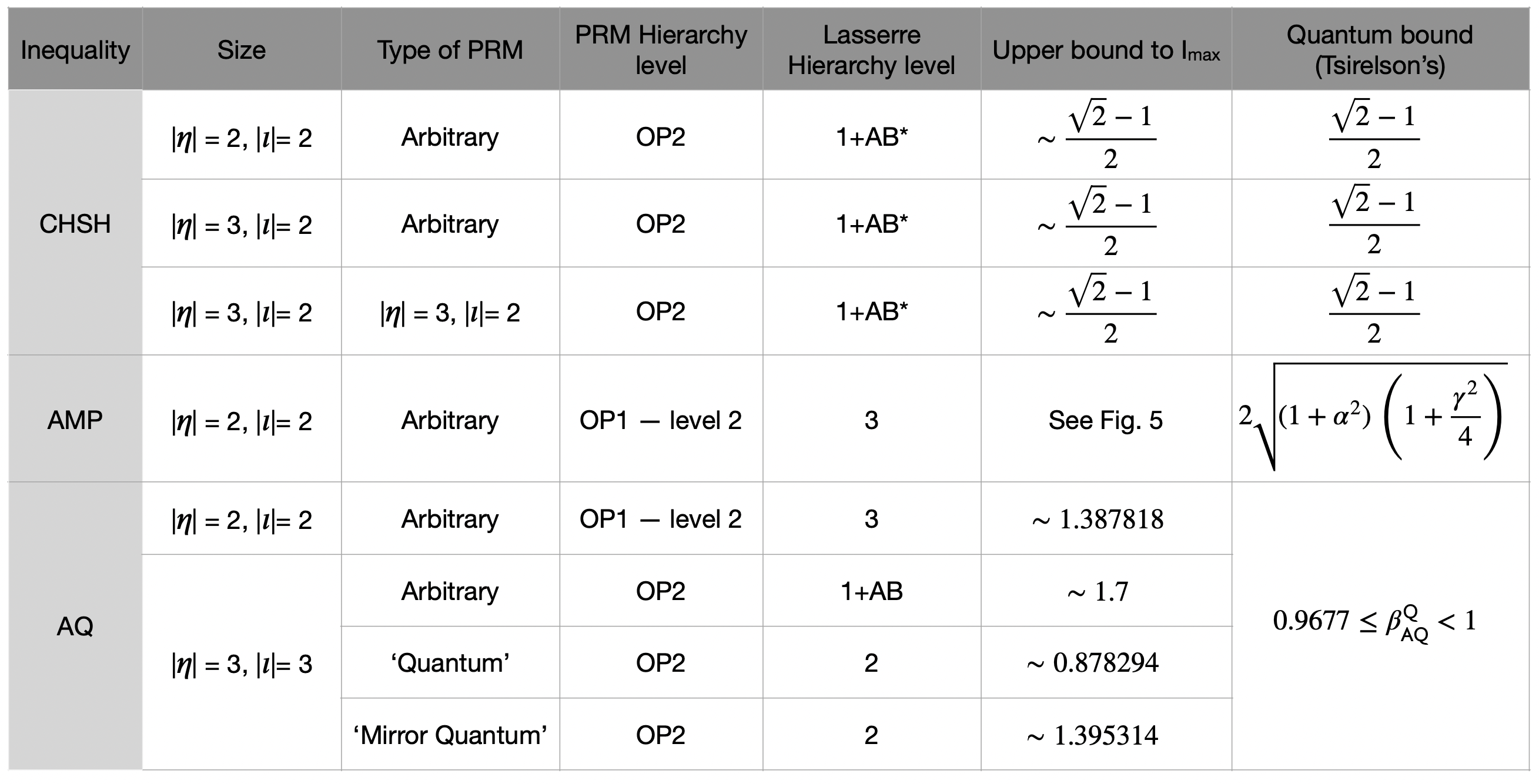}
	\caption{Bounds for Bell inequalities from parity reading measurement. ``Arbitrary PRM'' means that we do not restrict it in any way. In particular, for $|\iSet|=3$ it means that the PRM has 8 outcomes.} 
	\label{fig:table-Bell}
\end{figure}

\subsection{Necessity of local tomography}
{In this section we show that, if we give up on the assumption of local tomography in Optimization Problem \ref{op:rel},
then Tsirelson's bound is violated. Moreover, it is violated in an extreme way, namely, that PR-box correlations can be achieved.}
Recall that the PR-box is a nosignaling box that reaches the nosignaling bound, {that is, the algebraic maximum,} for the CHSH inequality (i.e., in our notation of Eq.~\eqref{eq:chsh} it {achieves the value of} $1/2$). The PR box {can be defined by the fact that it} exhibits perfect correlations for $XX$, $XZ$, and $ZX$ {observables,} and perfect anticorrelations for $ZZ$. 

We now assume that local tomography does not hold, and, {in particular, that} 
the states are described by one extra {non local, ``holistic'',} parameter, {which we denote by} $w_{NL}$. {Our parameterisation of the state, that is, the equivalent of Eq.~\eqref{eq:LTparam-ex1}, now} takes the form:
\begin{align}
    \label{eq:LTparam-f-2}
    \mathbf{p}_s &= (\mathbf{p}_s^{LT},w_{NL}) \nonumber \\
    &=
    \left( p_s(00|00)\,,\, p_s(00|10)\,,\, p^\mathrm{(2)}_s(0|0) \,,\, p_s(00|01)\,,\, p_s(00|11)\,,\,  p^\mathrm{(2)}_s(0|1)\,, \right. \nonumber \\ &\ \hspace{8cm}\left. \, p^\mathrm{(1)}_s(0|0)\,,\, p^\mathrm{(1)}_s(0|1)\,,\, 1 ,w_{NL}\right)\,.
\end{align}
{The parameter $w_{NL}$ is described as a holistic degree of freedom, as products of local observables are independent of its value. In general, however, a PRM will not be simply a product of local observables, and hence, it is possible that it will indeed depend on this holistic parameter.}

{For the remaining of this section it is more convenient to use a more compact matrix notation for bipartite states and effects, {given by:}}
\begin{equation}
    \label{eq:LTparam-f-nl}
    \mathbf{p_s}=\left[ \begin{array}{ccc|c}
p_s(00|00)&p_s(00|01)   & p_s^{(1)}(0|0) &  \\
p_s(00|10)& p_s(00|11)  & p_s^{(1)}(0|1) &  \\
p_s^{(2)}(0|0)& p_s^{(2)}(0|1)  & 1 & 
\\
\hline
&   &  &   w_{NL} 
    \end{array}
    \right].
\end{equation}
In this notation, {the state which realises a PR box}
looks as follows
\begin{equation}
    \label{eq:LTparam-f}
    \mathbf{p}_s^{\mathrm{PR}}=\left[ \begin{array}{ccc|c}
\frac12&\frac12   & \frac12 &  \\
\frac12& 0  & \frac12 &  \\
\frac12& \frac12  & 1 & 
\\
\hline
&   &  &   w_{NL}^{\mathrm{PR}}
    \end{array}
    \right],
\end{equation}
{where $w_{NL}^{\mathrm{PR}}$ can be an arbitrary value.}
By definition of a PRM, the sums  
$\mathcal{P}^{00}+  \mathcal{P}^{01}$
and 
$\mathcal{P}^{00}+  \mathcal{P}^{10}$
depend only on local parameters {-- hence, their} {holistic} parameter is zero and we get 
\begin{equation}
    \mathcal{P}^{00}+  \mathcal{P}^{01}=\left[ \begin{array}{ccc|c}
2&0   & -1 &  \\
0& 0  & 0 &  \\
-1& 0  & 1 & 
\\
\hline
&   &  &   0
    \end{array}
    \right]\equiv \mathcal{R}^0,\quad
    \mathcal{P}^{00}+  \mathcal{P}^{10}=\left[ \begin{array}{ccc|c}
0&0   & 0 &  \\
0& 2  & -1 &  \\
0& -1  & 1 & 
\\
\hline
&   &  &   0
    \end{array}
    \right]\equiv\mathcal{R}^1.
\end{equation}
Preserving probability by PRM reads as 
\begin{align}
    \mathcal{P}^{00}+  \mathcal{P}^{01} +
    \mathcal{P}^{10}+  \mathcal{P}^{11}=
    \left[ \begin{array}{ccc|c}
0&0   & 0 &  \\
0& 0  & 0 &  \\
0& 0  & 1 & 
\\
\hline
&   &  &   0
    \end{array}
    \right]\equiv \Id \,.
\end{align}
Using this, we can write {the free parameters for our optimisation problem as follows:}
\begin{equation}
    \mathcal{P}^{00}-  \mathcal{P}^{01} -
    \mathcal{P}^{10}+  \mathcal{P}^{11}
    =(\mathcal{C}_{LT},c_{NL})=
    \left[ \begin{array}{ccc|c}
c_{11}&c_{12}   & c_{13} &  \\
c_{21}& c_{22}  & c_{23} &  \\
c_{31}& c_{32}  & c_{33} & 
\\
\hline
&   &  &   c_{NL}
    \end{array}
    \right] {\equiv \mathcal{C}.}
\end{equation}
{Here,} $c_{NL}$ {corresponds to} the holistic parameter.

We can then express our parity reading effects in terms of these matrices $\mathcal{R}^0$, $\mathcal{R}^1$ and $\Id$ and the free parameters $\mathcal{C}$, as:
\begin{eqnarray}
\label{eq:PRM-effects-RC}
&&\mathcal{P}^{00}=
    \frac14\left(
    2 \mathcal{R}^0+
    2 \mathcal{R}^1 
    +\mathcal{C}- \Id \right), \nonumber \\
&&\mathcal{P}^{01}=
    \frac14\left(
    2 \mathcal{R}^0-
    2 \mathcal{R}^1 
    -\mathcal{C}
    + \Id \right),
    \nonumber \\
&&\mathcal{P}^{10}=
    \frac14\left(
    -2 \mathcal{R}^0+
    2 \mathcal{R}^1 
    -\mathcal{C}
    + \Id \right),
    \nonumber \\
&&\mathcal{P}^{11}=
    \frac14\left(
    -2 \mathcal{R}^0-
    2 \mathcal{R}^1 
    +\mathcal{C}
    + 3 \Id \right).
\end{eqnarray}
In particular, the nonlocal parameter for PRM effects amounts to
\begin{equation}
\label{eq:nlPRM}
    \mathcal{P}^{00}_{NL}=\mathcal{P}^{11}_{NL}=\frac{c_{NL}}{4},\quad
    \mathcal{P}^{01}_{NL}=\mathcal{P}^{10}_{NL}=-\frac{c_{NL}}{4}.
\end{equation}
Let us now write the requirement of PRM effects to be positive on the {PR-box state.}
{First, notice} that 
\begin{equation}
    \mathcal{R}^0 \cdot \mathbf{p}_s^{\mathrm{PR}}= 
    1, \quad \mathcal{R}^1 \cdot \mathbf{p}_s^{\mathrm{PR}}= 
    0\,,
\end{equation}
(as it should be, since PR box has perfect XX correlations and perfect ZZ anticorrelations).
We thus get 
\begin{eqnarray}
&&\mathcal{P}^{00}\cdot
\mathbf{p}_s^{PR}=
    \frac14\left(1  
    +\mathcal{C}\cdot \mathbf{p}_s^{PR} \right), \nonumber \\
&&\mathcal{P}^{01}\cdot
\mathbf{p}_s^{PR}=
    \frac14\left(3  
    -\mathcal{C}\cdot \mathbf{p}_s^{PR} \right), \nonumber \\
&&\mathcal{P}^{10}\cdot
\mathbf{p}_s^{PR}=
    \frac14\left(-1  
    -\mathcal{C}\cdot \mathbf{p}_s^{PR} \right), \nonumber \\
&&\mathcal{P}^{11}\cdot
\mathbf{p}_s^{PR}=
    \frac14\left(1  
    +\mathcal{C}\cdot \mathbf{p}_s^{PR} \right).
\end{eqnarray}
We see that the positivity of a PRM effect on the PR-box is equivalent to the following condition:
\begin{equation}
\mathcal{C}\cdot \mathbf{p}_s^{PR} = -1.
\end{equation}
Let us now {explore the conditions that follow from} products of steered states. 
The form of the unnormalized steered state is given by Eq.~\eqref{eq:steered-ex1}. 
Thus, the normalized ones
arising from {the PR-box state} (for each party) are given by
\begin{equation}
    s_1=(0,0,1)^T, \quad
    s_2=(0,1,1)^T, \quad
    s_3=(1,0,1)^T, \quad
    s_4=(1,1,1)^T.
\end{equation}
We see that these {define the vertices of} the so-called square bit, as can be seen in 
 Fig.~\ref{fig:square-bit}.
\begin{figure}
    \centering
    \includegraphics[width=0.4\linewidth]{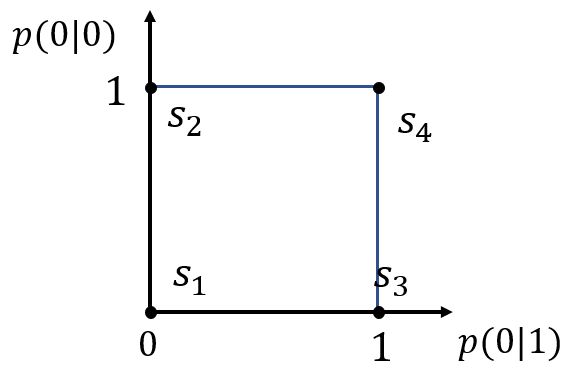}
    \caption{Steered states coming from PR box. {Here we are drawing the hyperplane of normalised vectors defined by $(0,0,1)\cdot v = 1$.}}
    \label{fig:square-bit}
\end{figure}
The {tomographically local degrees of freedom for the} products of steered states are $\mathbf{p}_{s,LT}^{ij}=s^{(1)}_i \otimes s^{(2)}_j$, $i,j=1,\ldots 4$. 
We then denote:
\begin{equation}
    \mathbf{p}_s^{ij} =
    (\mathbf{p}_{s,LT}^{ij},w_{NL}^{ij})
    =(s_i^{(1)}\otimes s_j^{(2)},w_{NL}^{ij}).
\end{equation}
Note that, here, the  bracket does not mean scalar product, {but rather indicates} two groups of parameters: the group of locally tomographic ones, and the group consisting of one nonlocal parameter. 
For steered states, the nonlocal parameter $w_{NL}^{ij}$ must be a linear combination of the local parameters:
\begin{equation}
 w_{NL}^{ij}= \mathbf{h}\cdot    
 \mathbf{p}_{s,LT}^{ij}.
\end{equation}
{This {follows from} noting that: }
{
\begin{compactitem}
\item[i)] the way that local states are combined to give product states must be given by a bilinear function from the local vector spaces into the global vector space; 
\item[ii)] the universal property of the tensor product means that this can be written as a linear function from the tensor product space of the local vector spaces into the global vector space;
\item[iii)] the local parameters are simply the tensor product space;
\item[iv)] this means that the {value of the} non-local parameter  is  {given by a} linear functional on the local parameters (i.e., a linear map from the local vector spaces into the reals); 
\item[v)] finally, the Riesz representation theorem means that we can write this as the dot product with some vector $\mathbf{h}$ in the local parameter space.
\end{compactitem}
}  
We thus have 
\begin{equation}
    \mathcal{P}^{qr}  \cdot
    \mathbf{p}_s^{ij}  = 
    \mathcal{P}^{qr}_{LT}  \cdot \mathbf{p}_{s,LT}^{ij}
    + \mathcal{P}^{qr}_{NL}\, \mathbf{p}_{s,NL}^{ij}=
    (\mathcal{P}^{qr}_{LT}  
    +\mathcal{P}^{qr}_{NL}\, \mathbf{h})\cdot \mathbf{p}_{s,LT}^{ij}\,,
\end{equation}
where we recall, that 
$\mathcal{P}^{qr}_{NL}$ are numbers (the values of the nonlocal parameter for PRM effects) given by Eq.~\eqref{eq:nlPRM}.
Using the above equation {together with} Eq.~\eqref{eq:PRM-effects-RC}, 
we obtain 
\begin{eqnarray}
&&\mathcal{P}^{00}\cdot \mathbf{p}_s^{ij}=
\frac12(\mathcal{R}^0+ \mathcal{R}^1) \mathbf{p}_s^{ij}
+\frac14(-1+\mathbf{g}\cdot \mathbf{p}_s^{ij})\nonumber \\
&&\mathcal{P}^{01}\cdot \mathbf{p}_s^{ij}=
\frac12(\mathcal{R}^0-\mathcal{R}^1) \mathbf{p}_s^{ij}
+\frac14(1-\mathbf{g}\cdot \mathbf{p}_s^{ij})\nonumber \\
&&\mathcal{P}^{10}\cdot \mathbf{p}_s^{ij}=
\frac12(-\mathcal{R}^0+ \mathcal{R}^1) \mathbf{p}_s^{ij}
+\frac14(1-\mathbf{g}\cdot \mathbf{p}_s^{ij})\nonumber \\
&&\mathcal{P}^{11}\cdot \mathbf{p}_s^{ij}=
\frac12(-\mathcal{R}^0- \mathcal{R}^1) \mathbf{p}_s^{ij}
+\frac14(3+\mathbf{g}\cdot \mathbf{p}_s^{ij})\,,
\end{eqnarray}
where we have denoted \begin{equation}
    \mathbf{g}=\mathcal{C}_{LT} +c_{NL} \mathbf{h}\,,
\end{equation}
with $\mathcal{C}_{LT}$  
being {the locally tomographic} part of $\mathcal C$, 
and $c_{NL}$ {the holistic} 
part of $\mathcal{C}$.
Now, the positivity of PRM effects on products of steered states means that we require all four terms to be positive 
for all $i,j=1,\ldots,4$.
By using Mathematica \cite{MathematicaSoft} we find that positivity is satisfied for only one choice of $\mathbf{g}$:
\begin{equation}
\mathbf{g}=\left[
    \begin{array}{ccc}
4 & 0 & -2 \\
0 &  4& -2 \\
-2 & -2 & 4
    \end{array}
    \right]\,.
\end{equation}
{To summarise, positivity conditions of the PRM on the PR-box states and products of its steered states reduce to:}
\begin{eqnarray}
&&\text{positivity on PR box state:}  \quad \mathcal{C}_{LT}\cdot \mathbf{p}_{s,LT}^{PR}+ c_{NL} w_{NL}^{PR}= -1 \\
&&\text{positivity on steered states: }\quad  \mathcal{C}_{LT} + c_{NL} \mathbf{h}= 
\left[
    \begin{array}{ccc}
4 & 0 & -2 \\
0 &  4& -2 \\
-2 & -2 & 4
    \end{array}
    \right].
\end{eqnarray}
To {prove our original claim, the idea is to choose values for}  $\mathcal{C}_{LT}, c_{NL}, \mathbf{h}$ and $w^{PR}_{NL}$ such that the above two constraints hold. 
{Our choice is the following:}
\begin{equation}
    \mathcal{C}_{LT}=\mathbf{0},\quad c_{NL}=1, \quad w_{NL}^{PR}=-1, \quad \mathbf{h}=
\left[
    \begin{array}{ccc}
4 & 0 & -2 \\
0 &  4& -2 \\
-2 & -2 & 4
    \end{array}
    \right].
\end{equation}
{These values for $\mathcal{C}_{LT}$ and $c_{NL}$ fix the PRM to take the form:}

\begin{eqnarray}
   &&\mathcal{P}^{00}=\frac14\left[ \begin{array}{ccc|c}
4&0   & -2 &  \\
0& 4  & -2 &  \\
-2& -2  & -1 & 
\\
\hline
&   &  &  1
    \end{array}
    \right],\quad
    \mathcal{P}^{01}  =\left[ \begin{array}{ccc|c}
4&0   & -2 &  \\
0& -4  & 2 &  \\
-2& 2  & 1 & 
\\
\hline
&   &  &   -1
    \end{array}
    \right]  \nonumber \\
&&    \mathcal{P}^{10}=\frac14\left[ \begin{array}{ccc|c}
-4&0   & 2 &  \\
0& 4  & -2 &  \\
2& -2  & 1 & 
\\
\hline
&   &  &  -1
    \end{array}
    \right],\quad
    \mathcal{P}^{11}  =\left[ \begin{array}{ccc|c}
-4&0   & 2 &  \\
0& -4  & 2 &  \\
2& 2  & 3 & 
\\
\hline
&   &  &   1
    \end{array}
    \right]. 
\end{eqnarray}
{In addition, our choice of $\mathbf{h}$ defines the value of the nonlocal parameter for steered states to be}
\begin{equation}
    w^{ij}_{NL}={\mathbf{h}\cdot     \mathbf{p}_{s,LT}^{ij}}.
\end{equation}
{We see then that the PR-box state is consistent with the existence of a PRM that satisfies the constraints of OP\ref{op:rel}. Since performing fiducial measurements on a PR-box state yields PR-box correlations, this shows that $\mathcal{I}^\mathrm{R}_{\max} = \frac{1}{2}$ for the CHSH inequality, as per Eq.~\eqref{eq:chsh}. With this we conclude the proof of our claim.}

\bigskip

{Let us make a final comment on an interesting interpretation for the values of the nonlocal parameter for the pairs of steered states:}
 they count the number of correlations. If both observables have the same value for a given pair of steered states (which happens when Alice and Bob's steered states are the same) then the parameter takes the value $2$. When 
only one of the observables has the same value, then it takes the value $1$, and when both observables have the opposite value, then it takes the value $0$. This is depicted in Fig.~\ref{fig:pairs-of-steered}.

\begin{figure}
	\centering
\begin{tikzpicture}[scale = 5]

    \draw [->, color=gray] (0,0) -- (0,1.5) node (zaxis) [above] {$p(0|0)$};
    \draw [->, color=gray] (0,0) -- (1.5,0) node (xaxis) [right] {$p(0|1)$};


\node at (1.33,1.15) {\textcolor{carmine}{2}};
\node at (1.1,-0.1) {1};
\node at (-0.1,1.1) {1};
\node at (-0.1,-0.1) {0};

\node[draw,fill, color=black, scale=.3, shape=circle] (s1) at (0,0) {};
\node at (0.2,0.07) {$s_1$};

\node[draw,fill, color=black, scale=.3, shape=circle] (s2) at (0,1) {};
\node at (0.2,0.93) {$s_2$};

\node[draw,fill, color=black, scale=.3, shape=circle] (s3) at (1,0) {};
\node at (0.8,0.07) {$s_3$};

\node[draw,fill, color=black, scale=.3, shape=circle] (s4) at (1,1) {};
\node at (0.8,0.93) {$s_4$};

\draw[color=gray] (s2) -- (s4);
\draw[color=gray] (s3) -- (s4);


\draw[thick, dotted, color=darkgreen] (0.03,0.97) -- (0.97,0.03);
\draw[thick, dotted, color=darkgreen] (0.97,0.97) -- (0.03,0.03);
\node at (0.7,0.5) {\textcolor{darkgreen}{0}};


\draw[thick, dashed, color=panblue] (-0.03,0.03) [out=110, in=250] to (-0.03,0.97);
\draw[thick, dashed, color=panblue] (0.03,1.03) [out=20, in=160] to (0.97,1.03);
\draw[thick, dashed, color=panblue] (1.03,0.97) [out=-70, in=70] to (1.03,0.03);
\draw[thick, dashed, color=panblue] (0.97,-0.03) [out=200, in=-20] to (0.03,-0.03);

\node at (0.5,1.2) {\textcolor{panblue}{1}};

\draw[thick, dashdotdotted, color=carmine,rotate=0, x radius=0.15cm, y radius=0.15cm, delta angle=330] (-0.04,0.01) arc [start angle=70];
\draw[thick, dashdotdotted, color=carmine,rotate=0, x radius=0.15cm, y radius=0.15cm, delta angle=330] (0.02,1.05) arc [start angle=-25];
\draw[thick, dashdotdotted, color=carmine,rotate=0, x radius=0.15cm, y radius=0.15cm, delta angle=330] (1.05,0.99) arc [start angle=250];
\draw[thick, dashdotdotted, color=carmine,rotate=0, x radius=0.15cm, y radius=0.15cm, delta angle=330] (1.01,-0.04) arc [start angle=160];
\end{tikzpicture}
	\caption{The value of the nonlocal parameter for the pairs of steered states. Each  dashdotted-red loop denotes a  pair of the same steered states. Each  dotted-green or  dashed-blue line means two pairs depending on which state goes to Alice and which to Bob. This gives rise to 16 pairs of steered states.  Dashdotted-red pairs have the same value for both observables  and dashed-blue pairs have the same value for one observable,  whilst for dotted-green pairs both observables have opposite values.} 
	\label{fig:pairs-of-steered}
\end{figure}
\section{Discussion}

In this paper we have shown that postulating within a theory the existence of particular bipartite measurements has a surprisingly rich set of consequences for the structure of the theory itself. 
Indeed, we showed that this leads to an infinite hierarchy of constraints on the possible bipartite states. 
These conditions translate analogously  into constraints on the statistical correlations allowed by the theory. In other words, the maximum violation of any Bell inequality by the correlations  among fiducial measurements  featured by the theory will be subjected to an infinite hierarchy of constraints. 

We further explored the consequences of this rich structure for the particular case where there exists a bipartite measurement that can read out the parity of local fiducial measurements. For the case of tomographically-local GPTs, we found that these constraints on the structure of bipartite sates are enough to recover  (up to numerical precision) Tsirelson's bound for various inequalities in the CHSH scenario. {In addition, we also showed that non-tomographically local GPTs may still reach the maximum algebraic violation of such inequalities (i.e., go beyond Tsirelson's bound) when only the first levels of the hierarchy of constraints are considered.}  We also noticed that, for inequalities where the maximum quantum violation is not achieved by measuring complementary observables, our technique may also yield values below Tsirelson's bound. 

Our  initial  numerical results led us to formulate a conjecture on the constraints that the existence of a Parity Reading Measurement may yield for tomographically local GPTs: 
\setcounter{conj}{0}
\begin{conj}\label{conj:1}
{\it Under {the assumption of local tomography,}
the local observables that are {\bf parity-readable} satisfy Tsirelson's bound, i.e., they cannot violate Bell inequalities better than quantum mechanics does (with arbitrary measurements).} 
\end{conj}
It is worth mentioning that, after formulating the conjecture, we ran further numerics in other scenarios (all presented in this manuscript) which did not disprove the conjecture. 


From looking at Conjecture \ref{conj:1} one can take a step back and further conjecture that quantum theory, among the landscape of GPTs that are locally tomographic, is the theory that displays the necessary balance between its allowed states and effects to feature the following property:
\begin{conj}\label{conj:2}
{\it Quantum Theory yields the largest violation of any Bell inequality by parity-readable fiducial measurements, within the landscape of possible locally-tomographic physical theories.} 
\end{conj}
\noindent Notice that in this conjecture we are comparing correlations obtained from \emph{parity-readable} measurements in quantum theory vs.~in other more generic (yet locally-tomographic) GPTs, and state that quantum theory will always produce correlations that are more non-classical.  This is in contrast to Conjecture \ref{conj:1} which compared correlations obtained from \emph{arbitrary} measurements in quantum theory vs.~those that are achieved with parity-readable measurements in an arbtirary tomographically-local GPT. The significance of this is that not all quantum correlations can be achieved with parity-readable measurements alone.  However, if measurements beyond parity-readable ones may be used, then it is possible that other GPTs beyond quantum (e.g., Boxworld) can generate correlations that are more non-classical than any that quantum theory may produce. 
Now, whether Conjecture \ref{conj:2} is true, how to formally express it, and what its consequences are, comprise a topic for future work.

Going beyond the CHSH scenario or GPTs with affine local dimension $\geq 3$ is a computationally demanding task. Indeed, the complexity of the optimisation problems to be solved rises considerably with the number of settings and dimension. A complete understanding of the reach of the constraints imposed by parity reading measurements require the further development of analytical and numerical techniques, which are deferred to future work.

Moving forward, one may apply our technique to explore the constraints that entangled measurements beyond parity reading ones may impose. Indeed, Optimisation Problems \ref{op:rel} and \ref{op:relQuant} may be straightforwardly adapted to study other bipartite measurements. It would be interesting to see if there is a relation between the properties of bipartite entangled measurements and those of the Bell inequalities whose Tsirelson's bound they recover. 

More ambitiously, there is the natural question of multi-partite entangled measurements. Would the structure they impose on multi-partite state spaces have special features that we cannot envision from the phenomenology at the bipartite level? We hope such explorations will bring new insight into the structure of states and effect spaces in GPTs, and their non-classical properties.

\section*{Acknowledgements} 
JHS, ABS, and MH acknowledge support by the Foundation for Polish Science (IRAP project, ICTQT, contract no.2018/MAB/5, co-financed by EU within Smart Growth Operational Programme).
Victor Magron was supported by the EPOQCS grant funded by the LabEx CIMI (ANR-11-LABX-0040), the FastQI grant funded by the Institut Quantique Occitan, the PHC Proteus grant 46195TA, the European Union's Horizon 2020 research and innovation programme under the Marie Sk{\l}odowska-Curie Actions, grant agreement 813211 (POEMA), by the AI Interdisciplinary Institute ANITI funding, through the French ``Investing for the Future PIA3'' program under the Grant agreement n${}^\circ$ ANR-19-PI3A-0004 as well as by the National Research Foundation, Prime Minister's Office, Singapore under its Campus for Research Excellence and Technological Enterprise (CREATE) programme.

\bibliographystyle{quantum}
\bibliography{bibliography}

\appendix

\newpage
\section{Results for  AMP inequalities}
\label{sec:AMP-table}

In Table \eqref{table:alphagamma} 
we present the values we chose for the pairs of parameters $(\alpha,\gamma)$, used to define AMP inequalities that we examine in our numerical optimization. Here, the variable $x$ corresponds to the value of the $x$-axis in Fig.~\ref{fig:AMP-Table}.

 \beq \label{table:alphagamma}
 \small
 \begin{array}{l|ll}
x & \alpha & \gamma \\
 \hline
 1&1&	0	 \\
 2&1&	0,4	\\
 3&1&	0,8	\\
 4&1&	1,2	\\
 5&1&	1,6	\\
 6&1&	2	\\
 7&3&	0	\\
 8&3&	0,133333333	\\
 9&3&	0,266666667	\\
 10&3&	0,4	\\
 11&3&	0,533333333	\\
12&3&	0,666666667
\end{array}
\hspace{2cm}
 \begin{array}{l|ll}
 13&5&	0	\\
 14&5&	0,08	\\
15 &5&	0,16	\\
16 &5&  0,24	\\
17 &5&	0,32	\\
 18&5&	0,4	\\
19 &7&	0 \\
20 &7&	0,057142857	\\
 21&7&	0,114285714	\\
 22&7&	0,171428571	\\
23 &7&	0,228571429	\\
24 &7&	0,285714286	
\end{array}
\hspace{2cm}
 \begin{array}{l|ll}
 25 &9&	0\\
 26&9&	0,04444444	\\
 27&9&	0,088888889	\\
28 &9&	0,133333333	\\
29 &9&	0,177777778	\\
30 &9&	0,222222222	\\
31 &11&	0	\\
32 &11&	0,036363636	\\
33 &11&	0,072727273	\\
34 &11&	0,109090909	\\
35 &11&	0,145454545	\\
36 &11&	0,181818182
 \end{array}
\eeq

\section{Introduction our GPT formalism}\label{App:GPT}

In this paper we take a categorical approach to GPTs, which we summarise in this Section. The reader unfamiliar with the mathematics of category theory is referred to Ref.~\cite{coecke2010categories} for a physicist friendly introduction to the topic, and to Ref.~\cite{mac2013categories} as the classic mathematics textbook on the subject.

The particular formalism that we use here is based on the observation that any tomographically local GPT can be thought of as a particular symmetric monoidal subcategory of the symmetric monoidal category $\mathbf{Vect}_\mathds{R}$ of real vector spaces and linear maps (see, e.g., Ref.~\cite{schmid2020structure}).
In particular, such a subcategory has the following properties:
 \begin{enumerate}
 \itemsep -0.4em
    \item objects are finite dimensional;
   \item the scalars are the unit interval;
   \item the hom-sets\footnote{{A hom-set is the set of transformations from one object to another, in this case, this will be a subset of the set of linear maps from one vector space to another.}} are closed under convex combinations;
   \item the points (and copoints) for an object span the vector space (resp. dual vector space);
   \item there is a unique `deterministic' copoint, $u_V$, for each object $V$. This is defined by the constraint that for any other copoint $e$ for the object $V$ there exists a copoint $e^\perp$ such that $e+e^\perp = u_V$.
 \end{enumerate}

 One of the benefits of this categorical approach to generalised probabilistic theories is that there is a faithful diagrammatic representation using string diagrams, which we will use throughout the paper. This diagrammatic notation moreover immediately suggests the correct interpretation of the abstract categorical definition given above. For example, consider the following diagram (read bottom to top):

\begin{equation}
\begin{tikzpicture}
	\begin{pgfonlayer}{nodelayer}
		\node [style=none] (0) at (-1, 0) {};
		\node [style=none] (1) at (0, -1.25) {};
		\node [style=none] (2) at (1, 0) {};
		\node [style=none] (3) at (-0.5, 0) {};
		\node [style=none] (4) at (0.5, 0) {};
		\node [style=none] (5) at (0, 2) {};
		\node [style=none] (6) at (0, 1) {};
		\node [style=none] (7) at (2, 1) {};
		\node [style=none] (8) at (2, 2) {};
		\node [style=none] (9) at (0.5, 1) {};
		\node [style=none] (10) at (1.5, 1) {};
		\node [style=none] (11) at (1.5, -2) {};
		\node [style=none] (12) at (-0.5, 3) {};
		\node [style=none] (13) at (0.5, 2) {};
		\node [style=none] (14) at (0.5, 3) {};
		\node [style=none] (15) at (1, 3) {};
		\node [style=none] (16) at (0.5, 4) {};
		\node [style=none] (17) at (-0.5, 4) {};
		\node [style=none] (18) at (-1, 3) {};
		\node [style=none] (19) at (1.5, 2) {};
		\node [style=copoint] (20) at (1.5, 4.5) {$E$};
		\node [style=none] (21) at (0, 4) {};
		\node [style=none] (22) at (0, 6.25) {};
		\node [style=none] (26) at (0, -0.5) {$S$};
		\node [style=none] (27) at (1, 1.5) {$T_1$};
		\node [style=none] (28) at (0, 3.5) {$T_2$};
		\node [style=right label] (30) at (1.5, -1.5) {$U$};
		\node [style=right label] (31) at (0.5, 0.5) {$V$};
		\node [style=right label] (32) at (-0.5, 0.5) {$V$};
		\node [style=right label] (33) at (0.5, 2.5) {$V$};
		\node [style=right label] (34) at (1.5, 2.5) {$W$};
		\node [style=right label] (35) at (0, 5.75) {$W$};
	\end{pgfonlayer}
	\begin{pgfonlayer}{edgelayer}
		\draw (0.center) to (2.center);
		\draw (2.center) to (1.center);
		\draw (1.center) to (0.center);
		\draw (5.center) to (8.center);
		\draw (8.center) to (7.center);
		\draw (7.center) to (6.center);
		\draw (6.center) to (5.center);
		\draw (18.center) to (15.center);
		\draw (15.center) to (16.center);
		\draw (16.center) to (17.center);
		\draw (17.center) to (18.center);
		\draw [style=qWire] (22.center) to (21.center);
		\draw [style=qWire] (12.center) to (3.center);
		\draw [style=qWire] (13.center) to (14.center);
		\draw [style=qWire] (9.center) to (4.center);
		\draw [style=qWire] (10.center) to (11.center);
		\draw [style=qWire] (19.center) to (20);
	\end{pgfonlayer}
\end{tikzpicture}
\,.
\end{equation}
We view the wires in the above diagram, corresponding to objects (i.e. finite dimensional real vector spaces), as representing physical systems. Then, points, such as, $S:\mathds{R}\to V\otimes V$, represent physical states, general morphisms, such as $T_1:V\otimes U \to V\otimes W$ and $T_2:V\otimes V\to W$ represent physical transformations, and copoints, such as $E:W\to \mathds{R}$, correspond to physical effects. Closed diagrams, such as:
\beq
\begin{tikzpicture}
	\begin{pgfonlayer}{nodelayer}
		\node [style=point] (8) at (0, -1) {$S$};
		\node [style=copoint] (9) at (0, 1) {$E$};
		\node [style={right label}] (10) at (0, -0) {$V$};
	\end{pgfonlayer}
	\begin{pgfonlayer}{edgelayer}
		\draw [style=qWire] (8.center) to (9.center);
	\end{pgfonlayer}
\end{tikzpicture},
\eeq
that is, elements of the unit interval, are interpreted as the probability of observing effect $E$ given the system was prepared in state $S$. We denote the unique deterministic effect as:
\beq
\begin{tikzpicture}
	\begin{pgfonlayer}{nodelayer}
		\node [style=none] (0) at (0, -0.5) {};
		\node [style=upground] (1) at (0, 1) {};
		\node [style={right label}] (2) at (0, -0) {$V$};
	\end{pgfonlayer}
	\begin{pgfonlayer}{edgelayer}
		\draw [style=qWire] (1) to (0.center);
	\end{pgfonlayer}
\end{tikzpicture}\,,
\eeq
which defines the normalised states $S$ as those satisfying:
\beq
\begin{tikzpicture}
	\begin{pgfonlayer}{nodelayer}
		\node [style=none] (3) at (0, -0.5) {};
		\node [style=none] (4) at (0, 0.5) {};
		\node [style=point] (5) at (0, -0.75) {$S$};
		\node [style=right label] (6) at (0, 0) {$V$};
		\node [style=upground] (7) at (0, 0.75) {};
	\end{pgfonlayer}
	\begin{pgfonlayer}{edgelayer}
		\draw [style=qWire] (4.center) to (5);
	\end{pgfonlayer}
\end{tikzpicture}
 =1\,.
\eeq
Note that for a set of effects $\{E_i\}_{i\in I}$ to describe a measurement it must be the case that:
\beq
\sum_{i\in I}
\begin{tikzpicture}
	\begin{pgfonlayer}{nodelayer}
		\node [style=none] (0) at (0, -1.25) {};
		\node [style=copoint] (5) at (0, 0.25) {$E_i$};
		\node [style=right label] (6) at (0, -1) {$V$};
	\end{pgfonlayer}
	\begin{pgfonlayer}{edgelayer}
		\draw [style=qWire] (5) to (0.center);
	\end{pgfonlayer}
\end{tikzpicture}
=
\begin{tikzpicture}
	\begin{pgfonlayer}{nodelayer}
		\node [style=none] (0) at (0, -1.25) {};
		\node [style=upground] (1) at (0, 0.05) {};
		\node [style=right label] (2) at (0, -1) {$V$};
	\end{pgfonlayer}
	\begin{pgfonlayer}{edgelayer}
		\draw [style=qWire] (1) to (0.center);
	\end{pgfonlayer}
\end{tikzpicture}
\,.
\eeq
One can then see that (finite dimensional) quantum theory defines such a GPT by noting that the set of Hermitian operators for some Hilbert space $\mathcal{H}$ forms a real vector space $\mathcal{B(H)}$, and that completely positive trace non-increasing (CPTNI) maps between these spaces are a particular class of linear maps between these vector spaces. The other constraints are simple to verify. Similarly, classical stochastic dynamics can be represented as such a GPT. To see this note that stochastic dynamics from some (finite) set $X$ to another (finite) set $A$ can be represented as a particular class of linear maps from the finite dimensional vector space $\mathds{R}^X$ to the finite dimensional vector space $\mathds{R}^A$.

We will work with the representation of GPTs in which this classical GPT is included as a subtheory. To distinguish it, we will represent the classical systems by thin gray wires, 
and, for convenience, we will simply label them by the finite set $X$, $A$, ..., rather than the vector spaces $\mathds{R}^X$, $\mathds{R}^A$, ... . This is convenient because it allows us to explicitly represent measurement outcomes and setting variables within the diagrammatic representation. For example, a controlled measurement of system $V$ with setting variable $X$ and outcome variable $A$ is denoted as:
\beq
\begin{tikzpicture}
	\begin{pgfonlayer}{nodelayer}
		\node [style={right label}] (0) at (1.25, -1.5) {$V$};
		\node [style=none] (1) at (0.5, -0.5) {};
		\node [style=none] (2) at (1.25, -1.75) {};
		\node [style=none] (3) at (0.5, 0.5) {};
		\node [style=none] (4) at (-0.75, 0.5) {};
		\node [style=none] (5) at (-0.75, -0.5) {};
		\node [style=none] (6) at (0, -0) {$M$};
		\node [style=none] (7) at (-0.25, -0.5) {};
		\node [style=none] (8) at (-0.25, -1.75) {};
		\node [style=none] (9) at (-0.25, 0.5) {};
		\node [style=none] (10) at (-0.25, 1.75) {};
		\node [style={right label}] (11) at (-0.25, -1.5) {$X$};
		\node [style={right label}] (12) at (-0.25, 1.25) {$A$};
		\node [style=none] (13) at (1, -0.5) {};
	\end{pgfonlayer}
	\begin{pgfonlayer}{edgelayer}
		\draw [qWire, in=90, out=-90, looseness=1.00] (1.center) to (2.center);
		\draw (3.center) to (4.center);
		\draw (4.center) to (5.center);
		\draw [cWire, in=90, out=-90, looseness=1.00] (7.center) to (8.center);
		\draw [cWire, in=-90, out=90, looseness=1.00] (9.center) to (10.center);
		\draw (5.center) to (13.center);
		\draw (13.center) to (3.center);
	\end{pgfonlayer}
\end{tikzpicture}\ ,
\eeq
{which must satisfy the constraint:
\beq
\begin{tikzpicture}
	\begin{pgfonlayer}{nodelayer}
		\node [style={right label}] (0) at (1.25, -1.5) {$V$};
		\node [style=none] (1) at (0.5, -0.5) {};
		\node [style=none] (2) at (1.25, -1.75) {};
		\node [style=none] (3) at (0.5, 0.5) {};
		\node [style=none] (4) at (-0.75, 0.5) {};
		\node [style=none] (5) at (-0.75, -0.5) {};
		\node [style=none] (6) at (0, -0) {$M$};
		\node [style=none] (7) at (-0.25, -0.5) {};
		\node [style=none] (8) at (-0.25, -1.75) {};
		\node [style=none] (9) at (-0.25, 0.5) {};
		\node [style=none] (10) at (-0.25, 1.75) {};
		\node [style={right label}] (11) at (-0.25, -1.5) {$X$};
		\node [style={right label}] (12) at (-0.25, 1) {$A$};
		\node [style=none] (13) at (1, -0.5) {};
		\node [style=upground] (14) at (-0.25, 2) {};
	\end{pgfonlayer}
	\begin{pgfonlayer}{edgelayer}
		\draw [qWire, in=90, out=-90, looseness=1.00] (1.center) to (2.center);
		\draw (3.center) to (4.center);
		\draw (4.center) to (5.center);
		\draw [cWire, in=90, out=-90, looseness=1.00] (7.center) to (8.center);
		\draw [cWire, in=-90, out=90, looseness=1.00] (9.center) to (10.center);
		\draw (5.center) to (13.center);
		\draw (13.center) to (3.center);
	\end{pgfonlayer}
\end{tikzpicture}\quad=\quad \begin{tikzpicture}
	\begin{pgfonlayer}{nodelayer}
		\node [style={right label}] (0) at (0.75, -0.75) {$V$};
		\node [style=none] (1) at (0.75, 0.25) {};
		\node [style=none] (2) at (0.75, -1) {};
		\node [style=none] (3) at (-0.75, 0.25) {};
		\node [style=none] (4) at (-0.75, -1) {};
		\node [style={right label}] (5) at (-0.75, -0.75) {$X$};
		\node [style=upground] (6) at (-0.75, 0.5) {};
		\node [style=upground] (7) at (0.75, 0.5) {};
	\end{pgfonlayer}
	\begin{pgfonlayer}{edgelayer}
		\draw [qWire, in=90, out=-90, looseness=1.00] (1.center) to (2.center);
		\draw [cWire, in=90, out=-90, looseness=1.00] (3.center) to (4.center);
	\end{pgfonlayer}
\end{tikzpicture}\ .
\eeq}
The situation where we perform this measurement $M$ on the system $V$ prepared in some normalisted state $S$ is denoted by:
\beq
\begin{tikzpicture}
	\begin{pgfonlayer}{nodelayer}
		\node [style={right label}] (0) at (1, -1.25) {$V$};
		\node [style=none] (1) at (0.5, -0.5) {};
		\node [style=none] (2) at (1.25, -1.75) {};
		\node [style=none] (3) at (0.5, 0.5) {};
		\node [style=none] (4) at (-0.75, 0.5) {};
		\node [style=none] (5) at (-0.75, -0.5) {};
		\node [style=none] (6) at (0, -0) {$M$};
		\node [style=none] (7) at (-0.25, -0.5) {};
		\node [style=none] (8) at (-0.25, -3) {};
		\node [style=none] (9) at (-0.25, 0.5) {};
		\node [style=none] (10) at (-0.25, 1.75) {};
		\node [style={right label}] (11) at (-0.25, -2.75) {$X$};
		\node [style={right label}] (12) at (-0.25, 1.25) {$A$};
		\node [style=none] (13) at (1, -0.5) {};
		\node [style=point] (14) at (1.25, -2) {$S$};
	\end{pgfonlayer}
	\begin{pgfonlayer}{edgelayer}
		\draw [qWire, in=90, out=-90, looseness=1.00] (1.center) to (2.center);
		\draw (3.center) to (4.center);
		\draw (4.center) to (5.center);
		\draw [cWire, in=90, out=-90, looseness=1.00] (7.center) to (8.center);
		\draw [cWire, in=-90, out=90, looseness=1.00] (9.center) to (10.center);
		\draw (5.center) to (13.center);
		\draw (13.center) to (3.center);
	\end{pgfonlayer}
\end{tikzpicture}\ ,
\eeq
and is simply a stochastic map from the setting variable $X$ to the outcome variable $A$. The probabilities of obtaining a particular outcome $a\in A$ given a setting $x\in X$ can be extracted from this map via:
\beq\begin{tikzpicture}
	\begin{pgfonlayer}{nodelayer}
		\node [style={right label}] (0) at (1, -1.25) {$V$};
		\node [style=none] (1) at (0.5, -0.5) {};
		\node [style=none] (2) at (1.25, -1.75) {};
		\node [style=none] (3) at (0.5, 0.5) {};
		\node [style=none] (4) at (-0.75, 0.5) {};
		\node [style=none] (5) at (-0.75, -0.5) {};
		\node [style=none] (6) at (0, -0) {$M$};
		\node [style=none] (7) at (-0.25, -0.5) {};
		\node [style=none] (8) at (-0.25, -3) {};
		\node [style=none] (9) at (-0.25, 0.5) {};
		\node [style=none] (10) at (-0.25, 1.75) {};
		\node [style={right label}] (11) at (-0.25, -2.5) {$X$};
		\node [style={right label}] (12) at (-0.25, 1) {$A$};
		\node [style=none] (13) at (1, -0.5) {};
		\node [style=point] (14) at (1.25, -2) {$S$};
		\node [style=point] (15) at (-0.25, -3.25) {$x$};
		\node [style=copoint] (16) at (-0.25, 2) {$a$};
	\end{pgfonlayer}
	\begin{pgfonlayer}{edgelayer}
		\draw [qWire, in=90, out=-90, looseness=1.00] (1.center) to (2.center);
		\draw (3.center) to (4.center);
		\draw (4.center) to (5.center);
		\draw [cWire, in=90, out=-90, looseness=1.00] (7.center) to (8.center);
		\draw [cWire, in=-90, out=90, looseness=1.00] (9.center) to (10.center);
		\draw (5.center) to (13.center);
		\draw (13.center) to (3.center);
	\end{pgfonlayer}
\end{tikzpicture}
\ = \ p_S(a|x) \ .
\eeq

We will also find it useful to use certain processes which live in $\mathbf{Vect}_{\mathds{R}}$ but which are not part of the subtheory describing the GPT. To visually distinguish these `non-physical' processes we draw them as shaded objects:
\beq
\begin{tikzpicture}
	\begin{pgfonlayer}{nodelayer}
		\node [style=none] (0) at (0, -0.5) {};
		\node [style=none] (1) at (0, -1.5) {};
		\node [style=none] (2) at (0, 0.5) {};
		\node [style=none] (3) at (0, 1.5) {};
		\node [style=small box, fill=black!70] (4) at (0, -0) {$\color{white}L$};
		\node [style={right label}] (5) at (0, -1.25) {$U$};
		\node [style={right label}] (6) at (0, 1) {$V$};
	\end{pgfonlayer}
	\begin{pgfonlayer}{edgelayer}
		\draw [style=qWire] (2.center) to (3.center);
		\draw [style=qWire] (0.center) to (1.center);
	\end{pgfonlayer}
\end{tikzpicture}\,.
\eeq
Finally, we will define a particular type of linear functionals $I$. The objects these act on are linear maps from one vector space $U$ to a vector space $V$. We diagrammatically denote them as:
\beq
\begin{tikzpicture}
	\begin{pgfonlayer}{nodelayer}
		\node [style=none] (0) at (0, -0.5) {};
		\node [style=none] (1) at (0, -1.5) {};
		\node [style=none] (2) at (0, 0.5) {};
		\node [style=none] (3) at (0, 1.5) {};
		\node [style=none] (4) at (1.5, -0) {$\color{white}I$};
		\node [style={right label}] (5) at (0, -1) {$U$};
		\node [style={right label}] (6) at (0, 1) {$V$};
		\node [style=none] (7) at (-0.75, 1.75) {};
		\node [style=none] (8) at (-0.75, 1.5) {};
		\node [style=none] (9) at (1, 1.5) {};
		\node [style=none] (10) at (1, -1.5) {};
		\node [style=none] (11) at (-0.75, -1.5) {};
		\node [style=none] (12) at (-0.75, -1.75) {};
		\node [style=none] (13) at (2, -1.75) {};
		\node [style=none] (14) at (2, 1.75) {};
	\end{pgfonlayer}
	\begin{pgfonlayer}{edgelayer}
		\draw [style=qWire] (2.center) to (3.center);
		\draw [style=qWire] (0.center) to (1.center);
		\draw[fill=black!70] (7.center) to (8.center) to (9.center) to (10.center) to (11.center) to (12.center) to (13.center) to (14.center) to cycle;
	\end{pgfonlayer}
\end{tikzpicture}\,.
\eeq
Such a {linear functional}, $I$, maps some linear map $L:U\to V$ to a real number by:
\beq
\begin{tikzpicture}
	\begin{pgfonlayer}{nodelayer}
		\node [style=none] (0) at (0, -0.5) {};
		\node [style=none] (1) at (0, -1.5) {};
		\node [style=none] (2) at (0, 0.5) {};
		\node [style=none] (3) at (0, 1.5) {};
		\node [style=none] (4) at (1.5, -0) {$\color{white}I$};
		\node [style={right label}] (5) at (0, -1) {$U$};
		\node [style={right label}] (6) at (0, 1) {$V$};
		\node [style=none] (7) at (-0.75, 1.75) {};
		\node [style=none] (8) at (-0.75, 1.5) {};
		\node [style=none] (9) at (1, 1.5) {};
		\node [style=none] (10) at (1, -1.5) {};
		\node [style=none] (11) at (-0.75, -1.5) {};
		\node [style=none] (12) at (-0.75, -1.75) {};
		\node [style=none] (13) at (2, -1.75) {};
		\node [style=none] (14) at (2, 1.75) {};
	\end{pgfonlayer}
	\begin{pgfonlayer}{edgelayer}
		\draw [style=qWire] (2.center) to (3.center);
		\draw [style=qWire] (0.center) to (1.center);
		\draw[fill=black!70] (7.center) to (8.center) to (9.center) to (10.center) to (11.center) to (12.center) to (13.center) to (14.center) to cycle;
	\end{pgfonlayer}
\end{tikzpicture} :: \begin{tikzpicture}
	\begin{pgfonlayer}{nodelayer}
		\node [style=none] (0) at (0, -0.5) {};
		\node [style=none] (1) at (0, -1.5) {};
		\node [style=none] (2) at (0, 0.5) {};
		\node [style=none] (3) at (0, 1.5) {};
		\node [style=small box, fill=black!70] (4) at (0, -0) {$\color{white}L$};
		\node [style={right label}] (5) at (0, -1.25) {$U$};
		\node [style={right label}] (6) at (0, 1) {$V$};
	\end{pgfonlayer}
	\begin{pgfonlayer}{edgelayer}
		\draw [style=qWire] (2.center) to (3.center);
		\draw [style=qWire] (0.center) to (1.center);
	\end{pgfonlayer}
\end{tikzpicture} \mapsto \begin{tikzpicture}
	\begin{pgfonlayer}{nodelayer}
		\node [style=none] (0) at (0, -0.5) {};
		\node [style=none] (1) at (0, -1.5) {};
		\node [style=none] (2) at (0, 0.5) {};
		\node [style=none] (3) at (0, 1.5) {};
		\node [style=none] (4) at (1.5, -0) {$\color{white}I$};
		\node [style={right label}] (5) at (0, -1) {$U$};
		\node [style={right label}] (6) at (0, 1) {$V$};
		\node [style=none] (7) at (-0.75, 1.75) {};
		\node [style=none] (8) at (-0.75, 1.5) {};
		\node [style=none] (9) at (1, 1.5) {};
		\node [style=none] (10) at (1, -1.5) {};
		\node [style=none] (11) at (-0.75, -1.5) {};
		\node [style=none] (12) at (-0.75, -1.75) {};
		\node [style=none] (13) at (2, -1.75) {};
		\node [style=none] (14) at (2, 1.75) {};
		\node [style=small box, fill=black!70] (15) at (0, -0) {$\color{white}L$};
	\end{pgfonlayer}
	\begin{pgfonlayer}{edgelayer}
		\draw [style=qWire] (2.center) to (3.center);
		\draw [style=qWire] (0.center) to (1.center);
		\draw[fill=black!70] (7.center) to (8.center) to (9.center) to (10.center) to (11.center) to (12.center) to (13.center) to (14.center) to cycle;
	\end{pgfonlayer}
\end{tikzpicture}\,.
\eeq
Note that, as $\mathbf{Vect}_\mathds{R}$ is a compact closed category, it can be readily verified that these linear functionals can always be written as:
\beq 
\begin{tikzpicture}
	\begin{pgfonlayer}{nodelayer}
		\node [style=none] (0) at (0, -0.5) {};
		\node [style=none] (1) at (0, -1.5) {};
		\node [style=none] (2) at (0, 0.5) {};
		\node [style=none] (3) at (0, 1.5) {};
		\node [style=none] (4) at (1.5, -0) {$\color{white}I$};
		\node [style={right label}] (5) at (0, -1) {$U$};
		\node [style={right label}] (6) at (0, 1) {$V$};
		\node [style=none] (7) at (-0.75, 1.75) {};
		\node [style=none] (8) at (-0.75, 1.5) {};
		\node [style=none] (9) at (1, 1.5) {};
		\node [style=none] (10) at (1, -1.5) {};
		\node [style=none] (11) at (-0.75, -1.5) {};
		\node [style=none] (12) at (-0.75, -1.75) {};
		\node [style=none] (13) at (2, -1.75) {};
		\node [style=none] (14) at (2, 1.75) {};
	\end{pgfonlayer}
	\begin{pgfonlayer}{edgelayer}
		\draw [style=qWire] (2.center) to (3.center);
		\draw [style=qWire] (0.center) to (1.center);
		\draw[fill=black!70] (7.center) to (8.center) to (9.center) to (10.center) to (11.center) to (12.center) to (13.center) to (14.center) to cycle;
	\end{pgfonlayer}
\end{tikzpicture}
=
\begin{tikzpicture}
	\begin{pgfonlayer}{nodelayer}
		\node [style=none] (0) at (0.25, -0.5) {};
		\node [style=none] (1) at (0.25, -1.5) {};
		\node [style=none] (2) at (0.25, 0.5) {};
		\node [style=none] (3) at (0.25, 1.5) {};
		\node [style=none] (4) at (1, -2) {$\color{white}v_I$};
		\node [style={right label}] (5) at (0.25, -1) {$U$};
		\node [style={right label}] (6) at (0.25, 1) {$V$};
		\node [style=none] (7) at (-0.5, 1.5) {};
		\node [style=none] (8) at (1, 2.75) {};
		\node [style=none] (9) at (2.5, 1.5) {};
		\node [style=none] (10) at (-0.5, -1.5) {};
		\node [style=none] (11) at (1, -2.75) {};
		\node [style=none] (12) at (2.5, -1.5) {};
		\node [style=none] (13) at (1, 2) {$\color{white}c_I$};
		\node [style=none] (14) at (1.75, 1.5) {};
		\node [style=none] (15) at (1.75, -1.5) {};
		\node [style=right label] (16) at (1.75, -0) {$\zeta_I$};
	\end{pgfonlayer}
	\begin{pgfonlayer}{edgelayer}
		\draw [style=qWire] (2.center) to (3.center);
		\draw [style=qWire] (0.center) to (1.center);
		\draw [fill=black!70](7.center) to (8.center) to (9.center) to cycle;
		\draw[fill=black!70] (10.center) to (11.center) to (12.center)to cycle;
		\draw [qWire] (14.center) to (15.center);
	\end{pgfonlayer}
\end{tikzpicture}\,,
\eeq
for some vector space $\zeta_I$, vector $v_I$ and covector $c_I$.

\section{Geometric constraints on state and effect spaces}\label{App:GPTNew}
We define the dual of a set of vectors $\mathcal{V}\subseteq V$ by:
\beq
\mathcal{V}^* := \{w\in V^* | w(v)\in[0,1]\ \forall v\in\mathcal{V}\}\,.
\eeq
If we then denote the set of  states by $\Omega_V$ and the set of effects by $\mathcal{E}_V$ then the constraint on state-effect pairs implies\footnote{Where we identify ${V^*}^*\cong V$} the pair of constraints:
\beq
\mathcal{E}_V\subseteq \Omega_V^* \quad\text{and}\quad \Omega_V\subseteq \mathcal{E}_V^*.
\eeq
That is, the effect space is constrained by the state space and vice versa.

Now, if we consider the special case of bipartite systems $V\otimes W$ then this means that:
\beq\label{eq:bipartiteDuality}
\mathcal{E}_{V\otimes W}\subseteq \Omega_{V\otimes W}^* \quad\text{and}\quad \Omega_{V\otimes W}\subseteq \mathcal{E}_{V\otimes W}^*.
\eeq
Hence, introducing some bipartite effects for the theory (i.e., enlarging $\mathcal{E}_{V\otimes W}$) will induce a constraint on the bipartite state space (since $\mathcal{E}_{V\otimes W}^*$ will potentially be smaller).

This constraint, however, whilst necessary is not sufficient to ensure that we will end up with a valid GPT. Considerations of compositionality and convexity further constrain our state spaces. For example, it follows from compositionality and convexity, that any state of the form:
\beq
\sum_i p_i\ %
\InputIfFileExists{Diagrams/sepState.tikz}{}{\input{./figures/Diagrams/sepState.tikz}},
\eeq
where $s^{(i)}_v \in \Omega_V$ and $s^{(i)}_w\in \Omega_W$, $p_i\in \mathds{R}^+$, and $\sum_i p_i =1$, is a valid state for the composite system. This condition -- that the bipartite state space contains all separable states -- means that $\Omega_V\otimes_{\min}\Omega_W \subseteq \Omega_{V\otimes W}$, where, the so called `min tensor product' is defined as the set of separable states. The same is also true for effects -- compositionality and convexity mean that any effects of the form:
\beq
\sum_j q_j\ %
\InputIfFileExists{Diagrams/sepEffect.tikz}{}{\input{./figures/Diagrams/sepEffect.tikz}},
\eeq
where $e^{(j)}_v \in \mathcal{E}_V$ and $e^{(j)}_w\in \Omega_W$, $q_j\in \mathds{R}^+$, and $\sum_j q_j =1$, is a valid effect for the composite system. This means that $\mathcal{E}_V\otimes_{\min}\mathcal{E}_W \subseteq \mathcal{E}_{V\otimes W}$.

In conjunction with condition Eq.~\eqref{eq:bipartiteDuality}, we can use this to obtain an upper bound on the state space as follows:
\beq \label{eq7}
\Omega_V\otimes_{\min}\Omega_W \subseteq \Omega_{V\otimes W} \subseteq (\mathcal{E}_V\otimes_{\min} \mathcal{E}_W)^* =: \mathcal{E}_V^*\otimes_{\max} \mathcal{E}_W^*.
\eeq
That is, the bipartite state space is bound between the min-tensor product of the local state spaces and the max-tensor of the duals of the local effect spaces. Similarly for the bipartite effect space we obtain:
\beq \label{eq8}
\mathcal{E}_V\otimes_{\min}\mathcal{E}_W \subseteq \mathcal{E}_{V\otimes W} \subseteq (\Omega_V\otimes_{\min} \Omega_W)^* =: \Omega_V^*\otimes_{\max} \Omega_W^*.
\eeq

\section{Tensor representation}\label{sec:tensor}

 Essentially this representation boils down to picking a suitable basis (and dual basis) for each vector space.

We have already seen, via Eq.~\eqref{eq:vecrepst}, how a local state of $V$ can be represented as a $n+1$-dimensional vector. Next we will see how to extend this to arbitrary processes. The simplest way to do so is to introduce a decomposition of the identity into orthogonal rank-$1$ projectors for each system. There are actually only three relevant systems (and their composites) in the above problem, the two classical systems, $\bSet$ and $\nSet$, which decompose as:
\beq
\begin{tikzpicture}
	\begin{pgfonlayer}{nodelayer}
		\node [style=none] (0) at (0, 1) {};
		\node [style=none] (1) at (0, -1) {};
		\node [style=right label] (2) at (0, -0.5000001) {$\bSet$};
	\end{pgfonlayer}
	\begin{pgfonlayer}{edgelayer}
		\draw [cWire](0.center) to (1.center);
	\end{pgfonlayer}
\end{tikzpicture}\quad =\quad \begin{tikzpicture}
	\begin{pgfonlayer}{nodelayer}
		\node [style=copoint] (0) at (0, -0.7500001) {$0$};
		\node [style=none] (1) at (0, -1.5) {};
		\node [style={right label}] (2) at (0, -1.5) {$\bSet$};
		\node [style=point] (3) at (0, 0.7500001) {$0$};
		\node [style=none] (4) at (0, 1.5) {};
		\node [style={right label}] (5) at (0, 1.5) {$\bSet$};
	\end{pgfonlayer}
	\begin{pgfonlayer}{edgelayer}
		\draw [cWire] (0) to (1.center);
		\draw [cWire] (3) to (4.center);
	\end{pgfonlayer}
\end{tikzpicture} \quad +\quad \begin{tikzpicture}
	\begin{pgfonlayer}{nodelayer}
		\node [style=copoint] (0) at (0, -0.7500001) {$1$};
		\node [style=none] (1) at (0, -1.5) {};
		\node [style={right label}] (2) at (0, -1.5) {$\bSet$};
		\node [style=point] (3) at (0, 0.7500001) {$1$};
		\node [style=none] (4) at (0, 1.5) {};
		\node [style={right label}] (5) at (0, 1.5) {$\bSet$};
	\end{pgfonlayer}
	\begin{pgfonlayer}{edgelayer}
		\draw [cWire] (0) to (1.center);
		\draw [cWire] (3) to (4.center);
	\end{pgfonlayer}
\end{tikzpicture}\qquad\text{and}\qquad \begin{tikzpicture}
	\begin{pgfonlayer}{nodelayer}
		\node [style=none] (0) at (0, 1) {};
		\node [style=none] (1) at (0, -1) {};
		\node [style=right label] (2) at (0, -0.5000001) {$\nSet$};
	\end{pgfonlayer}
	\begin{pgfonlayer}{edgelayer}
		\draw [cWire](0.center) to (1.center);
	\end{pgfonlayer}
\end{tikzpicture}\quad =\quad \sum_{i\in\nSet}\ \ \begin{tikzpicture}
	\begin{pgfonlayer}{nodelayer}
		\node [style=copoint] (0) at (0, -0.7500001) {$i$};
		\node [style=none] (1) at (0, -1.5) {};
		\node [style={right label}] (2) at (0, -1.5) {$\nSet$};
		\node [style=point] (3) at (0, 0.7500001) {$i$};
		\node [style=none] (4) at (0, 1.5) {};
		\node [style={right label}] (5) at (0, 1.5) {$\nSet$};
	\end{pgfonlayer}
	\begin{pgfonlayer}{edgelayer}
		\draw [cWire] (0) to (1.center);
		\draw [cWire] (3) to (4.center);
	\end{pgfonlayer}
\end{tikzpicture}\,,
\eeq
and the GPT system $V$ which decomposes as:
\begin{align}
\begin{tikzpicture}
	\begin{pgfonlayer}{nodelayer}
		\node [style=none] (0) at (0, 1) {};
		\node [style=none] (1) at (0, -1) {};
		\node [style=right label] (2) at (0, -0.5000001) {$V$};
	\end{pgfonlayer}
	\begin{pgfonlayer}{edgelayer}
		\draw [qWire](0.center) to (1.center);
	\end{pgfonlayer}
\end{tikzpicture}\quad &=\quad
\sum_{i\in\nSet}\begin{tikzpicture}
	\begin{pgfonlayer}{nodelayer}
		\node [style={right label}] (0) at (0.25, -4.5) {$V$};
		\node [style=none] (1) at (0.25, -2.75) {};
		\node [style=none] (2) at (0.25, -4.75) {};
		\node [style=none] (3) at (-1, -2.75) {};
		\node [style=none] (4) at (0.7500001, -2.75) {};
		\node [style=none] (5) at (0.7500001, -1.75) {};
		\node [style=none] (6) at (-0.7500001, -1.75) {};
		\node [style=none] (7) at (0, -2.25) {$\mathcal{F}$};
		\node [style=none] (8) at (-0.5000001, -2.75) {};
		\node [style=point] (9) at (-0.5000001, -3.5) {$i$};
		\node [style=none] (10) at (0, -1.75) {};
		\node [style=copoint] (11) at (0, -1) {$0$};
		\node [style=point,fill=black!70] (12) at (0, 1) {$\color{white}v_0$};
		\node [style=none] (13) at (0, 2) {};
		\node [style={right label}] (14) at (0, 1.75) {$V$};
	\end{pgfonlayer}
	\begin{pgfonlayer}{edgelayer}
		\draw [qWire] (1.center) to (2.center);
		\draw (3.center) to (4.center);
		\draw (4.center) to (5.center);
		\draw (5.center) to (6.center);
		\draw (6.center) to (3.center);
		\draw [cWire] (11) to (10.center);
		\draw [cWire] (8.center) to (9);
		\draw [qWire] (12) to (13.center);
	\end{pgfonlayer}
\end{tikzpicture}
\quad+\quad
\begin{tikzpicture}
	\begin{pgfonlayer}{nodelayer}
		\node [style={right label}] (0) at (0, -2.75) {$V$};
		\node [style=none] (1) at (0, -1) {};
		\node [style=none] (2) at (0, -3) {};
		\node [style=upground] (3) at (0, -0.75) {};
		\node [style=point,fill=black!70] (4) at (0, 1) {$\color{white}v_2$};
		\node [style=none] (5) at (0, 2) {};
		\node [style={right label}] (6) at (0, 1.75) {$V$};
	\end{pgfonlayer}
	\begin{pgfonlayer}{edgelayer}
		\draw [qWire] (1.center) to (2.center);
		\draw [qWire] (4) to (5.center);
	\end{pgfonlayer}
\end{tikzpicture}
\\
&=\quad
\sum_{i\in\nSet}\ \ \begin{tikzpicture}
	\begin{pgfonlayer}{nodelayer}
		\node [style=copoint] (0) at (0, -0.7500001) {$e_i$};
		\node [style=none] (1) at (0, -1.5) {};
		\node [style={right label}] (2) at (0, -1.5) {$V$};
		\node [style=point,fill=black!70] (3) at (0, 0.7500001) {$\color{white}v_i$};
		\node [style=none] (4) at (0, 1.5) {};
		\node [style={right label}] (5) at (0, 1.5) {$V$};
	\end{pgfonlayer}
	\begin{pgfonlayer}{edgelayer}
		\draw [qWire] (0) to (1.center);
		\draw [qWire] (3) to (4.center);
	\end{pgfonlayer}
\end{tikzpicture}
\quad +\quad  \begin{tikzpicture}
	\begin{pgfonlayer}{nodelayer}
		\node [style=copoint] (0) at (0, -0.7500001) {$e_n$};
		\node [style=none] (1) at (0, -1.5) {};
		\node [style={right label}] (2) at (0, -1.5) {$V$};
		\node [style=point,fill=black!70] (3) at (0, 0.7500001) {$\color{white}v_n$};
		\node [style=none] (4) at (0, 1.5) {};
		\node [style={right label}] (5) at (0, 1.5) {$V$};
	\end{pgfonlayer}
	\begin{pgfonlayer}{edgelayer}
		\draw [qWire] (0) to (1.center);
		\draw [qWire] (3) to (4.center);
	\end{pgfonlayer}
\end{tikzpicture}\,.
\end{align}
The $e_i$ are physically realisable effects, however, the $v_i$ are simply vectors in $V$ which satisfy $e_i(v_j)=\delta_{ij}$. It is important that we do not demand that the $v_i$ are physically realisable states, as, for any non-classical GPT, there are insufficient perfectly distinguishable states to span the vector space.

A remark on notation: in the following sections we will also denote the unit effect for the $\bSet$ systems as $u_{\bSet}$, and their fiducial effects by $\vec{0}$ and $\vec{1}$.

Now, to obtain a tensorial representation of any diagram we simply decompose all of the internal identities in the diagram and attach $e_i$ and $v_j$ to the free inputs and outputs, for example:
\begin{align}
\begin{tikzpicture}
	\begin{pgfonlayer}{nodelayer}
		\node [style=none] (0) at (2, 2) {};
		\node [style=none] (1) at (1, -1.25) {};
		\node [style={right label}] (2) at (2, 1.5) {$V$};
		\node [style={right label}] (3) at (-1, -1) {$V$};
		\node [style=none] (4) at (0, -1.75) {$s$};
		\node [style=none] (5) at (1.75, -1.25) {};
		\node [style=none] (6) at (-1.75, -1.25) {};
		\node [style=none] (7) at (0, -2.5) {};
		\node [style=none] (8) at (-1, -0) {};
		\node [style=none] (9) at (-1, -1.25) {};
		\node [style=none] (10) at (-0.5, -0) {};
		\node [style=none] (11) at (-1, 1) {};
		\node [style=none] (12) at (-2.25, 1) {};
		\node [style=none] (13) at (-2.25, -0) {};
		\node [style=none] (14) at (-1.5, 0.5) {$\mathcal{F}$};
		\node [style=none] (15) at (-1.75, -0) {};
		\node [style=none] (16) at (-3, -2.75) {};
		\node [style=none] (17) at (-1.75, 1) {};
		\node [style=none] (18) at (-2.5, 2.25) {};
		\node [style={right label}] (19) at (-3, -2.5) {$\nSet$};
		\node [style={right label}] (20) at (-2.25, 1.75) {$\bSet$};
	\end{pgfonlayer}
	\begin{pgfonlayer}{edgelayer}
		\draw [qWire, in=-90, out=90, looseness=1.00] (1.center) to (0.center);
		\draw (5.center) to (7.center);
		\draw (7.center) to (6.center);
		\draw (6.center) to (5.center);
		\draw [qWire] (8.center) to (9.center);
		\draw (13.center) to (10.center);
		\draw (10.center) to (11.center);
		\draw (11.center) to (12.center);
		\draw (12.center) to (13.center);
		\draw [cWire, in=90, out=-90, looseness=1.00] (15.center) to (16.center);
		\draw [cWire, in=-90, out=90, looseness=1.00] (17.center) to (18.center);
	\end{pgfonlayer}
\end{tikzpicture}
\qquad &\mapsto \qquad
\left[\sum_l\begin{tikzpicture}
	\begin{pgfonlayer}{nodelayer}
		\node [style=copoint] (0) at (2, 3) {$e_k$};
		\node [style=none] (1) at (1.25, -2.25) {};
		\node [style={right label}] (2) at (2, 2.25) {$V$};
		\node [style={right label}] (3) at (-0.7500001, -2) {$V$};
		\node [style=none] (4) at (0.25, -2.75) {$s$};
		\node [style=none] (5) at (2, -2.25) {};
		\node [style=none] (6) at (-1.5, -2.25) {};
		\node [style=none] (7) at (0.25, -3.5) {};
		\node [style=copoint] (8) at (-0.7500001, -1.25) {$e_l$};
		\node [style=none] (9) at (-0.7500001, -2.25) {};
		\node [style=none] (10) at (-0.25, 1) {};
		\node [style=none] (11) at (-0.7500001, 2) {};
		\node [style=none] (12) at (-2, 2) {};
		\node [style=none] (13) at (-2, 1) {};
		\node [style=none] (14) at (-1.25, 1.5) {$\mathcal{F}$};
		\node [style=none] (15) at (-1.5, 1) {};
		\node [style=point] (16) at (-3, -3.5) {$j$};
		\node [style=none] (17) at (-1.5, 2) {};
		\node [style=copoint] (18) at (-2.25, 3.25) {$i$};
		\node [style={right label}] (19) at (-3, -3) {$\nSet$};
		\node [style={right label}] (20) at (-2, 2.5) {$\bSet$};
		\node [style=none] (21) at (-0.7500001, 1) {};
		\node [style=point,fill=black!70] (22) at (-0.7500001, -0) {$\color{white}v_l$};
		\node [style={right label}] (23) at (-0.7500001, 0.500001) {$V$};
	\end{pgfonlayer}
	\begin{pgfonlayer}{edgelayer}
		\draw [qWire, in=-90, out=90, looseness=1.00] (1.center) to (0);
		\draw (5.center) to (7.center);
		\draw (7.center) to (6.center);
		\draw (6.center) to (5.center);
		\draw [qWire] (8) to (9.center);
		\draw (13.center) to (10.center);
		\draw (10.center) to (11.center);
		\draw (11.center) to (12.center);
		\draw (12.center) to (13.center);
		\draw [cWire, in=90, out=-90, looseness=1.00] (15.center) to (16);
		\draw [cWire, in=-90, out=90, looseness=1.00] (17.center) to (18);
		\draw [qWire] (22) to (21.center);
	\end{pgfonlayer}
\end{tikzpicture}\right]_{j}^{ik}\\
\qquad &= \quad \sum_l \left[\begin{tikzpicture}
	\begin{pgfonlayer}{nodelayer}
		\node [style=none] (0) at (1, -0) {};
		\node [style=none] (1) at (0.5000002, 0.9999999) {};
		\node [style=none] (2) at (-0.7499998, 0.9999999) {};
		\node [style=none] (3) at (-0.7499998, -0) {};
		\node [style=none] (4) at (0, 0.5000002) {$\mathcal{F}$};
		\node [style=none] (5) at (-0.2499996, -0) {};
		\node [style=point] (6) at (-0.7499998, -2) {$j$};
		\node [style=none] (7) at (-0.2499996, 0.9999999) {};
		\node [style=copoint] (8) at (-1, 2.25) {$i$};
		\node [style={right label}] (9) at (-0.5000002, -1.25) {$\nSet$};
		\node [style={right label}] (10) at (-0.5000002, 1.5) {$\bSet$};
		\node [style=none] (11) at (0.5000002, -0) {};
		\node [style=point,fill=black!70] (12) at (0.5000002, -1.25) {$\color{white}v_l$};
		\node [style={right label}] (13) at (0.5000002, -0.5000002) {$V$};
	\end{pgfonlayer}
	\begin{pgfonlayer}{edgelayer}
		\draw (3.center) to (0.center);
		\draw (0.center) to (1.center);
		\draw (1.center) to (2.center);
		\draw (2.center) to (3.center);
		\draw [cWire, in=90, out=-90, looseness=1.00] (5.center) to (6);
		\draw [cWire, in=-90, out=90, looseness=1.00] (7.center) to (8);
		\draw [qWire] (12) to (11.center);
	\end{pgfonlayer}
\end{tikzpicture}\right]_{jl}^i\left[\begin{tikzpicture}
	\begin{pgfonlayer}{nodelayer}
		\node [style=copoint] (0) at (1.25, 0.9999999) {$e_k$};
		\node [style=none] (1) at (1, -0.5000002) {};
		\node [style={right label}] (2) at (1.25, 0.2500001) {$V$};
		\node [style={right label}] (3) at (-1, -0.2500001) {$V$};
		\node [style=none] (4) at (0, -0.9999999) {$s$};
		\node [style=none] (5) at (1.749999, -0.5000002) {};
		\node [style=none] (6) at (-1.75, -0.5000002) {};
		\node [style=none] (7) at (0, -1.75) {};
		\node [style=copoint] (8) at (-1, 0.5000002) {$e_l$};
		\node [style=none] (9) at (-1, -0.5000002) {};
	\end{pgfonlayer}
	\begin{pgfonlayer}{edgelayer}
		\draw [qWire, in=-90, out=90, looseness=1.00] (1.center) to (0);
		\draw (5.center) to (7.center);
		\draw (7.center) to (6.center);
		\draw (6.center) to (5.center);
		\draw [qWire] (8) to (9.center);
	\end{pgfonlayer}
\end{tikzpicture}\right]^{lk}\\
\qquad &=: \quad \sum_l \mathcal{F}_{jl}^i s^{lk}
\end{align}
{A bipartite state, such as $s$ in the above diagram, is therefore represented by a two-index tensor. If this bipartite state is a product state, then it is easy to see that this two-index tensor is simply the Kronecker product of the one-index tensors associated to the two components:
\beq
(s_1\otimes s_2)^{ij}=\left[\begin{tikzpicture}
	\begin{pgfonlayer}{nodelayer}
		\node [style=point] (0) at (-0.75, -0.5) {$s_1$};
		\node [style=copoint] (1) at (-0.75, 0.5) {$e_i$};
		\node [style=point] (2) at (0.75, -0.5) {$s_2$};
		\node [style=copoint] (3) at (0.75, 0.5) {$e_j$};
	\end{pgfonlayer}
	\begin{pgfonlayer}{edgelayer}
		\draw [qWire] (0) to (1);
		\draw [qWire] (2) to (3);
	\end{pgfonlayer}
\end{tikzpicture}\right]^{ij} = \left[\begin{tikzpicture}
	\begin{pgfonlayer}{nodelayer}
		\node [style=point] (0) at (0, -0.5) {$s_1$};
		\node [style=copoint] (1) at (0, 0.5) {$e_i$};
	\end{pgfonlayer}
	\begin{pgfonlayer}{edgelayer}
		\draw [qWire] (0) to (1);
	\end{pgfonlayer}
\end{tikzpicture}\right]^i\left[\begin{tikzpicture}
	\begin{pgfonlayer}{nodelayer}
		\node [style=point] (0) at (0.75, -0.5) {$s_2$};
		\node [style=copoint] (1) at (0.75, 0.5) {$e_j$};
	\end{pgfonlayer}
	\begin{pgfonlayer}{edgelayer}
		\draw [qWire] (0) to (1);
	\end{pgfonlayer}
\end{tikzpicture}\right]^j = s_1^i s_2^j\ .
\eeq}

\subsection{Example}

 Consider the case where $\bSet = \{0,1\}$, $\nSet = \{0,1\}$, and $|\iSet|=2$. Moreover, take the dimension of the local GPT system to be 3. 

Define $\mathcal{F}^{a}_{xv}$ real tensors where $a,x\in \{0,1\}$ $v\in\{0,1,2\}$ as follows:

\begin{equation}
\label{eq:F-table}
\mathcal{F}^a_{xv} :=
\begin{array}{c | c c c}
 & v=0 & v=1 & v=2  \\ \hline
a=0, x=0 & 1 &	0 & 0 \\	
a=1, x=0 &	-1 & 0	&	1\\
a=0, x=1 & 0	& 1	& 0 \\	
a=1, x=1 &	0 & -1	&	1 
\end{array}
\end{equation}

We then define our variables $s_{vw}$, $\mathcal{P}_{vw}^{pq}$ as real tensors with $v,w \in\{0,1,2\}$ $p,q \in \{0,1\}$.
The tensors $\mathcal{F}$, $\mathcal{P}$ and $s$ will determine local fiducial measurements,  parity reading measurement, and state, respectively.

Given a particular linear functional $I_{ab}^{xy}$ we optimise:
\begin{equation}
\beta := \mathsf{sup}_{s,\mathcal{P}} \ \  \sum _{abxy} I_{ab}^{xy}  \sum_{v_1,v_2} \mathcal{F}_{x v_1}^{a} \mathcal{F}_{y v_2}^{b} s_{v_{1} v_{2}}
\end{equation}
subject to the following constraints. Note that in these constraints any sum is implicitly taken over its whole range and there is an implict $\forall$ for any index which is not contracted: 
\begin{compactitem} 
\item Parity Reading:
\begin{align}
\sum_{p} \mathcal{P}^{pq}_{vw} = \mathcal{F}^0_{1v}\mathcal{F}^{q}_{1w} + \mathcal{F}^{1}_{1v}\mathcal{F}^{1\oplus q}_{1w} \\
\sum_{q} \mathcal{P}^{pq}_{vw} = \mathcal{F}^0_{0v}\mathcal{F}^{p}_{0w} + \mathcal{F}^{1}_{0v}\mathcal{F}^{1\oplus p}_{0w}
\end{align}

\item Probabilities for fiducial measurements:
\begin{align}
\label{eq:prob1}
\sum_{v_1,v_2} \mathcal{F}_{x v_1}^{a} \mathcal{F}_{y v_2}^{b} s_{v_{1} v_{2}} &\geq 0 \\
\label{eq:prob2}
1= \sum_{a,b,v_1,v_2} \mathcal{F}_{x v_1}^{a} \mathcal{F}_{y v_2}^{b} s_{v_{1} v_{2}}
\end{align}

\item Hierarchy L1:
\begin{equation}
\sum_{v_1,v_2} \mathcal{P}_{v_1,v_2}^{p_1p_2}  s_{v_{\pi(1)} v_{\pi(2)}} \geq 0 \end{equation}
for all $\pi$ where $\pi$ is a permutation of $\{1,2\}$.

\item Hierarchy L2:
\begin{align}
\sum_{v_1,v_2,v_3,v_4} \mathcal{P}_{v_1,v_2}^{p_1p_2} \mathcal{P}_{v_3,v_4}^{p_3p_4} s_{v_{\pi(1)} v_{\pi(2)}}s_{v_{\pi(3)} v_{\pi(4)}} &\geq 0 \\
\sum_{v_1,v_2,v_3,v_4} \mathcal{P}_{v_1,v_2}^{p_1p_2} \mathcal{F}_{x v_3}^{a} \mathcal{F}_{y v_3}^{b} s_{v_{\pi(1)} v_{\pi(2)}}s_{v_{\pi(3)} v_{\pi(4)}} &\geq 0
\end{align}
for all $\pi$ where now $\pi$ is a permutation of $\{1,2,3,4\}$

\item Hierarchy L3:
\begin{align}
\sum_{v_1,v_2,v_3,v_4} \mathcal{P}_{v_1,v_2}^{p_1p_2} \mathcal{P}_{v_3,v_4}^{p_3p_4} \mathcal{P}_{v_5,v_6}^{p_5p_6} s_{v_{\pi(1)} v_{\pi(2)}}s_{v_{\pi(3)} v_{\pi(4)}}s_{v_{\pi(5)} v_{\pi(6)}} &\geq 0 \\
\sum_{v_1,v_2,v_3,v_4,v_5,v_6} \mathcal{P}_{v_1,v_2}^{p_1p_2} \mathcal{P}_{v_3,v_4}^{p_3p_4} \mathcal{F}_{x v_5}^{a} \mathcal{F}_{y v_6}^{b} s_{v_{\pi(1)} v_{\pi(2)}}s_{v_{\pi(3)} v_{\pi(4)}} s_{v_{\pi(5)} v_{\pi(6)}} &\geq 0 \\
\end{align}
for all $\pi$ where now $\pi$ is a permutation of $\{1,2,3,4,5,6\}$.
\end{compactitem}

\bigskip

Notice that the constraints  \eqref{eq:prob1} and \eqref{eq:prob2} just say that the    
\begin{align}\label{eq:box}
    p(ab|xy)=
\sum_{v_1,v_2} \mathcal{F}_{x v_1}^{a} \mathcal{F}_{y v_2}^{b} s_{v_{1} v_{2}}
\end{align} are probability distributions
for each fixed $x,y$. 

Also, the form of tensor $\mathcal{F}$ of Eq.~\eqref{eq:F-table} further says that  the 
parametrization state  
\begin{align}
    s=(s_{00},s_{10},s_{2,0},s_{01},s_{11},s_{21}, s_{02},s_{12},s_{22})
\end{align}
has the following interpretation in terms of the box  $p(ab|xy)$:
\begin{align}
\label{eq:LTparam-ex1-2}
    s = \left( p_s(00|00)\,,\, p_s(00|10)\,,\, p^\mathrm{(2)}_s(0|0) \,,\, p_s(00|01)\,,\, p_s(00|11)\,,\, p^\mathrm{(2)}_s(0|1) \,,\, p^\mathrm{(1)}_s(0|0)\,,\, p^\mathrm{(1)}_s(0|1)\,,\, 1 \right).
\end{align}
Here $p^{(1)}$ and 
$p^{(2)}$ are marginals obtained from $p(ab|xy)$, e.g., $p^{(1)}(a|x)=\sum_b p(ab|xy)$.  Note that the latter  does not depend on $y$ due to no-signaling of $p(ab|xy)$, which in turn is  enforced by the form of Eq.~\eqref{eq:box} 
and the definition of $\mathcal{F}$  given in Eq.~\eqref{eq:F-table}.

\section{Convergence of state cone hierarchy}
\label{App:Convergence}

In this appendix we demonstrate that the hierarchy that we define does indeed converge to the cone $K[\mathcal{P}]$.

The cone $K[\mathcal{P}]$ is characterised by the condition: $s\in K[\mathcal{P}]$ if and only if every diagram with only classical inputs and outputs formed from a finite number of processes must be non-negative.

We now show that this condition is equivalent to our hierarchy.

To begin with, note that any diagram in our theory is constructed by wiring together a finite number of each of: i) the bipartite state, $s$, ii) the controlled fiducial measurement, $\mathcal{F}$, and iii) the parity reading measurement, $\mathcal{P}$. We call these the generating processes. We can therefore classify diagrams by first representing them in terms of the generating processes, and then counting the number of copies, $k$, of $s$ that appear.

Next, note that if a diagram has only classical inputs and outputs, then any copy of $V$ that appears in the diagram must have a start point and an end point in the diagram. There is only one generating process which can serve as a start point, namely, the bipartite state $s$, and either $\mathcal{F}$ or $\mathcal{P}$ can serve as the end point. 

If we have $k$ copies of the state $s$ within the diagram, then these must therefore be wired into the measurements $\mathcal{F}$ and $\mathcal{P}$. Every such diagram will factorise into totally connected subdiagrams. Note then, that  nonnegativity of the full diagram is guaranteed by nonnegativity of the component subdiagrams. That is, to ensure nonnegativity for every diagram (with only classical inputs and outputs) we must only demand nonnegativity of totally connected diagrams (with only classical inputs and outputs). 

It is then simple to see that the totally connected diagrams with $k$ copies of $s$ come in two forms. Firstly, those in which there are $k$ copies of $\mathcal{P}$ which the states $s$ are wired to, and secondly, those in which there are $k-1$ copies of $\mathcal{P}$ and two copies of $\mathcal{F}$. If there were more than two copies of $\mathcal{F}$ then the diagram would necessarily not be totally connected. Clearly, the first of these is captured by condition \eqref{eq:hie2} and the second by the condition \eqref{eq:hie1} of level $k$ in the hierarchy. Therefore, our hierarchy of constraints fully charaterises the cone $K[\mathcal{P}]$.

\end{document}